\documentclass[fleqn,usenatbib]{mnras}
\usepackage{newtxtext,newtxmath}
\DeclareRobustCommand{\VAN}[3]{#2}
\let\VANthebibliography\thebibliography
\def\thebibliography{\DeclareRobustCommand{\VAN}[3]{##3}\VANthebibliography}

\usepackage{adjustbox}
\usepackage{graphicx}
\usepackage[normalem]{ulem}
\usepackage{float}

\title[Low-luminosity compact RLAGN]{The nature of compact radio-loud AGN: a systematic look at the LOFAR AGN population}
\author[J. Chilufya et al.]{
J. Chilufya,$^{1}$
\thanks{E-mail: \href{mailto:j.chilufya@herts.ac.uk}{j.chilufya@herts.ac.uk} (JC)}
M. J. Hardcastle,$^{1}$
J. C. S. Pierce,$^{1}$
J.H. Croston,$^{2}$
B. Mingo,$^{2}$
X. Zheng,$^{3,5}$
R. D. Baldi,$^{4}$
\newauthor and H. J. A. R\"ottgering$^{3}$\\
$^{1}$Centre for Astrophysics Research, Department of Physics, Astronomy and Mathematics, University of Hertfordshire, Hatfield AL10 9AB, UK\\
$^{2}$School of Physical Sciences, The Open University, Walton Hall, Milton Keynes, MK7 6AA, UK\\
$^{3}$Leiden Observatory, Leiden University, PO Box 9513, 2300 RA Leiden, The Netherlands\\
$^{4}$INAF - Instituto di Radioastronomia, via Gobetti 101 40129 Bologna\\
{$^{5}$Key Laboratory for Research in Galaxies and Cosmology, Shanghai Astronomical Observatory, Chinese Academy of Sciences, 80 Nandan Road,}\\
Shanghai 200030, China\\
}
\date{Accepted 2024 February 26. Received 2024 January 29; in original form 2023 November 06}
\pubyear{2023}
\begin{document}
\label{firstpage}
\pagerange{\pageref{firstpage}--\pageref{lastpage}}
\maketitle

%%%Abstract
\begin{abstract}
We investigate the nature of low-luminosity radio-loud active galactic nuclei (RLAGN) selected from the LOFAR Two-metre Sky Survey (LoTSS) first data release (DR1). Using optical, mid-infrared, and radio data, we have conservatively selected 55 radiative AGN candidates from DR1 within the redshift range $0.03<z<0.1$. We show using high-frequency {\it Karl G. Jansky} Very Large Array (VLA) observations that 10 out of 55 objects show radio emission on scales $>$$1-3$ kpc, 42 are compact at the limiting resolution of 0.35 arcsec (taking an upper limit on the projected physical size, this corresponds to less than 1 kpc), and three are undetected. The extended objects display a wide range of radio morphologies: two-jet (5), one-jet (4), and double-lobed (1). We present the radio spectra of all detected radio sources which range from steep to flat/inverted and span the range seen for other compact radio sources such as compact symmetric objects (CSOs), compact steep spectrum (CSS) sources, and gigahertz peaked-spectrum (GPS) sources. Assuming synchrotron self-absorption (SSA) for flat/inverted radio spectrum sources, we predict small physical sizes for compact objects to range between $2-53$ pc. Alternatively, using free-free absorption (FFA) models, we have estimated the free electron column depth for all compact objects, assuming a homogeneous absorber. We find that these objects do not occupy a special position on the power/linear size ($P-D$) diagram but some share a region with radio-quiet quasars (RQQs) and so-called `FR0' sources in terms of radio luminosity and linear size.
\end{abstract}
%%%%%
%%%%%BODY OF PAPER
\section{Introduction}
\label{sec:intro}

\defcitealias{2019A&A...622A..12H}{H19}
\defcitealias{2016AN....337..114B}{B16/}
\defcitealias{2019MNRAS.482.2294B}{19}
\defcitealias{2020MNRAS.494.2053P}{P20}
%%%% 
Theoretical studies have proposed the existence of two main types of active galactic nuclei (AGN) feedback, the radiative (or quasar) mode and the kinetic (or jet) mode \citep[e.g.][]{2006MNRAS.365...11C,2007ARA&A..45..117M,2008A&A...491..407N,2012MNRAS.424.1774H, 2012ARA&A..50..455F,2015ARA&A..53...51S,2020NewAR..8801539H}. 

The quasar mode is often implemented in simulations \citep[e.g.][]{2012MNRAS.425..605F,2015ARA&A..53...51S,2018MNRAS.479.2079C} and is associated with periods of intense accretion unto a supermassive black hole (SMBH), leading to the formation of a quasar. The intense radiation emitted by quasars \citep[i.e. radiatively efficient AGN; e.g.][]{2020NewAR..8801539H} can heat and ionise the surrounding gas, creating strong winds of hot gas that can sweep material out of the host galaxy hence regulating the growth of the SMBH itself, and influencing star formation rates \citep[SFRs; see e.g.][]{2017NatAs...1E.165H}.

The jet mode, often associated with radiatively inefficient AGN \citep[e.g.][]{2007MNRAS.376.1849H}, is well supported by observational evidence \citep[e.g.][]{2006MNRAS.368L..67B,2012ARA&A..50..455F,2012MNRAS.424.1774H,2020NewAR..8801539H} which makes it the preferred feedback mechanism in the local Universe. The large-scale emission (jets and sometimes lobes) from powerful radio-loud AGN (RLAGN; blazars, radio galaxies, and radio quasars) often lead to heating of the surrounding gas, preventing cooling flows in galaxy clusters. They can also regulate star formation (SF) in galaxies by heating and disrupting the gas required for SF \citep[e.g.][]{2012ARA&A..50..455F}.

These feedback processes regulate SF, which has a direct consequence on the growth of massive galaxies. However, the extent to which these jets affect the interstellar medium (ISM) of the host galaxy, and how much they affect SFRs is as yet poorly understood. A better understanding of AGN feedback requires the study of RLAGN populations at both high and low luminosities and from the local Universe out to high redshifts.

Recently, an interest in studying physically small (less than 100 kpc) RLAGN and their impact on their surrounding environments has emerged \citep[e.g.][]{2018MNRAS.475.3493B,2019MNRAS.485.2710J,2021MNRAS.503.4627U}. These objects range from powerful yet compact RLAGN such as compact symmetric objects (CSOs), compact steep spectrum (CSS) sources, and gigahertz peaked-spectrum (GPS) sources \citep[e.g][]{1990A&AS...84..549O,1990A&A...231..333F,1998PASP..110..493O,2000MNRAS.319..445S} to low-radio-power yet resolved galaxy scale jets \citep[GSJs; e.g.][]{2021MNRAS.500.4921W,2021MNRAS.508.5972W}, and low-radio-power yet unresolved or slightly resolved RLAGN, the so-called `FR0s' \citep[e.g.][]{2010A&A...519A..48B,2014MNRAS.438..796S,2015A&A...576A..38B}.

In the radio regime, the identification criteria for these compact objects depends on physical size. Thanks to large radio sky surveys such as the LOw Frequency ARray \citep[LOFAR;][]{2013A&A...556A...2V} Two-metre Sky Survey \citep[LoTSS;][]{2017A&A...598A.104S,2019A&A...622A...1S,2022A&A...659A...1S}, we can detect them in large numbers at low frequencies. LoTSS, currently the largest radio survey ever conducted, offers high sensitivity and high-resolution imaging of extragalactic sources which include star-forming galaxies (SFGs) and RLAGN. 

For radio sources where the observed spectral index is typically $\alpha\approx0.7$ (where $S_v\propto v^{-\alpha}$), the LoTSS survey is about ten times deeper than previous radio surveys, e.g. the Faint Images of the Radio Sky at Twenty centimetres \citep[FIRST;][]{1995ApJ...450..559B} and the NRAO VLA Sky Survey \citep[NVSS;][]{1998AJ....115.1693C}. The combination of short and long baselines also makes LOFAR sensitive to both compact and extended sources in the radio sky. However, as has been shown in previous studies \citep[e.g.][]{2016A&ARv..24...13P}, deep radio surveys such as LoTSS introduce new challenges, particularly at low frequencies (e.g. 150 MHz). Some of these include (1) separating SFGs from RLAGN \citep[mostly because the 150 MHz radio luminosities of SFGs and RLAGN can overlap;][]{1992ARA&A..30..575C} and (2) understanding the nature of the emerging population of low-luminosity compact RLAGN that numerically dominates the local Universe. 

To address the first point, several AGN-SF separation techniques have been developed. The best AGN-SF separation method would make use of the radio excess of RLAGN over the emission expected from the measured SFR \citep[e.g.][]{2016MNRAS.462.1910H}. However, where reliable SFR information is missing, other AGN-SF separation methods can be employed and we summarise some of these below.

The first and most common AGN-SF separation model was developed by \citet{1981PASP...93....5B}. In this method, extragalactic sources are separated using their optical emission-line properties. However, this method requires sources to have significant emission line luminosity, and not all low-luminosity RLAGN such as low excitation radio galaxies (LERGs) would show as AGN on such diagrams. In the second method, proposed by \citet{1999ApJS..123...41M}, the radio and far-infrared luminosities of sources are compared and AGN sources are those that are more radio-luminous than the tight far-infrared radio correlation (FIRC) of SFGs. The third method uses the comparison of the ratio of the radio power per unit of stellar mass against the 4000$\mathrm{\mathring{A}}$ break strength \citep{2005MNRAS.362...25B}. \citet{2008MNRAS.384..953K} further built upon this method by considering the ratio of the radio luminosity against the H$\alpha$ luminosity. Using these models, \citet{2019A&A...622A..17S} modified and extended the works of \citet{2012MNRAS.421.1569B} resulting in the WISE \citep[Wide-field Infrared Survey Explorer;][]{2010AJ....140.1868W} colour-colour method. This proposes that SFGs separate from hosts of RLAGN in their WISE colours in particular, in W2-W3 \citep[e.g.][]{2010AJ....140.1868W,2013AJ....145...55Y}. 

These AGN-SF separation methods have limitations and selection effects which can be reduced by applying as many methods as possible and then cross-matching them to obtain a purer sample of AGN; we return to this point in Sec.~\ref{sec:subsample}. 

To address the second challenge, which is understanding the nature of an emerging large population of faint (and physically small) RLAGN in the local Universe, \citet{2019A&A...622A..12H} (hereafter \citetalias{2019A&A...622A..12H}), constructed a sample of 23 344 RLAGN using spectroscopic and photometric techniques based on LoTSS DR1, extending the work of \citet{2005MNRAS.362...25B} to the much lower radio luminosity levels (greater than 10$\rm{^{21}~W~Hz^{-1}}$) detectable at 150 MHz. Using analytical models \citep{2018MNRAS.475.2768H}, \citetalias{2019A&A...622A..12H} used the RLAGN sample to construct a radio-power/linear size \citep[$P-D$;][]{1982IAUS...97...21B} diagram in which they showed that while the distribution of the projected physical sizes at the more luminous end of the AGN population is consistent with a uniform lifetime function distribution (distribution of possible jet active times), the class of low-luminosity RLAGN (which are numerically dominant and many of which have very small physical sizes less than 100 kpc) may require a different model from the more powerful and better-understood large, resolved objects. 

This class of low-luminosity objects have previously been studied \citep[e.g.][]{2010A&A...519A..48B,2011AIPC.1381..180G,2014MNRAS.438..796S}, and a model in which low-luminosity objects are the young counterparts of the more powerful sources is rejected \citep[e.g.][]{2016AN....337..114B} unless most objects at these luminosities do not survive to old age. A plausible speculation however is that the low-luminosity objects could be the low-jet-power continuation of the more luminous and larger RLAGN and if this is the case, we may be underestimating the input energy that they may be injecting into massive galaxies. These physically small objects that are compact on galaxy scales could play a crucial role in AGN feedback, but they are still poorly understood. 

The short baselines provided by LOFAR give us an added advantage in selecting these objects in a simple and homogeneous way without missing significant extended emission and we can robustly remove SF sources thanks to high-quality ancillary data. We aim to investigate the nature of these ubiquitous low-luminosity compact RLAGN numerically dominating the local Universe. Are these low-luminosity RLAGN scaled-down versions of the better understood AGN populations that we see at high luminosities with double lobes inflating bubbles in the ISM, or are they morphologically different? On what scales do they affect the ISM of the host galaxy?

This paper aims to answer these questions using a set of new VLA observations of low-luminosity RLAGN selected from DR1 and high-quality DR2 images \citep{2022A&A...659A...1S}. The rest of the paper is structured as follows: the LOFAR sub-sample construction, their VLA observations, and data reduction are presented in Sec.~\ref{sec:obs}, while Sec.~\ref{sec:analysis} presents the analysis and results of our findings. A discussion is given in Sec.~\ref{sec:disc} and drawn conclusions are presented in Sec.~\ref{sec:conclusion}. Throughout this paper, we have assumed a $\Lambda$CDM cosmological model with cosmological parameters: $H_0=70~\rm{km~s^{-1}~Mpc^{-1}}$; $\Omega_{\rm{m0}}=0.3$; $\Omega_{\Lambda0}=0.7$. The definition of the radio spectrum is taken to be $S_v\propto v^{-\alpha}$.

\section{Sample selection, observations, and data reduction}
\label{sec:obs}
\subsection{LOFAR sample}
\label{sec:subsample}
We are using radio data from the LOw Frequency Array (LOFAR) Two-metre Sky Survey (LoTSS) first data release \cite[DR1;][]{2019A&A...622A...3D,2019A&A...622A...1S,2019A&A...622A...2W} which consists of 318 520 radio sources that include SFG and RLAGN. LoTSS DR1 consists of 424 deg$^2$ of data in the region of the Hobby-Eberly Telescope Dark Energy Experiment \citep[HETDEX;][]{2008ASPC..399..115H} Spring field and lies between RA $\rm{=10^h45^m00^s}$ to $\rm{15^h30^m00^s}$ and Dec $\rm{=45^o00^{\prime}00^{\prime\prime}}$ to $\rm{57^o00^{\prime}00^{\prime\prime}}$. The limiting resolution of LoTSS is 6 arcsec, with a median sensitivity of around 70$\rm{~\mu Jy~beam^{-1}}$ at a central frequency of 150 MHz.

\begin{figure}
    \includegraphics[width=\columnwidth]{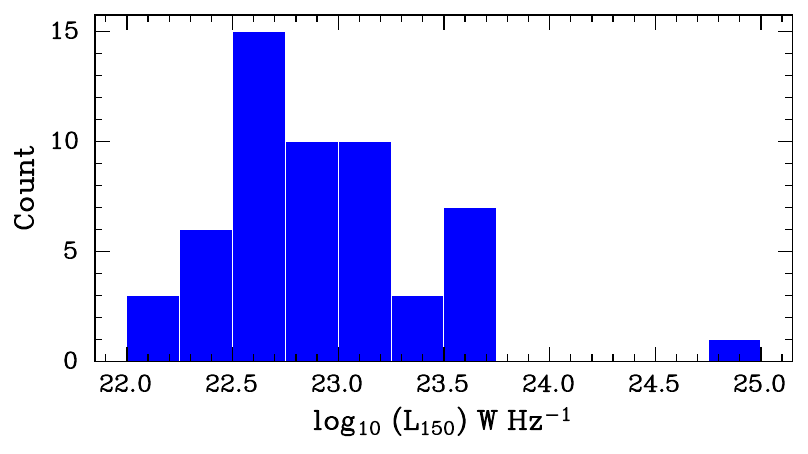}
    
    \includegraphics[width=\columnwidth]{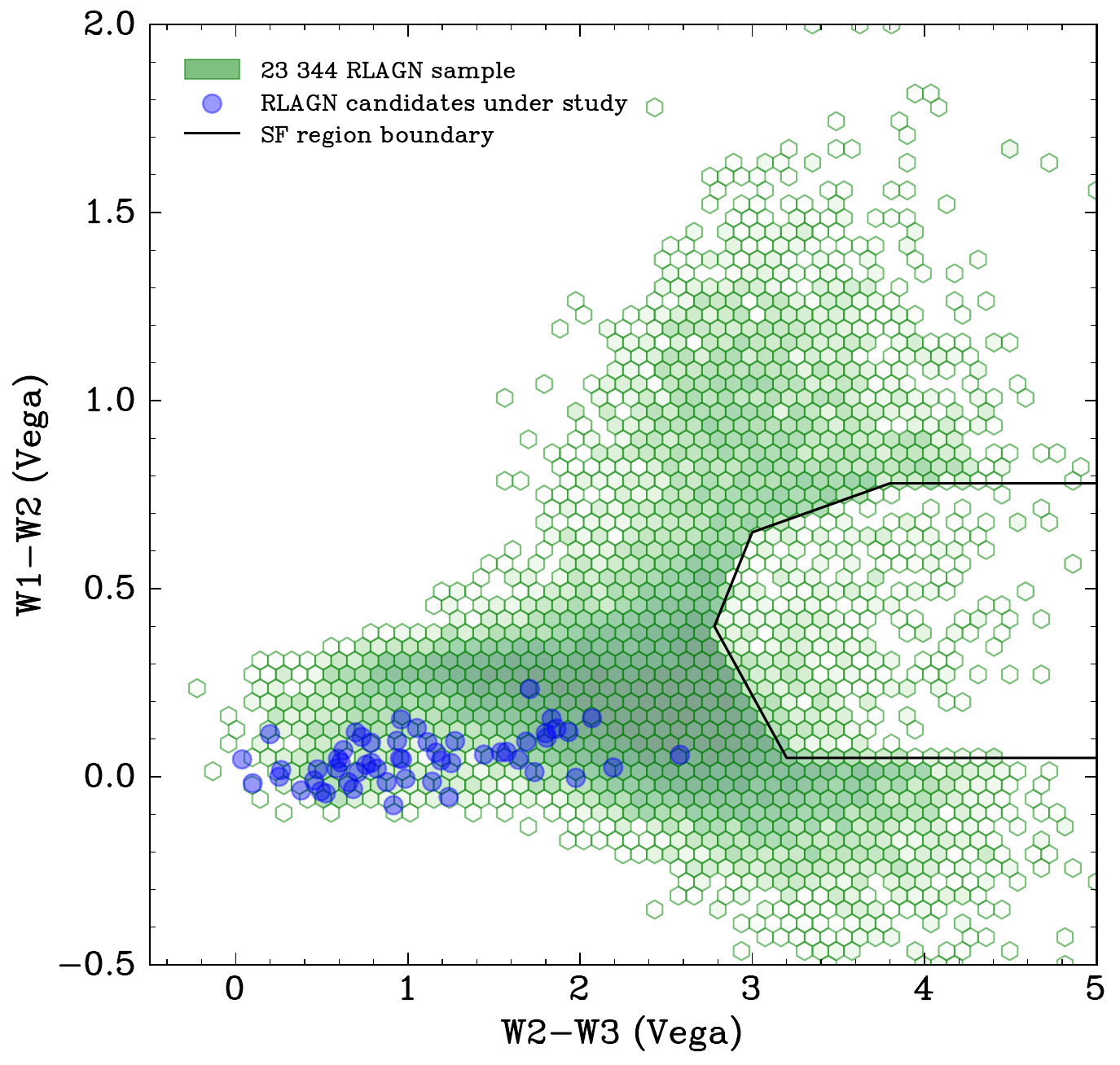}

    \includegraphics[width=\columnwidth]{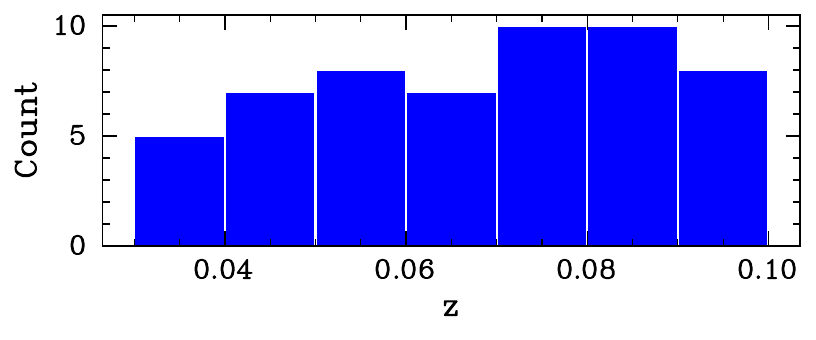}
    
    \caption{{\it Top panel:} distribution of radio luminosities for our candidates. {\it Middle panel:} green points show the final RLAGN sample (23 344 objects) constructed by \citet{2019A&A...622A..12H} and based on WISE colour AGN-SF separation techniques. This sample is overlaid with the position of the RLAGN sub-sample (in blue; 55 candidates) under study. The line (black) boundary indicates the locus of points where SF objects dominate in WISE colours. WISE AB magnitudes have been converted to Vega magnitudes. Note that most of these sources have an upper limit placed on their WISE W3 measurement and we expect them to be located further to the left-side than shown. Nonetheless, this does not affect the selection of these sources considering the fact that their physical measurements cannot be arbitrarily lower than the set limits. {\it Bottom panel:} the distribution of spectroscopic redshifts for our sources ($0.03<z<0.1$).}
    \label{fig:wise_color}
\end{figure}

Due to the absence of reliable SFR information in the HETDEX field for LoTSS DR1 sources to distinguish AGN from SF, we utilise the large RLAGN sample selected from LoTSS DR1 (see \citetalias{2019A&A...622A..12H} for details). This sample consists of 23 344 RLAGN constructed using spectroscopic and photometric data \citep{2010AJ....140.1868W, 2016arXiv161205560C, 2019A&A...622A...2W, 2019A&A...622A...3D}. The distribution of this RLAGN sample in WISE colour-colour space is shown in Fig.~\ref{fig:wise_color} (middle panel). It consists of 6 850 resolved sources \citep[studied in detail by e.g.][]{2019MNRAS.488.2701M} and 16 494 unresolved sources; our work aims to investigate the nature of these unresolved objects. To allow for a follow up with high-resolution observations, we carefully and conservatively constructed a sub-sample of 55 RLAGN candidates by applying the following criteria:

\begin{itemize}
    \item[1.] selecting all unresolved low-luminosity (less than $10^{25}$ W Hz$^{-1}$ at 150 MHz) DR1 sources cross matched with the MPA-JHU value-added SDSS catalogue \citep{2004MNRAS.351.1151B}; 
    \item[2.] selecting all objects that show as AGN on all the four AGN-SF separation methods from \citet{2019A&A...622A..17S} $-$ the \citeauthor{2005MNRAS.362...25B} $D_{4000}$ vs $L_{150}/M*$ method, the \citeauthor{1981PASP...93....5B} `BPT' diagnostic diagram method ([O{\scriptsize{III}}]/H$\beta$ vs [N{\scriptsize{II}}]/H$\alpha$), the \citeauthor{1999ApJS..123...41M} $L_{\rm{H}\alpha}$ vs $L_{150}$ method, and the W1$-$W2 vs W2$-$W3 WISE colour method. These objects are classified as high excitation radio galaxies (HERGs) using the BPT diagram;
    \item[3.] selecting all objects within redshift range $0.03<z<0.1$ to get the best spatial resolution and to ensure that we have no ambiguities in their AGN classification (i.e. no contamination from SF) and;
    \item[4.] selecting all sources with a peak flux density greater than 2~mJy~beam$^{-1}$, which has the dual effect of ensuring a high SNR-LOFAR detection and permitting good VLA maps.
\end{itemize}

These criteria ($1-4$) resulted in 55 low-$z$ (bottom panel of Fig.~\ref{fig:wise_color}) low-$L$ (top panel of Fig.~\ref{fig:wise_color}) compact RLAGN candidates (overlaid on the RLAGN WISE colour-colour plot from \citetalias{2019A&A...622A..12H}, middle panel of Fig.~\ref{fig:wise_color}); their basic properties are presented in Table~\ref{tb1:data}. We have carried out a new set of high resolution (sub-arcsecond) VLA observations of this sample with the aim of resolving the DR1 candidates on sub-kpc scales ($\leq$100 kpc); VLA images give us the opportunity to detect cores, jets, and lobes (if present) on these scales.

%%%% Table 1: 
\begin{table*}
     \caption{Basic information about the 55 DR1 radio-loud AGN candidates.}
    \label{tb1:data}
    \begin{adjustbox}{width=1\textwidth}
    \begin{tabular}{lcccccccc}
    \hline\hline
         LOFAR\,source\,name&RA&Dec&$z\rm{_{spec}}$&$S\rm{^{int}_{0.15}}$&$S\rm{^{int}_{1.4}}$&$\alpha_{0.15}^{1.4}$&log$_{10}(L_{0.15})$&log$_{10}(L_{1.4})$\\
         \hline
ILTJ104852.93+480314.8~~~~(J1048+48)	&		$\rm{10^h48^m52.93^s}$ &   $\rm{+48^{\circ}03^{\prime}14.8^{\prime\prime}}$	&	0.041	&	66.88	$\pm$	0.33	&	18.85	$\pm$	1.00	&	0.56	$\pm$	0.02	&	23.41	&	22.86	\\
ILTJ105535.82+462903.3~~~~(J1055+46)	&		$\rm{10^h55^m35.82^s}$ &   $\rm{+46^{\circ}29^{\prime}03.3^{\prime\prime}}$	&	0.089	&	3.86	$\pm$	0.12	&	2.14	$\pm$	0.40	&	0.26	$\pm$	0.08	&	22.87	&	22.62	\\
ILTJ110233.62+513124.3~~~~(J1102+51)	&		$\rm{11^h02^m33.62^s}$ &   $\rm{+51^{\circ}31^{\prime}24.3^{\prime\prime}}$	&	0.069	&	6.40	$\pm$	0.10	&	2.38	$\pm$	0.40	&	0.44	$\pm$	0.07	&	22.86	&	22.43	\\
ILTJ111156.65+555449.9~~~~(J1111+55)	&		$\rm{11^h11^m56.65^s}$ &   $\rm{+55^{\circ}54^{\prime}49.9^{\prime\prime}}$	&	0.071	&	5.40	$\pm$	0.15	&	1.18	$\pm$	0.50	&	0.67	$\pm$	0.19	&	22.81	&	22.15	\\
ILTJ111712.10+465135.6~~~~(J1117+46)	&		$\rm{11^h17^m12.10^s}$ &   $\rm{+46^{\circ}51^{\prime}35.6^{\prime\prime}}$	&	0.061	&	12.57	$\pm$	0.12	&	4.52	$\pm$	0.40	&	0.45	$\pm$	0.04	&	23.04	&	22.60	\\
ILTJ111745.92+473327.5~~~~(J1117+47)	&		$\rm{11^h17^m45.92^s}$ &   $\rm{+47^{\circ}33^{\prime}27.5^{\prime\prime}}$	&	0.075	&	27.76	$\pm$	0.16	&	11.31	$\pm$	0.50	&	0.39	$\pm$	0.02	&	23.57	&	23.18	\\
ILTJ112625.15+520503.6~~~~(J1126+52)	&		$\rm{11^h26^m25.15^s}$ &   $\rm{+52^{\circ}05^{\prime}03.6^{\prime\prime}}$	&	0.048	&	25.32	$\pm$	0.31	&	7.79	$\pm$	0.50	&	0.52	$\pm$	0.03	&	23.13	&	22.62	\\
ILTJ113545.33+491619.6~~~~(J1135+49)	&		$\rm{11^h35^m45.33^s}$ &   $\rm{+49^{\circ}16^{\prime}19.6^{\prime\prime}}$	&	0.053	&	4.78	$\pm$	0.13	&	2.10	$\pm$	1.10	&	0.36	$\pm$	0.23	&	22.49	&	22.14	\\
ILTJ113757.19+554207.5~~~~(J1137+55)	&		$\rm{11^h37^m57.19^s}$ &   $\rm{+55^{\circ}42^{\prime}07.5^{\prime\prime}}$	&	0.062	&	3.61	$\pm$	0.12	&	1.61	$\pm$	0.40	&	0.36	$\pm$	0.11	&	22.52	&	22.17	\\
ILTJ114047.35+463225.5~~~~(J1140+46)	&		$\rm{11^h40^m47.35^s}$ &   $\rm{+46^{\circ}32^{\prime}25.5^{\prime\prime}}$	&	0.054	&	8.14	$\pm$	0.13	&	18.42	$\pm$	0.60	&	-0.36	$\pm$	0.02	&	22.74	&	23.09	\\
ILTJ114316.26+551639.9~~~~(J1143+55)	&		$\rm{11^h43^m16.26^s}$ &   $\rm{+55^{\circ}16^{\prime}39.9^{\prime\prime}}$	&	0.055	&	25.05	$\pm$	0.25	&	11.72	$\pm$	0.50	&	0.33	$\pm$	0.02	&	23.25	&	22.92	\\
ILTJ114721.40+554348.4~~~~(J1147+55)	&		$\rm{11^h47^m21.40^s}$ &   $\rm{+55^{\circ}43^{\prime}48.4^{\prime\prime}}$	&	0.052	&	3.29	$\pm$	0.13	&	4.80	$\pm$	0.40	&	-0.17	$\pm$	0.04	&	22.31	&	22.47	\\
ILTJ115113.46+550659.0~~~~(J1151+55)	&		$\rm{11^h51^m13.46^s}$ &   $\rm{+55^{\circ}06^{\prime}59.0^{\prime\prime}}$	&	0.079	&	8.18	$\pm$	0.25	&	3.72	$\pm$	0.01	&	0.35	$\pm$	0.01	&	23.09	&	22.75	\\
ILTJ115157.87+532845.5~~~~(J1151+53)	&		$\rm{11^h51^m57.87^s}$ &   $\rm{+53^{\circ}28^{\prime}45.5^{\prime\prime}}$	&	0.060	&	3.91	$\pm$	0.11	&	2.45	$\pm$	0.50	&	0.21	$\pm$	0.09	&	22.51	&	22.31	\\
ILTJ115205.79+545817.1~~~~(J1152+54)	&		$\rm{11^h52^m05.79^s}$ &   $\rm{+54^{\circ}58^{\prime}17.1^{\prime\prime}}$	&	0.060	&	11.16	$\pm$	0.16	&	6.82	$\pm$	0.50	&	0.22	$\pm$	0.03	&	22.98	&	22.76	\\
ILTJ115330.55+524121.8~~~~(J1153+52)	&		$\rm{11^h53^m30.55^s}$ &   $\rm{+52^{\circ}41^{\prime}21.8^{\prime\prime}}$	&	0.072	&	33.95	$\pm$	0.33	&	22.92	$\pm$	0.80	&	0.17	$\pm$	0.02	&	23.62	&	23.45	\\
ILTJ115438.10+491851.4~~~~(J1154+49)	&		$\rm{11^h54^m38.10^s}$ &   $\rm{+49^{\circ}18^{\prime}51.4^{\prime\prime}}$	&	0.054	&	4.85	$\pm$	0.10	&	4.32	$\pm$	0.40	&	0.05	$\pm$	0.04	&	22.52	&	22.47	\\
ILTJ115531.40+545200.4~~~~(J1155+54)	&		$\rm{11^h55^m31.40^s}$ &   $\rm{+54^{\circ}52^{\prime}00.4^{\prime\prime}}$	&	0.050	&	12.93	$\pm$	1.35	&	31.76	$\pm$	1.00	&	-0.40	$\pm$	0.05	&	22.87	&	23.26	\\
ILTJ115952.14+553205.5~~~~(J1159+55)	&		$\rm{11^h59^m52.14^s}$ &   $\rm{+55^{\circ}32^{\prime}05.5^{\prime\prime}}$	&	0.081	&	5.34	$\pm$	0.12	&	$<$0.44	&	$>$1.10	&	22.92	&	21.84	\\
ILTJ120328.53+514256.3~~~~(J1203+51)	&		$\rm{12^h03^m28.53^s}$ &   $\rm{+51^{\circ}42^{\prime}56.3^{\prime\prime}}$	&	0.061	&	4.86	$\pm$	0.12	&	$<$0.51    &	$>$0.99&	22.63	&	21.65	\\
ILTJ120801.00+523635.3~~~~(J1208+52)	&		$\rm{12^h08^m01.00^s}$ &   $\rm{+52^{\circ}36^{\prime}35.3^{\prime\prime}}$	&	0.082	&	2.93	$\pm$	0.13	&	$<$0.41   	&	$>$0.87	&	22.67	&	21.82	\\
ILTJ121329.29+504429.4~~~~(J1213+50)	&		$\rm{12^h13^m29.29^s}$ &   $\rm{+50^{\circ}44^{\prime}29.4^{\prime\prime}}$	&	0.031	&	187.00	$\pm$	0.78	&	102.73	$\pm$	2.90	&	0.26	$\pm$	0.01	&	23.61	&	23.35	\\
ILTJ123011.86+470022.7~~~~(J1230+47)	&		$\rm{12^h30^m11.86^s}$ &   $\rm{+47^{\circ}00^{\prime}22.7^{\prime\prime}}$	&	0.039	&	102.66	$\pm$	0.45	&	87.47	$\pm$	2.80	&	0.07	$\pm$	0.01	&	23.56	&	23.49	\\
ILTJ123301.22+561013.7~~~~(J1233+56)	&		$\rm{12^h33^m01.22^s}$ &   $\rm{+56^{\circ}10^{\prime}13.7^{\prime\prime}}$	&	0.081	&	2.78	$\pm$	0.16	&	6.40	$\pm$	0.50	&	-0.37	$\pm$	0.04	&	22.65	&	23.01	\\
ILTJ124021.48+475143.6~~~~(J1240+47)	&		$\rm{12^h40^m21.48^s}$ &   $\rm{+47^{\circ}51^{\prime}43.6^{\prime\prime}}$	&	0.098	&	10.26	$\pm$	0.15	&	3.15	$\pm$	0.40	&	0.52	$\pm$	0.06	&	23.38	&	22.87	\\
ILTJ124343.52+544902.0~~~~(J1243+54)	&		$\rm{12^h43^m43.52^s}$ &   $\rm{+54^{\circ}49^{\prime}02.0^{\prime\prime}}$	&	0.085	&	6.82	$\pm$	0.18	&	1.85	$\pm$	0.40	&	0.57	$\pm$	0.10	&	23.07	&	22.51	\\
ILTJ125827.27+501650.6~~~~(J1258+50)	&		$\rm{12^h58^m27.27^s}$ &   $\rm{+50^{\circ}16^{\prime}50.6^{\prime\prime}}$	&	0.099	&	2.83	$\pm$	0.11	&	$<$0.44	&$>$0.82		&	22.84	&	22.03	\\
ILTJ130325.87+523522.5~~~~(J1303+52)	&		$\rm{13^h03^m25.87^s}$ &   $\rm{+52^{\circ}35^{\prime}22.5^{\prime\prime}}$	&	0.096	&	9.79	$\pm$	0.15	&	4.77	$\pm$	0.40	&	0.32	$\pm$	0.04	&	23.34	&	23.03	\\
ILTJ130404.58+492728.5~~~~(J1304+49)	&		$\rm{13^h04^m04.58^s}$ &   $\rm{+49^{\circ}27^{\prime}28.5^{\prime\prime}}$	&	0.033	&	5.64	$\pm$	0.11	&	$<$0.39	&$>$1.17		&	22.15	&	20.99	\\
ILTJ130410.78+552839.0~~~~(J1304+55)	&		$\rm{13^h04^m10.78^s}$ &   $\rm{+55^{\circ}28^{\prime}39.0^{\prime\prime}}$	&	0.082	&	3.90	$\pm$	0.12	&	$<$0.41	&$>$0.99	&	22.80	&	21.82	\\
ILTJ130535.92+540142.5~~~~(J1305+54)	&		$\rm{13^h05^m35.92^s}$ &   $\rm{+54^{\circ}01^{\prime}42.5^{\prime\prime}}$	&	0.091	&	23.84	$\pm$	0.24	&	23.14	$\pm$	1.00	&	0.01	$\pm$	0.02	&	23.68	&	23.67	\\
ILTJ131002.06+540005.2~~~~(J1310+54)	&		$\rm{13^h10^m02.06^s}$ &   $\rm{+54^{\circ}00^{\prime}05.2^{\prime\prime}}$	&	0.089	&	2.68	$\pm$	0.11	&	1.29	$\pm$	0.01	&	0.32	$\pm$	0.02	&	22.71	&	22.39	\\
ILTJ131015.19+544004.8~~~~(J1310+544)	&		$\rm{13^h10^m15.19^s}$ &   $\rm{+54^{\circ}40^{\prime}04.8^{\prime\prime}}$	&	0.064	&	5.59	$\pm$	0.11	&	1.90	$\pm$	0.01	&	0.47	$\pm$	0.01	&	22.73	&	22.26	\\
ILTJ132745.67+544756.8~~~~(J1327+54)	&		$\rm{13^h27^m45.67^s}$ &   $\rm{+54^{\circ}47^{\prime}56.8^{\prime\prime}}$	&	0.033	&	10.73	$\pm$	0.18	&	3.06	$\pm$	0.40	&	0.55	$\pm$	0.06	&	22.43	&	21.88	\\
ILTJ133430.04+502718.3~~~~(J1334+50)	&		$\rm{13^h34^m30.04^s}$ &   $\rm{+50^{\circ}27^{\prime}18.3^{\prime\prime}}$	&	0.085	&	3.05	$\pm$	0.11	&	1.76	$\pm$	0.40	&	0.24	$\pm$	0.10	&	22.73	&	22.49	\\
ILTJ134227.94+554939.3~~~~(J1342+55)	&		$\rm{13^h42^m27.94^s}$ &   $\rm{+55^{\circ}49^{\prime}39.3^{\prime\prime}}$	&	0.072	&	4.97	$\pm$	0.12	&	1.37	$\pm$	0.50	&	0.57	$\pm$	0.16	&	22.79	&	22.23	\\
ILTJ134233.86+485316.2~~~~(J1342+48)	&		$\rm{13^h42^m33.86^s}$ &   $\rm{+48^{\circ}53^{\prime}16.2^{\prime\prime}}$	&	0.091	&	7.45	$\pm$	0.11	&	3.67	$\pm$	0.40	&	0.31	$\pm$	0.05	&	23.18	&	22.87	\\
ILTJ134345.04+553801.3~~~~(J1343+55)	&		$\rm{13^h43^m45.04^s}$ &   $\rm{+55^{\circ}38^{\prime}01.3^{\prime\prime}}$	&	0.068	&	3.03	$\pm$	0.12	&	$<$0.31	&	$>$1.00	&	22.53	&	21.54	\\
ILTJ135632.65+493710.9~~~~(J1356+49)	&		$\rm{13^h56^m32.65^s}$ &   $\rm{+49^{\circ}37^{\prime}10.9^{\prime\prime}}$	&	0.066	&	3.89	$\pm$	0.11	&	8.59	$\pm$	0.40	&	-0.35	$\pm$	0.02	&	22.60	&	22.95	\\
ILTJ140013.20+462556.0~~~~(J1400+46)	&		$\rm{14^h00^m13.20^s}$ &   $\rm{+46^{\circ}25^{\prime}56.0^{\prime\prime}}$	&	0.053	&	3.79	$\pm$	0.10	&	$<$0.70	&	$>$0.74	&	22.40	&	21.67	\\
ILTJ140906.86+532749.1~~~~(J1409+53)	&		$\rm{14^h09^m06.86^s}$ &   $\rm{+53^{\circ}27^{\prime}49.1^{\prime\prime}}$	&	0.078	&	22.28	$\pm$	0.28	&	3.87	$\pm$	0.50	&	0.77	$\pm$	0.06	&	23.51	&	22.75	\\
ILTJ141118.46+551053.2~~~~(J1411+55)	&		$\rm{14^h11^m18.46^s}$ &   $\rm{+55^{\circ}10^{\prime}53.2^{\prime\prime}}$	&	0.042	&	26.93	$\pm$	0.21	&	8.51	$\pm$	0.50	&	0.51	$\pm$	0.03	&	23.04	&	22.54	\\
ILTJ141149.57+545733.1~~~~(J1411+54)	&		$\rm{14^h11^m49.57^s}$ &   $\rm{+54^{\circ}57^{\prime}33.1^{\prime\prime}}$	&	0.083	&	4.13	$\pm$	0.16	&	5.85	$\pm$	0.40	&	-0.15	$\pm$	0.03	&	22.84	&	22.99	\\
ILTJ141234.88+494611.5~~~~(J1412+49)	&		$\rm{14^h12^m34.88^s}$ &   $\rm{+49^{\circ}46^{\prime}11.5^{\prime\prime}}$	&	0.071	&	4.33	$\pm$	0.17	&	3.65	$\pm$	0.40	&	0.08	$\pm$	0.05	&	22.72	&	22.64	\\
ILTJ141422.37+553520.5~~~~(J1414+55)	&		$\rm{14^h14^m22.37^s}$ &   $\rm{+55^{\circ}35^{\prime}20.5^{\prime\prime}}$	&	0.077	&	2.72	$\pm$	0.13	&	$<$0.41 &$>$0.83	&	22.59	&	21.76	\\
ILTJ142025.23+544016.1~~~~(J1420+54)	&		$\rm{14^h20^m25.23^s}$ &   $\rm{+54^{\circ}40^{\prime}16.1^{\prime\prime}}$	&	0.042	&	2.88	$\pm$	0.12	&	1.91	$\pm$	0.50	&	0.18	$\pm$	0.12	&	22.06	&	21.88	\\
ILTJ142125.51+482933.3~~~~(J1421+48)	&		$\rm{14^h21^m25.51^s}$ &   $\rm{+48^{\circ}29^{\prime}33.3^{\prime\prime}}$	&	0.072	&	8.49	$\pm$	0.16	&	25.19	$\pm$	0.80	&	-0.48	$\pm$	0.02	&	23.02	&	23.49	\\
ILTJ142255.56+474713.9~~~~(J1422+47)	&		$\rm{14^h22^m55.56^s}$ &   $\rm{+47^{\circ}47^{\prime}13.9^{\prime\prime}}$	&	0.072	&	3.42	$\pm$	0.13	&	2.25	$\pm$	0.01	&	0.18	$\pm$	0.02	&	22.62	&	22.44	\\
ILTJ142707.30+525001.5~~~~(J1427+52)	&		$\rm{14^h27^m07.30^s}$ &   $\rm{+52^{\circ}50^{\prime}01.5^{\prime\prime}}$	&	0.083	&	8.37	$\pm$	0.18	&	3.56	$\pm$	0.50	&	0.38	$\pm$	0.06	&	23.15	&	22.78	\\
ILTJ142948.48+535754.2~~~~(J1429+53)	&		$\rm{14^h29^m48.48^s}$ &   $\rm{+53^{\circ}57^{\prime}54.2^{\prime\prime}}$	&	0.043	&	3.55	$\pm$	0.19	&	$<$0.91	&	$>$0.60	&	22.18	&	21.59	\\
ILTJ143334.77+524310.2~~~~(J1433+52)	&		$\rm{14^h33^m34.77^s}$ &   $\rm{+52^{\circ}43^{\prime}10.2^{\prime\prime}}$	&	0.045	&	3.84	$\pm$	0.14	&	2.98	$\pm$	0.01	&	0.11	$\pm$	0.02	&	22.26	&	22.15	\\
ILTJ143521.75+505123.0~~~~(J1435+50)	&		$\rm{14^h35^m21.75^s}$ &   $\rm{+50^{\circ}51^{\prime}23.0^{\prime\prime}}$	&	0.100	&	339.00	$\pm$	2.29	&	140.96	$\pm$	4.40	&	0.39	$\pm$	0.01	&	24.92	&	24.54	\\
ILTJ145350.18+540440.9~~~~(J1453+54)	&		$\rm{14^h53^m50.18^s}$ &   $\rm{+54^{\circ}04^{\prime}40.9^{\prime\prime}}$	&	0.099	&	5.23	$\pm$	0.19	&	4.00	$\pm$	0.40	&	0.12	$\pm$	0.05	&	23.10	&	22.99	\\
ILTJ145415.10+492549.9~~~~(J1454+49)	&		$\rm{14^h54^m15.10^s}$ &   $\rm{+49^{\circ}25^{\prime}49.9^{\prime\prime}}$	&	0.037	&	6.26	$\pm$	0.20	&	9.65	$\pm$	0.40	&	-0.19	$\pm$	0.02	&	22.30	&	22.49	\\
ILTJ150422.16+474111.2~~~~(J1504+47)	&		$\rm{15^h04^m22.16^s}$ &   $\rm{+47^{\circ}41^{\prime}11.2^{\prime\prime}}$	&	0.093	&	20.11	$\pm$	1.09	&	9.54	$\pm$	0.90	&	0.33	$\pm$	0.05	&	23.63	&	23.31	\\
         \hline
    \end{tabular}
    \end{adjustbox}
    {\raggedright \textit{Notes:} Column description: (1) LOFAR source name with `short names' shown in brackets for in-text discussion; (2) Right ascension (RA); (3) Declination (Dec); (4) Spectroscopic redshifts for our sources obtained from \citet{2019A&A...622A..12H}; (5) LOFAR integrated flux density (mJy); (6) FIRST integrated flux density (mJy); (7) Radio spectral index estimated between 150 MHz and 1.4 GHz; (8) 150 MHz radio luminosity (W Hz$^{-1}$); (9) 1.4 GHz radio luminosity (W Hz$^{-1}$). Upper limits are represented by less than ($<$) and greater than ($>$) signs. Objects are arranged in order of increasing right ascension (RA) and we adopt the listed shortened names (in brackets) hereafter, for brevity. \par}
\end{table*}

\subsection{VLA observations and reduction}
\label{sec:vla}
The VLA observations were taken in the A configuration across two identical four-hour observation sessions during September 2019 at C-band frequencies (i.e. 4$-$8 GHz). For brevity's sake, we refer to these observations as epochs 1 and 2. Allowing for calibration and slewing overheads, the two identical four-hour sessions were scheduled such that the observations for each target were achieved in two three-minute observing blocks to improve the snapshot $uv$ coverage. A summary of these observations is given in Table~\ref{tab:data_epochs}.

\begin{table*}
  \caption{Details of the VLA data across the two observing epochs. `ant' and `chan' refer to the number of antennas and the number of channels used during the observations respectively.}
   \label{tab:data_epochs}
    \begin{tabular}{lccccccccc}
    \hline\hline
        Label&Project\,code&Date\,of\,observation&Start\,time&End\,time&Duration&ant&chan&Bandwidth&Array\\
        &&&&&$\rm{(s)}$&&&$\rm{(GHz)}$&\\
     \hline    
         Epoch 1&19A-264&15-Sep-2019&15:22:56.0&19:22:12.0&14356&27&64&4$-$8&A\\
         Epoch 2&19A-264&24-Sep-2019&14:07:20.0&18:06:38.0&14358&27&64&4$-$8&A\\
     \hline
    \end{tabular}
\end{table*}

The observations were calibrated using the primary calibrator 3C286 (flux calibrator) and five phase calibrators chosen based on their proximity to the targets. A summary of calibrator sources for the data used in this paper is presented in Table~\ref{tab:data_calibrator}.

\begin{table}
\centering
    \caption{Summary of calibrator sources.}
     \label{tab:data_calibrator}
      \begin{tabular}{c c}
    \hline \hline
        Calibrator Name & Description \\
    \hline \hline
        1331+305 = 3C286 & flux calibrator\\
    \hline
        J1035+5628 & \\
        J1146+5356 & \\
        J1219+4829 & phase calibrators \\
        J1349+5341 & \\
        J1419+5423 & \\
    \hline
    J1407+2827 & polarization calibrator\\
    \hline
      \end{tabular}
\end{table}

We used the Common Astronomy Software Applications \citep[CASA;][]{2007ASPC..376..127M} software to perform VLA data reduction. Before calibration, we inspected the visibility data for any issues reported in the observer's log. Following basic data inspection and editing based on a priori information, the standard pipeline calibration process was performed on the measurement set ({\sevensize{MS}}). Throughout the calibration process, we used a suitable reference antenna from the middle of the array with good calibration solutions. After correcting antenna positions, we set a known flux density scale for our primary calibrators using the VLA standard model at C-band. This was applied using the \citet{2017ApJS..230....7P} flux density standard model. Once phase and delay calibration solutions were derived and corrected for, we corrected for the bandpass, and then derived `true' flux densities for secondary calibrators. All the derived calibration solutions were applied to calibrator sources, and then to the target sources.

Any traces of bad data were later flagged in the calibrated visibilities. We then split each science target from the {\sevensize{MS}} of epochs 1 and 2 and combined them into a single {\sevensize{MS}}. Initial 6 GHz images were produced using {\sevensize{TCLEAN}}; this was done by using the Clark algorithm \citep{1980A&A....89..377C} and setting a Briggs weighting to a robust parameter value of 0.5 \citep{1995AAS...18711202B}. For target sources with a peak flux density value greater than $\rm{2~mJy}$, a few rounds of phase-only self-calibration were performed. In some special cases (targets with peak flux densities above $\rm{6~mJy}$), a final round considering phase and amplitude calibration solutions was performed using a solution interval between $\rm{30-60s}$ and a $3\sigma$ threshold. A summary of the imaging parameters is shown in Table~\ref{tab:data_imaging} with the rest of the imaging parameters left on their default settings.

Additionally, we split the bandwidth of the calibrated visibilities from epochs 1 and 2 to obtain two datasets centred at 5 and 7 GHz. Following the same procedure as described above, we obtained 5 and 7 GHz image parameters. This information is tabulated in Table~\ref{tb:appendix1}. 

\begin{table}
\centering
  \caption{Imaging summary. Note that the image size shown is the minimum size allocated. We adjusted this to account for surrounding sources to mitigate artefacts in the final images during self-calibration.}
  \label{tab:data_imaging}
    \begin{tabular}{ccccc}
     \hline\hline
     CASA parameter && Value && Unit \\
     \hline
        imsize && 3200 $\times$ 3200&& pixels \\
        cell && 0.08 $\times$ 0.08 && $\rm{arcsec}$ \\
        gridder && standard &&\\
        deconvolver && clark &&\\
        weighting && briggs &&\\
        robust && 0.5 &&\\
        niter && 1000 &&\\
        threshold && 0.03 &&$\rm{mJy}$ \\
     \hline    
    \end{tabular}
\end{table}

\begin{table*}
     \caption{Final properties about the 42 VLA compact objects provided by {\scriptsize{IMFIT}} at 6 GHz.}
     \begin{adjustbox}{width=1\textwidth}
    \begin{tabular}{lcccccccccc}
    \hline\hline
Source&$\theta^{r}_{6.0}$&$\vartheta_{6.0}$&$\theta^{d}_{6.0}$&$S\rm{^{int}_{6.0}}$&$S\rm{^{p}_{6.0}}$&$\sigma\rm{_{6.0}^{rms}}$&$\rm{\alpha^{6000}_{150}}$&$\Theta$&$D$&log$_{10} (L_{6.0})$\\
         \hline
J1055+46	&	0.35	$\times$	0.31	&	79.1	&	0.05	$\times$	0.02	&	2.18	$\pm$	0.04	&	2.11	$\pm$	0.02	&	14.1	&	0.15	$\pm$	0.01	&	0.10	$\pm$	0.00	&	0.17	$\pm$	0.00	&	22.59	\\
J1102+51	&	0.35	$\times$	0.30	&	72.1	&	0.13	$\times$	0.04	&	2.65	$\pm$	0.06	&	2.46	$\pm$	0.03	&	13.7	&	0.24	$\pm$	0.01	&	0.26	$\pm$	0.04	&	0.34	$\pm$	0.05	&	22.45	\\
J1117+47	&	0.35	$\times$	0.32	&	62.1	&	0.09	$\times$	0.05	&	6.62	$\pm$	0.04	&	6.35	$\pm$	0.02	&	19.0	&	0.38	$\pm$	0.00	&	0.18	$\pm$	0.01	&	0.25	$\pm$	0.02	&	22.91	\\
J1135+49	&	0.43	$\times$	0.36	&	81.8	&	0.12	$\times$	0.09	&	2.96	$\pm$	0.05	&	2.75	$\pm$	0.03	&	13.6	&	0.13	$\pm$	0.01	&	0.24	$\pm$	0.04	&	0.24	$\pm$	0.04	&	22.27	\\
J1137+55	&	0.34	$\times$	0.30	&	60.5	&	0.09	$\times$	0.04	&	2.67	$\pm$	0.05	&	2.56	$\pm$	0.03	&	13.5	&	0.08	$\pm$	0.01	&	0.18	$\pm$	0.04	&	0.22	$\pm$	0.04	&	22.37	\\
J1140+46	&	0.32	$\times$	0.29	&	38.9	&	0.05	$\times$	0.04	&	30.75	$\pm$	0.13	&	30.08	$\pm$	0.07	&	20.0	&	-0.36	$\pm$	0.00	&	0.09	$\pm$	0.02	&	0.10	$\pm$	0.02	&	23.31	\\
J1143+55	&	0.33	$\times$	0.29	&	59.0	&	0.07	$\times$	0.06	&	22.62	$\pm$	0.10	&	21.77	$\pm$	0.06	&	13.9	&	0.03	$\pm$	0.00	&	0.13	$\pm$	0.01	&	0.14	$\pm$	0.01	&	23.19	\\
J1147+55	&	0.35	$\times$	0.31	&	66.9	&	0.06	$\times$	0.03	&	3.50	$\pm$	0.04	&	3.42	$\pm$	0.02	&	14.8	&	-0.02	$\pm$	0.01	&	0.13	$\pm$	0.04	&	0.13	$\pm$	0.04	&	22.32	\\
J1151+55	&	0.35	$\times$	0.31	&	71.5	&	0.08	$\times$	0.02	&	0.38	$\pm$	0.02	&	0.37	$\pm$	0.01	&	13.9	&	0.82	$\pm$	0.02	&	0.15	$\pm$	0.00	&	0.22	$\pm$	0.00	&	21.71	\\
J1151+53	&	0.38	$\times$	0.30	&	75.8	&	0.11	$\times$	0.03	&	3.23	$\pm$	0.03	&	3.07	$\pm$	0.02	&	17.0	&	0.05	$\pm$	0.01	&	0.22	$\pm$	0.02	&	0.25	$\pm$	0.03	&	22.41	\\
J1152+54	&	0.36	$\times$	0.32	&	73.5	&	0.07	$\times$	0.04	&	5.30	$\pm$	0.03	&	5.16	$\pm$	0.02	&	12.5	&	0.20	$\pm$	0.00	&	0.14	$\pm$	0.02	&	0.16	$\pm$	0.02	&	22.63	\\
J1153+52	&	0.44	$\times$	0.36	&	84.4	&	0.07	$\times$	0.05	&	15.99	$\pm$	0.05	&	15.60	$\pm$	0.03	&	20.6	&	0.20	$\pm$	0.00	&	0.14	$\pm$	0.01	&	0.20	$\pm$	0.02	&	23.27	\\
J1154+49	&	0.34	$\times$	0.32	&	61.6	&	0.05	$\times$	0.03	&	7.00	$\pm$	0.07	&	6.88	$\pm$	0.04	&	13.6	&	-0.10	$\pm$	0.01	&	0.10	$\pm$	0.03	&	0.11	$\pm$	0.03	&	22.66	\\
J1155+54	&	0.37	$\times$	0.31	&	68.9	&	0.07	$\times$	0.02	&	29.53	$\pm$	0.19	&	28.97	$\pm$	0.10	&	39.3	&	-0.22	$\pm$	0.03	&	0.14	$\pm$	0.02	&	0.13	$\pm$	0.02	&	23.22	\\
J1208+52	&	0.40	$\times$	0.30	&	83.8	&	$-$			&	0.41	$\pm$	0.04	&	0.26	$\pm$	0.02	&	14.6	&	0.53	$\pm$	0.03	&	$-$			&	$-$	&	21.77	\\
J1230+47	&	0.35	$\times$	0.30	&	73.1	&	0.09	$\times$	0.06	&	68.40	$\pm$	0.20	&	64.87	$\pm$	0.11	&	33.3	&	0.11	$\pm$	0.00	&	0.17	$\pm$	0.01	&	0.13	$\pm$	0.01	&	23.37	\\
J1233+56	&	0.39	$\times$	0.29	&	79.3	&	0.12	$\times$	0.05	&	1.68	$\pm$	0.10	&	1.56	$\pm$	0.05	&	14.5	&	0.14	$\pm$	0.02	&	0.24	$\pm$	0.11	&	0.37	$\pm$	0.16	&	22.40	\\
J1240+47	&	0.37	$\times$	0.32	&	78.1	&	0.15	$\times$	0.15	&	2.34	$\pm$	0.08	&	1.97	$\pm$	0.04	&	13.5	&	0.40	$\pm$	0.01	&	0.30	$\pm$	0.06	&	0.55	$\pm$	0.11	&	22.70	\\
J1258+50	&	0.39	$\times$	0.31	&	84.3	&	0.15	$\times$	0.09	&	0.32	$\pm$	0.03	&	0.26	$\pm$	0.02	&	13.8	&	0.58	$\pm$	0.03	&	0.29	$\pm$	0.00	&	0.53	$\pm$	0.00	&	21.84	\\
J1303+52	&	0.40	$\times$	0.34	&	80.3	&	0.08	$\times$	0.03	&	7.44	$\pm$	0.04	&	7.25	$\pm$	0.02	&	17.0	&	0.07	$\pm$	0.00	&	0.17	$\pm$	0.02	&	0.29	$\pm$	0.03	&	23.19	\\
J1305+54	&	0.39	$\times$	0.32	&	78.2	&	0.06	$\times$	0.05	&	17.07	$\pm$	0.06	&	16.66	$\pm$	0.03	&	19.4	&	0.09	$\pm$	0.00	&	0.13	$\pm$	0.01	&	0.21	$\pm$	0.02	&	23.50	\\
J1310+54	&	0.39	$\times$	0.30	&	77.8	&	0.09	$\times$	0.01	&	1.85	$\pm$	0.07	&	1.75	$\pm$	0.04	&	15.1	&	0.10	$\pm$	0.01	&	0.18	$\pm$	0.00	&	0.30	$\pm$	0.00	&	22.52	\\
J1327+54	&	0.36	$\times$	0.32	&	69.6	&	0.18	$\times$	0.15	&	2.32	$\pm$	0.12	&	1.88	$\pm$	0.06	&	14.4	&	0.41	$\pm$	0.01	&	0.36	$\pm$	0.08	&	0.24	$\pm$	0.05	&	21.75	\\
J1334+50	&	0.39	$\times$	0.31	&	-90.0	&	0.11	$\times$	0.07	&	0.80	$\pm$	0.02	&	0.75	$\pm$	0.01	&	13.5	&	0.36	$\pm$	0.01	&	0.23	$\pm$	0.07	&	0.36	$\pm$	0.11	&	22.11	\\
J1342+55	&	0.36	$\times$	0.31	&	74.9	&	0.24	$\times$	0.13	&	0.46	$\pm$	0.02	&	0.34	$\pm$	0.10	&	13.6	&	0.64	$\pm$	0.01	&	0.47	$\pm$	0.06	&	0.65	$\pm$	0.08	&	21.71	\\
J1342+48	&	0.39	$\times$	0.31	&	-87.2	&	0.22	$\times$	0.06	&	1.75	$\pm$	0.07	&	1.51	$\pm$	0.04	&	15.4	&	0.39	$\pm$	0.01	&	0.43	$\pm$	0.06	&	0.73	$\pm$	0.09	&	22.51	\\
J1343+55	&	0.36	$\times$	0.31	&	73.3	&	0.29	$\times$	0.13	&	0.41	$\pm$	0.02	&	0.28	$\pm$	0.01	&	12.3	&	0.54	$\pm$	0.02	&	0.58	$\pm$	0.06	&	0.76	$\pm$	0.08	&	21.62	\\
J1356+49	&	0.41	$\times$	0.33	&	-86.0	&	0.07	$\times$	0.04	&	6.61	$\pm$	0.03	&	6.45	$\pm$	0.02	&	15.4&	-0.14	$\pm$	0.01	&	0.15	$\pm$	0.01	&	0.19	$\pm$	0.02	&	22.82	\\
J1400+46	&	0.41	$\times$	0.31	&	-81.0	&	0.13	$\times$	0.09	&	0.69	$\pm$	0.04	&	0.63	$\pm$	0.02	&	14.2	&	0.46	$\pm$	0.02	&	0.25	$\pm$	0.11	&	0.26	$\pm$	0.11	&	21.63	\\
J1409+53	&	0.38	$\times$	0.30	&	78.8	&	0.18	$\times$	0.09	&	1.98	$\pm$	0.07	&	1.72	$\pm$	0.04	&	13.1	&	0.65	$\pm$	0.01	&	0.36	$\pm$	0.05	&	0.53	$\pm$	0.07	&	22.42	\\
J1411+54	&	0.36	$\times$	0.32	&	63.2	&	0.07	$\times$	0.05	&	9.78	$\pm$	0.05	&	9.52	$\pm$	0.03	&	14.0	&	-0.23	$\pm$	0.01	&	0.13	$\pm$	0.02	&	0.21	$\pm$	0.03	&	23.20	\\
J1412+49	&	0.37	$\times$	0.31	&	85.0	&	0.09	$\times$	0.07	&	3.08	$\pm$	0.11	&	2.92	$\pm$	0.06	&	13.3	&	0.09	$\pm$	0.01	&	0.18	$\pm$	0.08	&	0.24	$\pm$	0.11	&	22.54	\\
J1414+55	&	0.35	$\times$	0.30	&	61.0	&	0.09	$\times$	0.03	&	0.24	$\pm$	0.02	&	0.23	$\pm$	0.01	&	13.7	&	0.65	$\pm$	0.03	&	0.18	$\pm$	0.00	&	0.26	$\pm$	0.00	&	21.49	\\
J1420+54	&	0.37	$\times$	0.30	&	69.2	&	0.09	$\times$	0.02	&	1.87	$\pm$	0.05	&	1.74	$\pm$	0.03	&	13.0	&	0.12	$\pm$	0.01	&	0.17	$\pm$	0.00	&	0.14	$\pm$	0.00	&	21.86	\\
J1421+48	&	0.38	$\times$	0.30	&	-89.8	&	0.08	$\times$	0.05	&	23.00	$\pm$	0.06	&	22.27	$\pm$	0.03	&	16.6	&	-0.27	$\pm$	0.01	&	0.15	$\pm$	0.01	&	0.21	$\pm$	0.01	&	23.44	\\
J1427+52	&	0.37	$\times$	0.31	&	80.8	&	0.10	$\times$	0.04	&	3.41	$\pm$	0.03	&	3.25	$\pm$	0.02	&	13.6	&	0.24	$\pm$	0.01	&	0.20	$\pm$	0.02	&	0.31	$\pm$	0.03	&	22.73	\\
J1429+53	&	0.36	$\times$	0.29	&	74.2	&	0.09	$\times$	0.01	&	1.65	$\pm$	0.12	&	1.58	$\pm$	0.07	&	15.3	&	0.21	$\pm$	0.02	&	0.18	$\pm$	0.13	&	0.15	$\pm$	0.11	&	21.83	\\
J1433+52${\textbf{*}}$	&	0.38	$\times$	0.31	&	83.0	&	0.14	$\times$	0.05	&	1.95	$\pm$	0.04	&	1.82	$\pm$	0.02	&	13.9	&	0.18	$\pm$	0.01	&	0.27	$\pm$	0.04	&	0.24	$\pm$	0.03	&	21.95	\\
J1435+50	&	0.39	$\times$	0.31	&	86.2	&	0.23	$\times$	0.10	&	66.47	$\pm$	0.60	&	54.22	$\pm$	0.30	&	32.4	&	0.44	$\pm$	0.00	&	0.46	$\pm$	0.01	&	0.85	$\pm$	0.02	&	24.16	\\
J1453+54	&	0.37	$\times$	0.28	&	86.9	&	0.02	$\times$	0.01	&	5.92	$\pm$	0.04	&	5.91	$\pm$	0.02	&	13.4	&	-0.03	$\pm$	0.01	&	0.03	$\pm$	0.00	&	0.06	$\pm$	0.00	&	23.13	\\
J1454+49	&	0.42	$\times$	0.34	&	-64.0	&	0.05	$\times$	0.04	&	7.46	$\pm$	0.14	&	7.35	$\pm$	0.08	&	26.1	&	-0.05	$\pm$	0.01	&	0.11	$\pm$	0.07	&	0.08	$\pm$	0.05	&	22.37	\\
J1504+47${\textbf{*}}$	&	0.42	$\times$	0.30	&	-84.1	&	0.08	$\times$	0.05	&	2.20	$\pm$	0.11	&	2.13	$\pm$	0.06	&	14.3	&	0.59	$\pm$	0.02	&	0.15	$\pm$	0.00	&	0.26	$\pm$	0.00	&	22.62	\\
         \hline
    \end{tabular}
    \end{adjustbox}
    {\raggedright \textit{Notes:} Column description: (1) Source; (2) Restoring beam (arcsec); (3) Position angle (PA; in degrees); (4) Deconvolved size (arcsec); (5) Integrated flux density (mJy); (6) Peak flux density (mJy); (7) Image RMS ($\mu$Jy beam$^{-1}$); (8) Radio spectral index estimated between 150 MHz and 6 GHz; (9) Source angular size (arcsec) estimated from twice the deconvolved beam's major axis; (10) Source physical size (kpc); (11) Radio luminosity in W Hz$^{-1}$. Sources that are compact here but show extended emission in DR2 images are indicated by an asterisk (${\textbf{*}}$) while (-) indicates undetermined source parameters. \par}
\label{tab:unResolved}
\end{table*}

\section{ANALYSIS \& RESULTS}
\label{sec:analysis}
\subsection{Radio emission categories}
In order to determine and analyze the nature of the images produced, source parameters in the final images were computed using the CASA {\sevensize{IMFIT}} task. This tool fits one or more two-dimensional elliptical Gaussian components to a region in the image. This was achieved by interactively drawing boxes or tracing polygons around the edge of the object's emission. The final output returned the integrated flux, peak flux, convolved and deconvolved source parameters, that is, the angular Full Width at Half Maximum (FWHM) major and minor axes (arcseconds), and the position angles (PA; degrees). This information is listed in Tables~\ref{tab:unResolved} and \ref{tab:extended} together with associated errors \citep[derived from uncertainty estimates in the {\sevensize{IMFIT}} model fitting and taking into consideration the 5\% limit in the flux calibration;][]{2017ApJS..230....7P}. We also estimated the root mean square ($\sigma$) value of the image by selecting a region free of emission close to the source. The integrated spectral indices listed in Tables~\ref{tab:unResolved} and \ref{tab:extended} computed between the observing frequencies of LOFAR (150 MHz) and the VLA (6 GHz), using the power law definition in Sec~\ref{sec:intro}, are expected to be robust given the fact that we selected LOFAR candidates with angular scales below the angular sensitivity of the VLA. A visual inspection of the VLA final images together with the information derived from {\sevensize{IMFIT}} resulted in the construction of three categories of RLAGN: undetected, compact, and extended objects. Further analyses and discussions of their morphologies and scales are presented in Sec.~\ref{sec:und},~\ref{sec:unR}, and~\ref{sec:Re}.

\subsection{Undetected radio emission}
\label{sec:und}
A visual inspection of VLA images shows no radio emission in three ($3/55;\,\sim$5\%) objects, namely: J1159+55, J1203+51, and J1304+49. These objects, however, clearly show compact radio emission in LoTSS DR2 images \citep{2022A&A...659A...1S}. A discussion of their radio spectra is given in Sec.~\ref{sec:radio_spectra}; a non-detection implies a steep radio spectrum in all three cases (see Fig.~\ref{fig:spec}). The fact that these objects are undetected in the FIRST and VLA images at 1.4 and 6 GHz, respectively, and have steeper spectral indices may indicate that these sources are `switched off' $-$ a detection only at lower frequencies is a characteristic that is often displayed in remnant RLAGN \citep[e.g.][]{2018MNRAS.475.4557M}. This finding is in line with remnant sources constituting a small fraction in large RLAGN catalogues. According to e.g. \citet{2018MNRAS.475.4557M}, only 9\% of remnant sources were identified, a percentage comparable to the potential remnant sources observed in our study.

\subsection{Compact radio emission}
\label{sec:unR}
In order to correctly classify sources as unresolved, we compared the image component sizes (deconvolved from the beam) against the restoring beam parameters provided by {\sevensize{IMFIT}}. We also took into account comparisons of the total and peak flux densities for each object, with a source being classified as compact if the deconvolved size is less than the restoring beam, and the peak flux density is within reasonable estimates of the integrated flux density. Using these criteria, we found 42 (out of 55) compact RLAGN, representing $\sim$76\% of the sources in our catalogue (see Table~\ref{tab:unResolved}). These objects are compact at the limiting angular resolution of the VLA observations (0.35 arcsec) which corresponds to $\sim$0.1$-$0.8 kpc in projected physical size (estimated from twice the deconvolved major axis scale). While we selected all our RLAGN candidates based on their compact nature in the LoTSS DR1 images available at the time, the newer and better quality LoTSS DR2 images have revealed some extended sources (J1433+52 and J1504+47). These objects (classed as compact in the VLA images) appear to be in fact the radio cores of large resolved sources, which were not identified as such in LoTSS DR1 because of its poorer surface brightness sensitivity. Appendix~\ref{sec:appendix} gives a brief description and images of their structures. A detailed discussion of the spectral index distribution for the whole sample is given in Sec.~\ref{sec:radio_spectra}.

\begin{figure*}
\includegraphics[keepaspectratio,width=1.0\columnwidth]{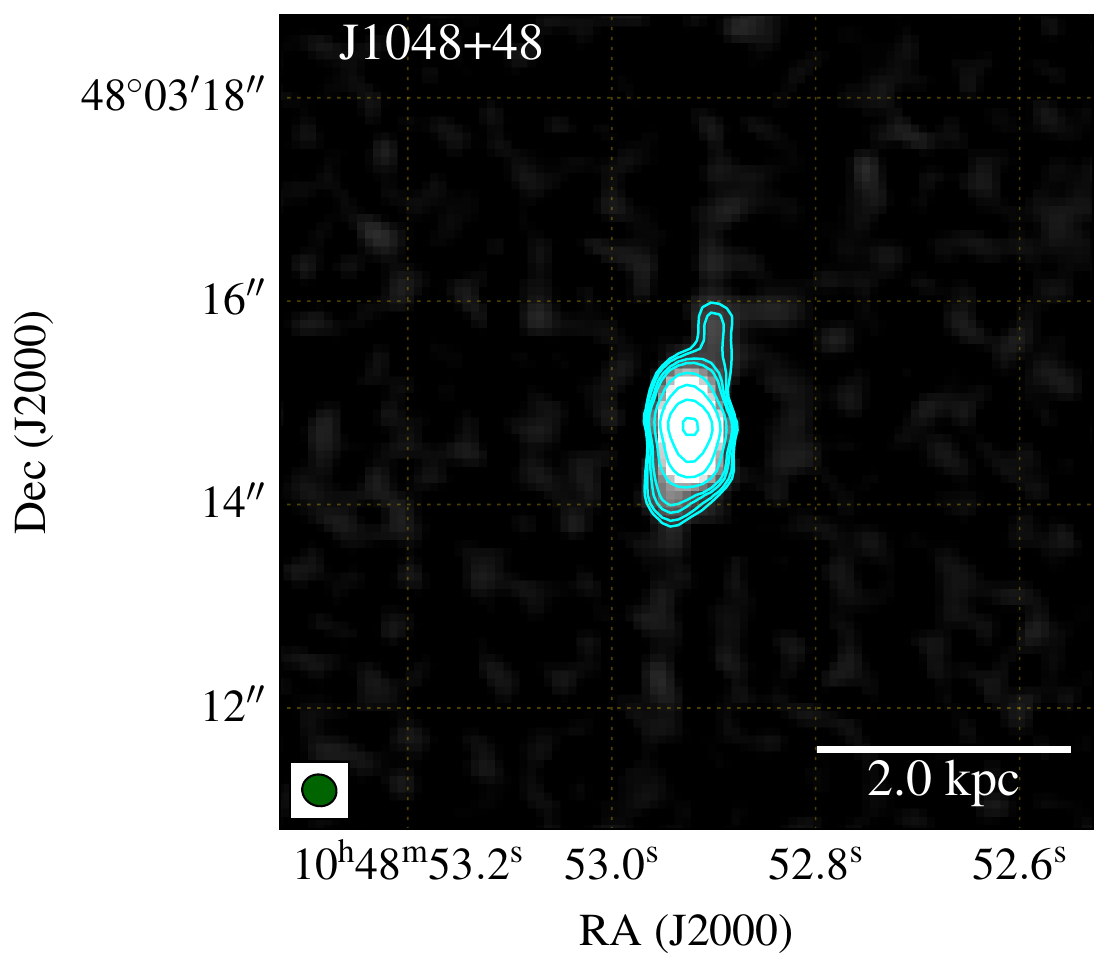}
\includegraphics[keepaspectratio,width=1.0\columnwidth]{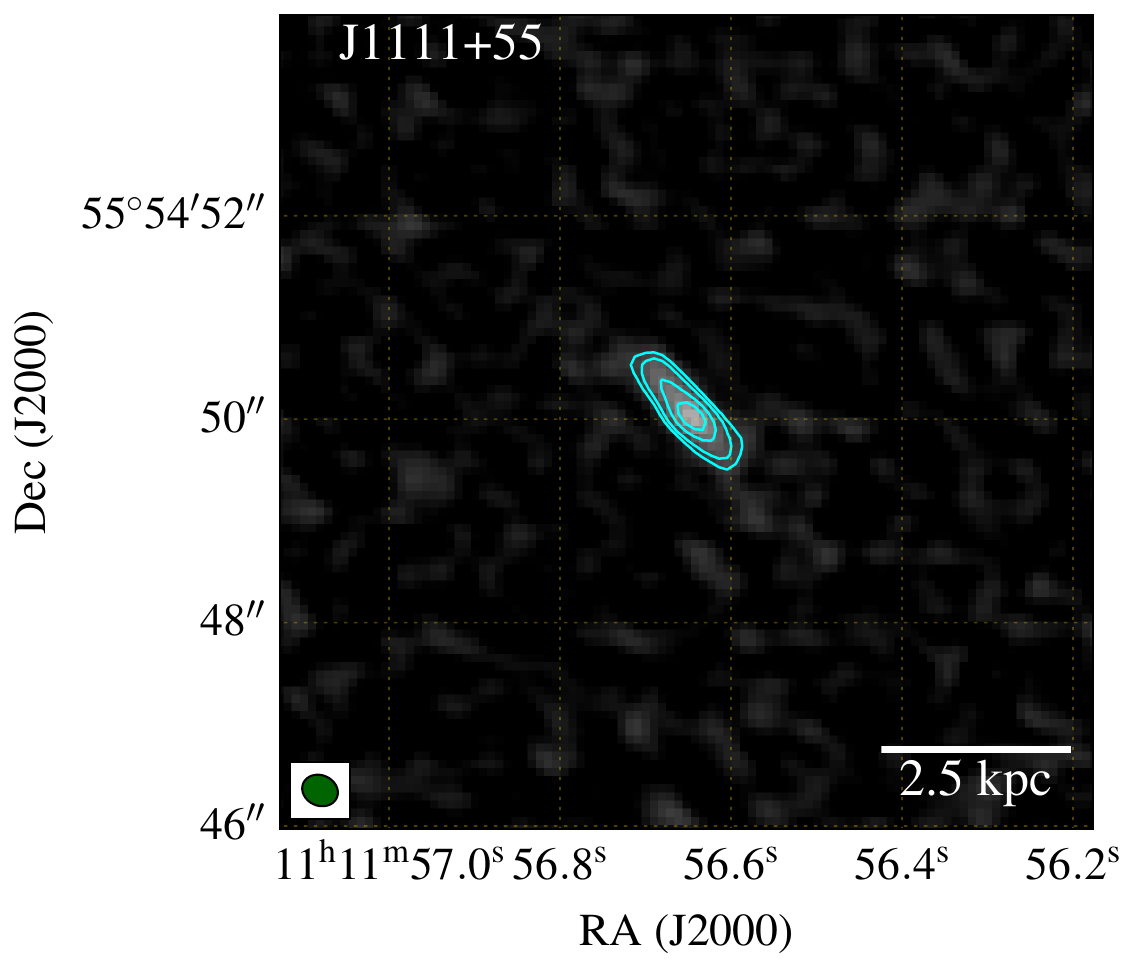}

\includegraphics[keepaspectratio,width=1.0\columnwidth]{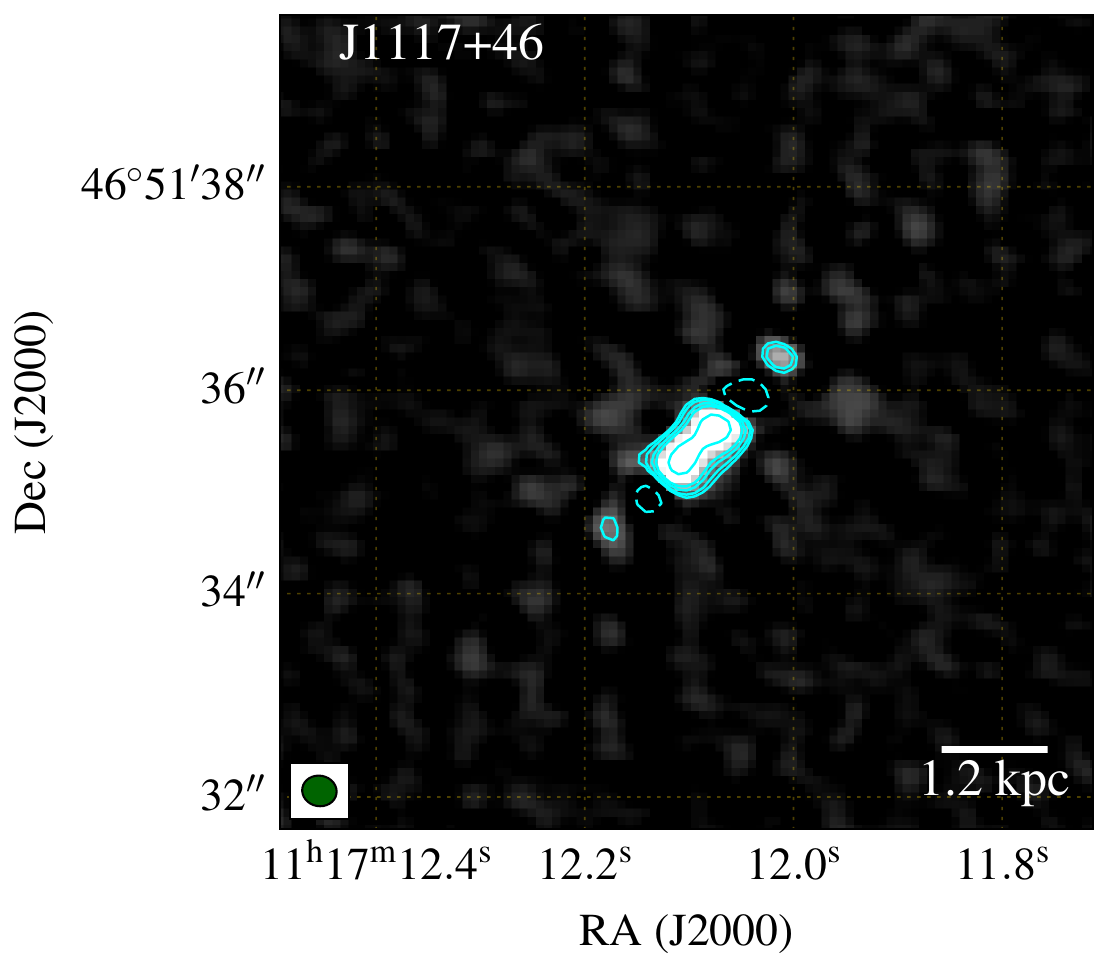}
\includegraphics[keepaspectratio,width=1.0\columnwidth]{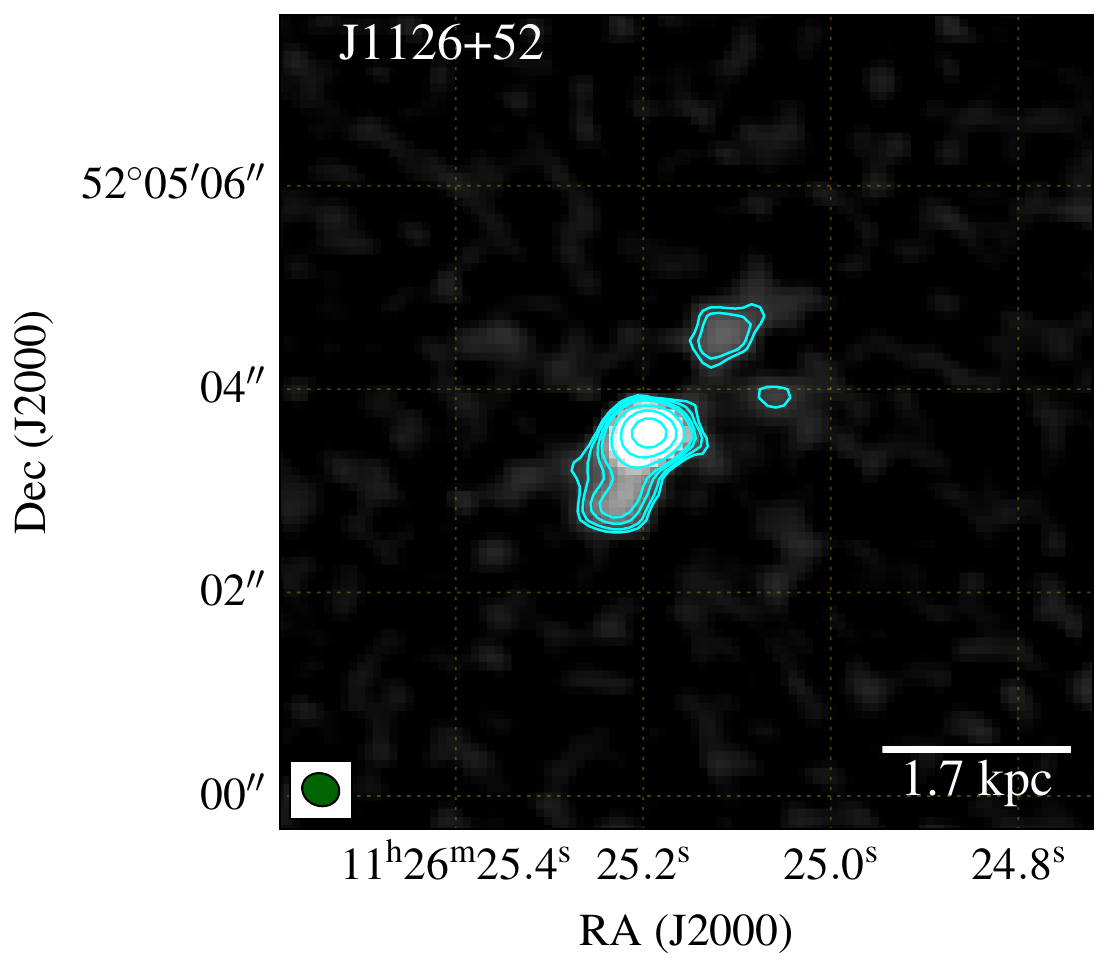}

\includegraphics[keepaspectratio,width=1.0\columnwidth]{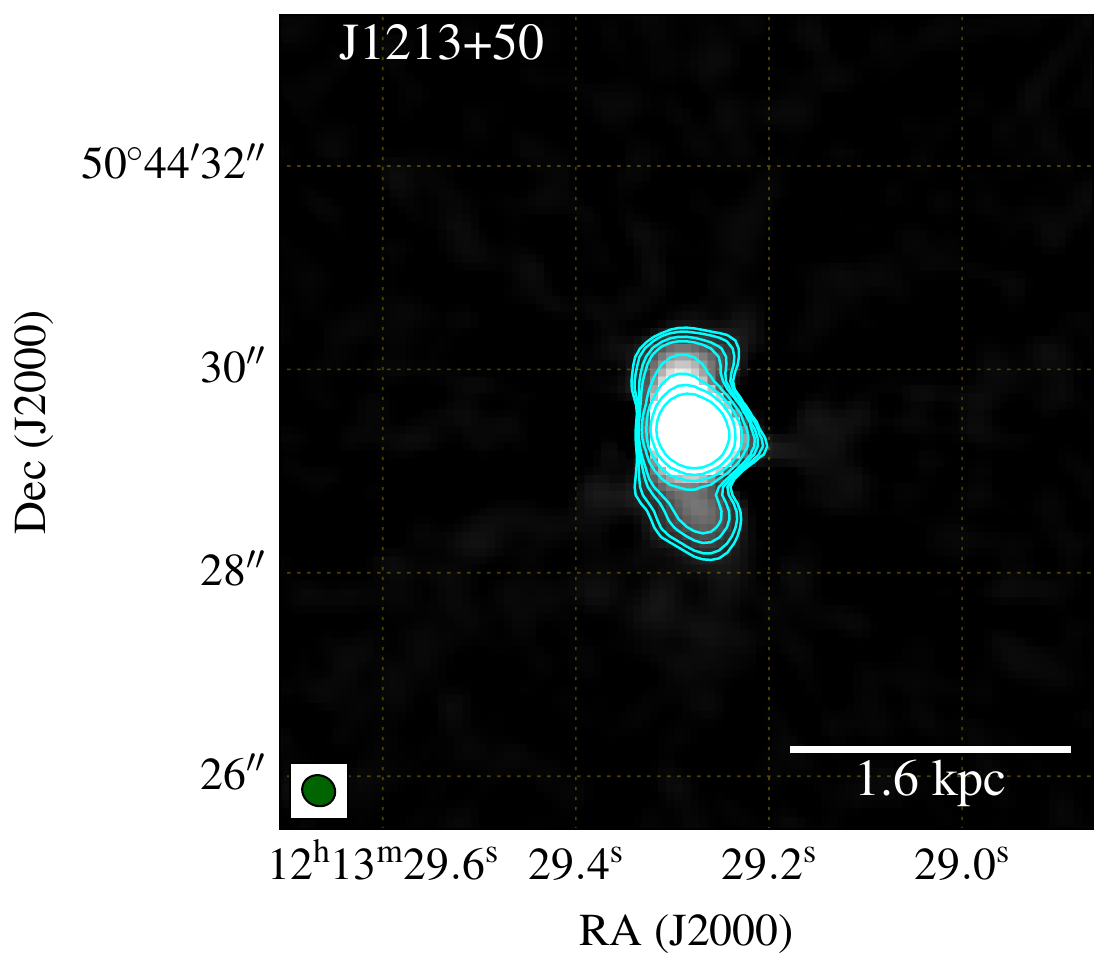}
\includegraphics[keepaspectratio,width=1.0\columnwidth]{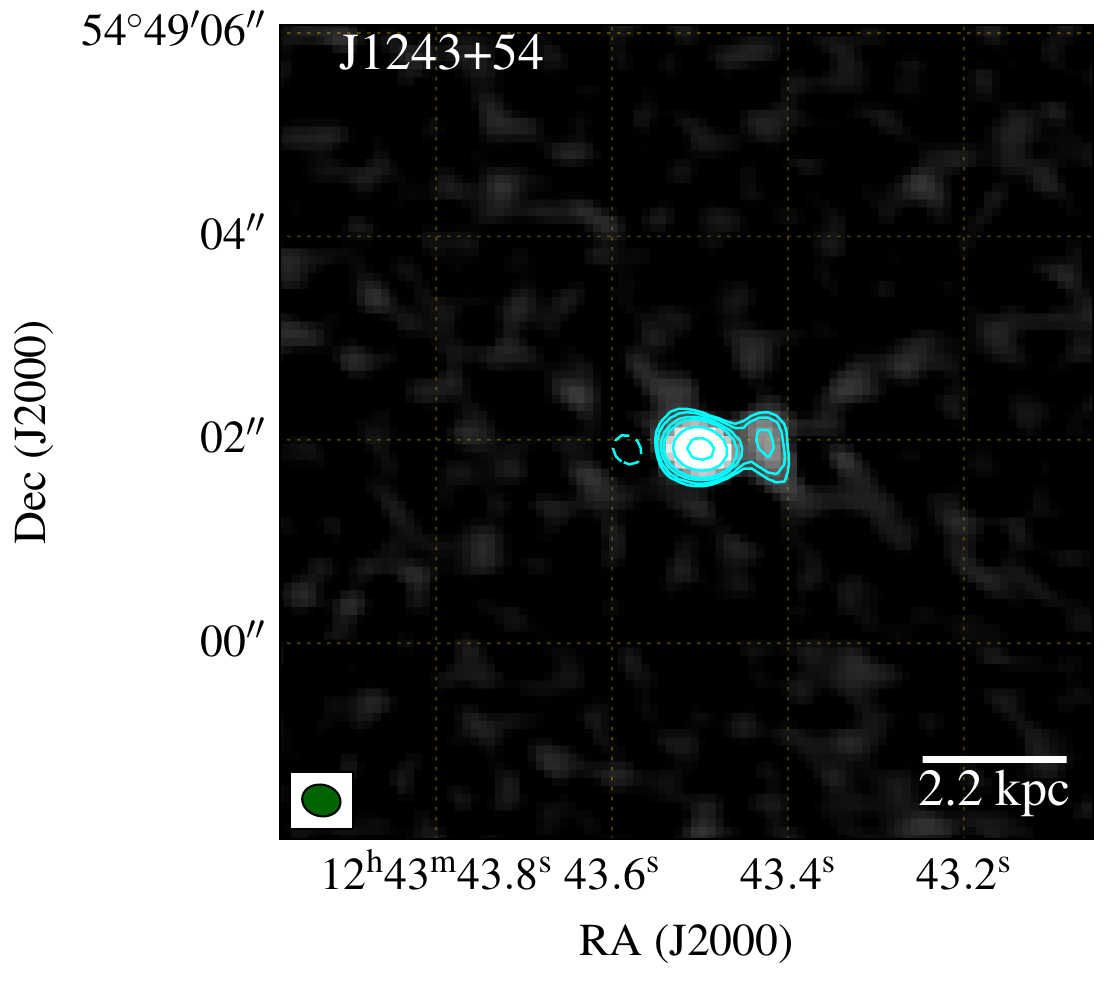}
\caption{VLA images overlaid with VLA 6 GHz contour maps (in cyan) at levels [-3, 3, 5, 10, 20, 40,...,640]$\times$ $\rm{\sigma^{rms}_{6.0}}$ where $\rm{\sigma^{rms}_{6.0}}$ is presented in Table~\ref{tab:extended} for each extended object. The beam of the VLA is shown in the bottom left corner (green ellipses).}
\label{fig:images}
\end{figure*}

\begin{figure*}
\includegraphics[keepaspectratio,width=1\columnwidth]{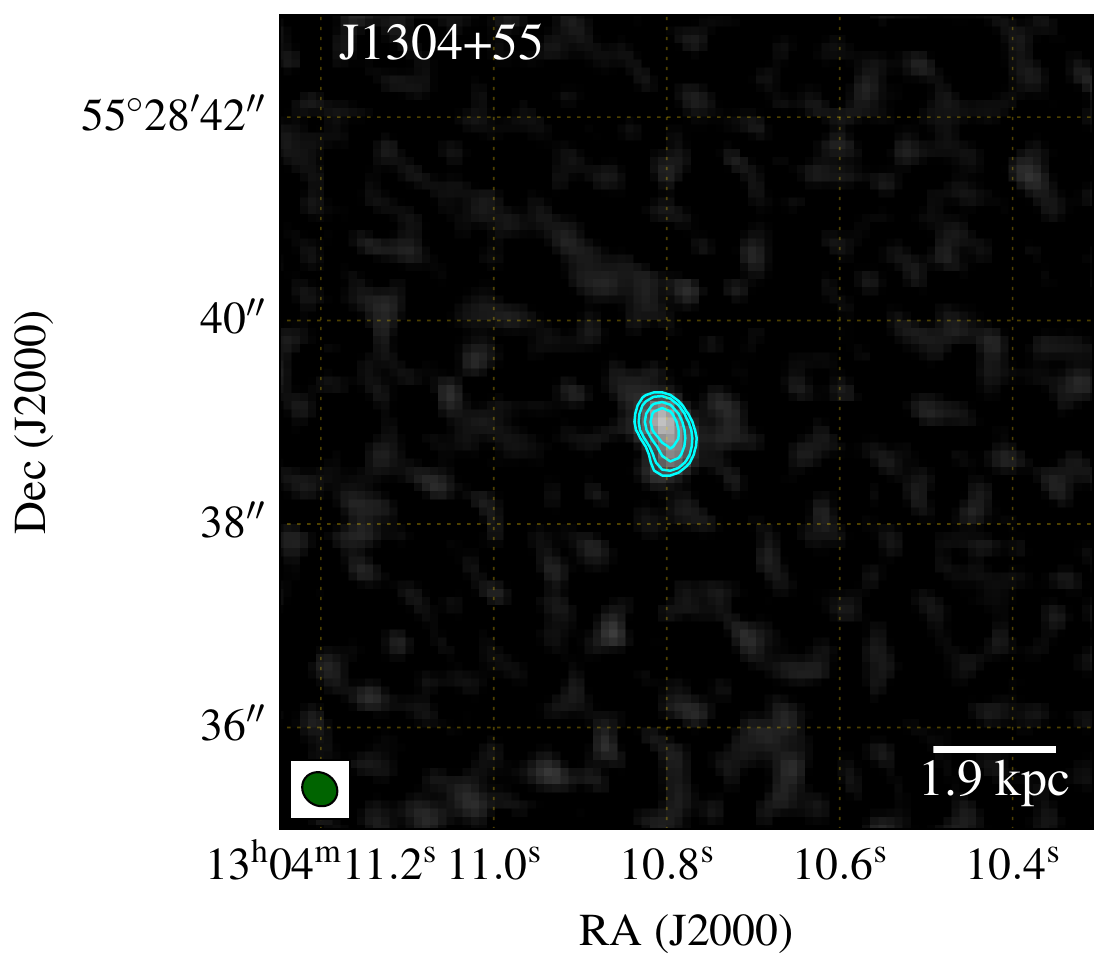}
\includegraphics[keepaspectratio,width=1\columnwidth]{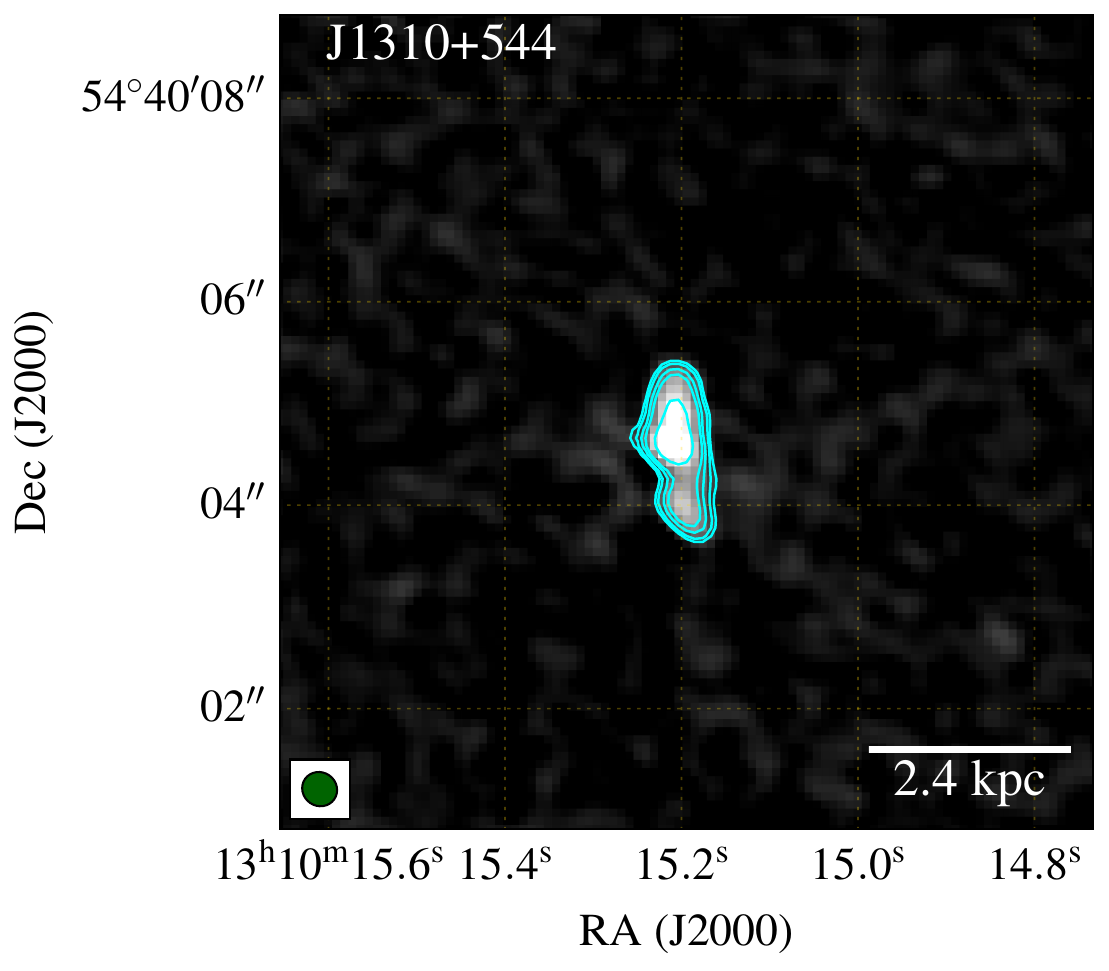}

\includegraphics[keepaspectratio,width=1.05\columnwidth]{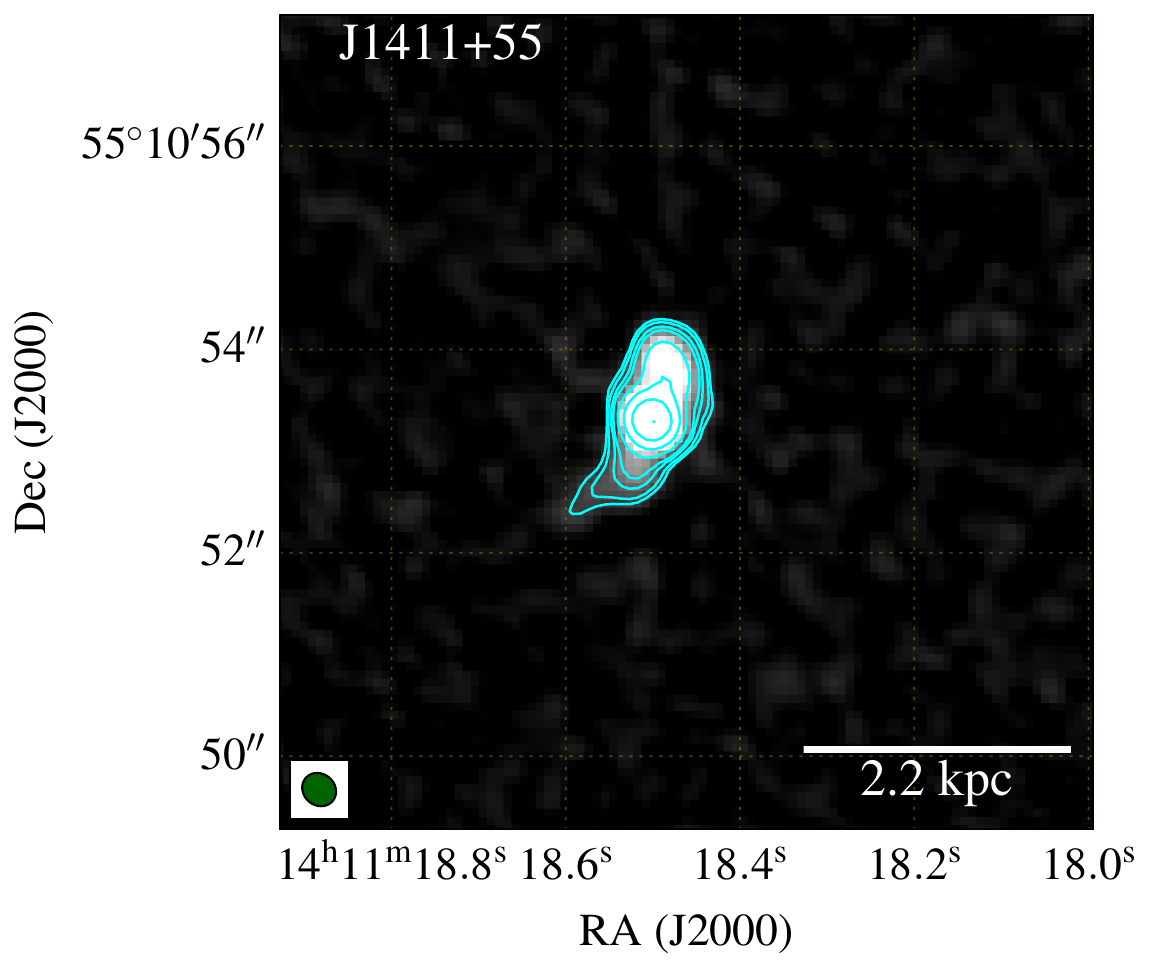}
\includegraphics[keepaspectratio,width=1\columnwidth]{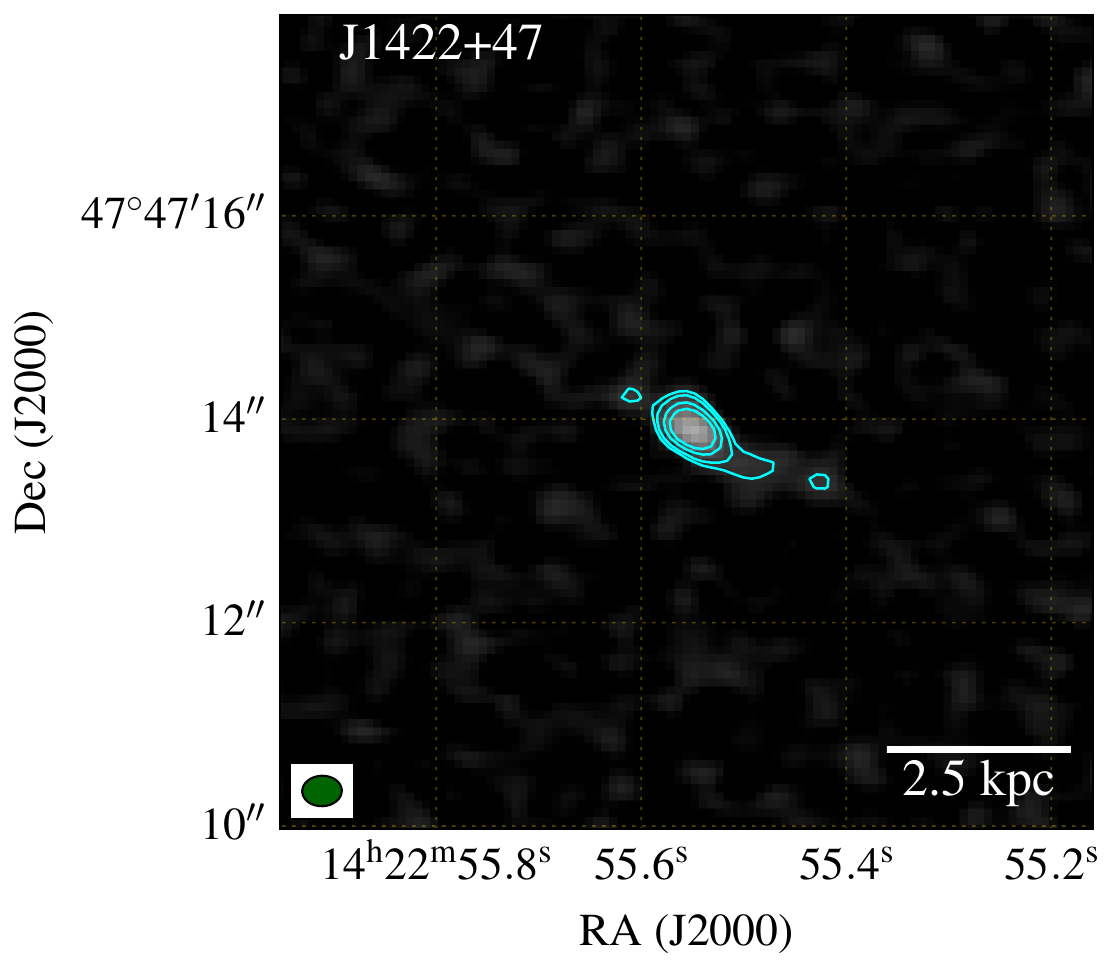}
\caption{VLA images overlaid with VLA 6 GHz contour maps (in cyan) at levels [-3, 3, 5, 10, 20, 40,...,640]$\times$ $\rm{\sigma^{rms}_{6.0}}$ where $\rm{\sigma^{rms}_{6.0}}$ is presented in Table~\ref{tab:extended} for each extended object. The beam of the VLA is shown in the bottom left corner (green ellipses).}
\label{fig:images1}
\end{figure*}

\subsection{Extended radio emission and radio maps}
\label{sec:Re}
The remaining 10/55 ($\sim$18\%) objects show extended radio emission and are resolved or at least marginally resolved in the VLA images. Information on these 10 objects is presented in Table~\ref{tab:extended} and VLA images overlaid with contour maps for each of these sources are shown in Fig.~\ref{fig:images} and \ref{fig:images1}. A brief description of their radio morphologies and projected physical scales (measured at the $3\sigma$ boundary of contour maps) follows here.

\begin{table*}
\caption{Source properties of the 10 extended VLA objects at 6 GHz.}
    \begin{tabular}{lccccccccc}
    \hline\hline
         Source&$\theta^{r}_{6.0}$&$\vartheta_{6.0}$&$S\rm{^{int}_{6.0}}$&$S\rm{^{p}_{6.0}}$&$\sigma\rm{_{6.0}^{rms}}$&$\rm{\alpha^{6000}_{150}}$&$\Theta$&$D$&log$_{10} (L_{6.0})$\\
    \hline
        J1048+48& 0.34 $\times$ 0.31&73.0&6.37 $\pm$ 0.26 & 3.28 $\pm$ 0.09& 15.3 & 0.63 $\pm$ 0.01 & 2.5 & 2.0&22.37 \\
        J1111+55${\textbf{*}}$& 0.36 $\times$ 0.30&66.6&0.54 $\pm$ 0.04&0.15 $\pm$ 0.01& 15.2 & 0.62 $\pm$ 0.02 & 1.9 & 2.5&21.77 \\
        J1117+46& 0.34 $\times$ 0.30&79.7&2.65 $\pm$ 0.35& 1.08 $\pm$ 0.10 & 30.4& 0.42 $\pm$ 0.04 & 1.0 & 1.2 &22.34 \\
        J1126+52& 0.37 $\times$ 0.32&70.6&3.46 $\pm$ 0.25& 2.63 $\pm$ 0.12& 25.8 & 0.53 $\pm$ 0.02 & 1.9 & 1.7 &22.25 \\
        J1213+50& 0.33 $\times$ 0.30&64.0&74.63 $\pm$ 0.47& 69.58 $\pm$ 0.26& 45.2 & 0.25 $\pm$ 0.10 & 2.8 &1.6&23.19\\
        J1243+54& 0.38 $\times$ 0.31&77.1&1.93 $\pm$ 0.18& 1.70 $\pm$ 0.09& 23.2  & 0.34 $\pm$ 0.03 & 1.4 & 2.2 &22.49 \\
        J1304+55& 0.36 $\times$ 0.32&52.7&0.60 $\pm$ 0.06& 0.33 $\pm$ 0.02& 19.9 & 0.50 $\pm$ 0.03 & 1.2 & 1.9&21.94 \\
        J1310+544${\textbf{*}}$& 0.35 $\times$ 0.33&54.7&2.12 $\pm$ 0.82& 0.70 $\pm$ 0.07& 25.3 & 0.26 $\pm$ 0.10 & 2.0 & 2.4&22.28 \\
        J1411+55& 0.35 $\times$ 0.31&49.3&4.91 $\pm$ 0.42& 3.93 $\pm$ 0.21& 18.1 & 0.46 $\pm$ 0.02 & 2.6 & 2.2&22.28 \\
        J1422+47& 0.39 $\times$ 0.30&-88.0 &0.44 $\pm$ 0.05& 0.21 $\pm$ 0.02& 13.8& 0.55 $\pm$ 0.03 & 1.8 & 2.5 &21.70 \\
    \hline
    \end{tabular}\\
    {\raggedright \textit{Notes:} Column description: (1) Source name; (2) Restoring beam (arcsec); (3) Position angle (PA; degrees); (4) Integrated flux density (mJy); (5) Peak flux density (mJy); (6) RMS values estimated from a region free from emission and close to the source in the VLA images ($\mu$Jy$/$beam); (7) Radio spectral index estimated between 150 MHz and 6 GHz; (8) Source angular size estimated at the $3\sigma$ contour boundary (arcesc); (9) Source physical sizes (kpc); (10) Radio luminosity (W Hz$^{-1})$. Note that extended objects in both the VLA and LoTSS DR2 images are represented by an asterisk (${\textbf{*}}$).\par}
    \label{tab:extended}
\end{table*}

\subsubsection{J1048+48}
\label{sec:93+}
The VLA radio images and contours for this source show a bright slightly resolved radio core with a symmetric extension. The two-jetted morphology is projected at $\sim$2.0 kpc in physical size. This object is unresolved in the DR2 images.

\subsubsection{J1111+55}
\label{sec:J1111}
The VLA radio contours for this object show a two-sided jet with a projected length of $\sim$2.5 kpc. The emission seems to propagate from its centre to the north-east direction and there is no sign of a compact radio core. However, the object clearly shows extended radio emission in the DR2 image extending in the same direction as in the VLA image. In the DR2 image, this object shows a bright unresolved central structure with double symmetric jets/plumes faintly seen propagating away from the central component. However, the $3\sigma$ radio contours overlaid on the DR2 image do not fully capture this radio emission (see Fig.~\ref{fig:appndx}).

\subsubsection{J1117+46}
\label{sec:10+}
This object shows a double-lobed radio structure with two bright edges, with a FRII radio morphology. The blobs on either side of the object appear to be real, however, future observations of this source will help in confirming if they may be image artefacts. The projected length for this source is estimated at $\sim$1.2 kpc. The object lacks any extended radio emission in the DR2 images.

\subsubsection{J1126+52}
\label{sec:15+}
The VLA radio image and contours for this object show a slightly resolved bright radio core with a one-sided jet extending south-east from the centre. It is uncertain whether the blob in the north-west of the source is an image artefact or real emission. The projected length of this source is $\sim$1.7 kpc. This object is also unresolved in the DR2 images.

\subsubsection{J1213+50}
This source (NGC 4187) shows extended radio emission ($\sim$1.6 kpc) in the VLA images with what appears to be a symmetrically two-jetted morphology. This is consistent with the findings of \citet{2019MNRAS.482.2294B} at a similar resolution. The object is unresolved in DR2 images. 

\subsubsection{J1243+54}
\label{sec:52+}
The VLA radio image and contours for this object show a bright and slightly resolved radio core with a radio jet extending to the west. The one-jetted structure is projected length at $\sim$2.2 kpc in physical size. The object is unresolved in the DR2 images.

\subsubsection{J1304+55}
\label{sec:78+}
This object shows a slightly resolved structure in the VLA images, in what appears to be a one-sided jet. The projected physical size is estimated at $\sim$1.9 kpc. The object is unresolved in the DR2 images.

\subsubsection{J1310+544}
\label{sec:J1310}
The VLA radio image and contours for this object show a slightly resolved radio core and radio structures/lobes extending symmetrically on either side of the radio core (two-jetted). The radio morphological classification of this object is of an FRI. The projected length of the radio emission is estimated at $\sim$2.4 kpc. In the DR2 images, this source shows faint and diffuse symmetric extended emission (Fig.~\ref{fig:appndx}). The fact that a similar radio morphology is seen in the VLA images but on smaller scales does suggest that the emission (represented by arrows, Fig.~\ref{fig:appndx}) seen in the DR2 is associated with jets rather than image artefacts.

\subsubsection{J1411+55}
\label{sec:46+}
The VLA radio image and contours for this object show a bright radio core with a two-sided symmetric structure estimated at $\sim$2.2 kpc in projected physical size. The object is unresolved in DR2 images.

\subsubsection{J1422+47}
\label{sec:56+}
The VLA radio contours for this object show a slightly resolved one-jetted structure extending to the south-west direction. The projected physical size of this object is estimated at $\sim$2.5 kpc. The object is unresolved in DR2 images.

\subsection{Radio spectral index}\label{sec:radio_spectra}
Investigating the shape of the radio spectrum of RLAGN can provide clues about the physics of radio sources \citep{1970ranp.book.....P,1964ApJ...140..969K}. A flat/inverted spectral index ($\alpha<0.5$) indicates recently accelerated relativistic particle populations, often associated with compact regions (e.g. cores/hotspots), while a steep spectral index ($\alpha>0.5$) indicates aged electron populations and is often associated with large-scale radio structures such as the jets and lobes of resolved RLAGN.

The radio spectral index is also important for classifying RLAGN to investigate if they are CSO \citep[e.g.][]{1994ApJ...432L..87W}, GPS (a very steep radio spectrum $\alpha\geq0.8$ at high frequencies and a very inverted $\alpha\leq-0.8$ spectrum at low frequencies) or CSS ($\alpha \geq 0.5$ at high frequencies) sources \citep[e.g.][]{1990A&A...231..333F,1998PASP..110..493O}.

\begin{figure}
	\includegraphics[width=\columnwidth]{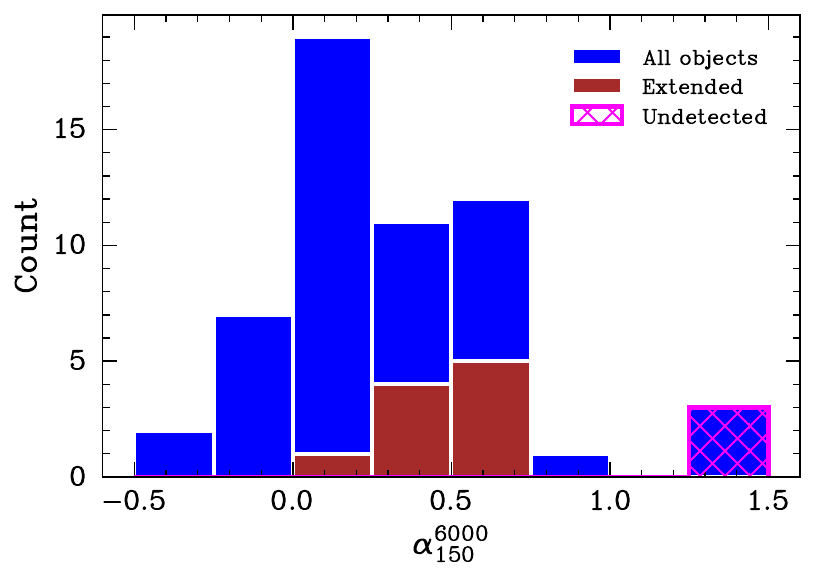}
    \caption{A histogram distribution of spectral indices computed between 150 MHz and 6 GHz. Blue bins represent the distribution of all sources in our sample overlaid with VLA extended objects (in brown) and upper limits of undetected sources in magenta.}
    \label{fig:spec}
\end{figure}

The spectral index distribution in Fig.~\ref{fig:spec} shows a broad range of values ranging from steep to flat and/or inverted radio spectra ($-0.36<\alpha_{150}^{6000}<1.31$). This implies that our RLAGN sample contains various types of compact radio sources such as CSOs, GPSs, and CSSs, and is not homogeneous in its properties. About 29\% (16 out of 55) of these objects show steep radio spectral indices ($\alpha>0.5$) which is consistent with standard models for optically thin synchrotron radio emission; 8 of these objects are compact in the VLA images at 6 GHz (0.35 arcsec resolution) while 5 show extended radio structures. The remaining objects (5 extended and 34 compact) show flat and/or inverted radio spectra ($\alpha<0.5$); this flattening in radio spectral indices could suggest that the emission is affected by synchrotron self-absorption (SSA), which is a key feature for the morphology of radio jets on sub-kpc scales \citep[e.g.][]{2021A&ARv..29....3O}.

\begin{figure}
    \centering
    \includegraphics[width=\columnwidth]{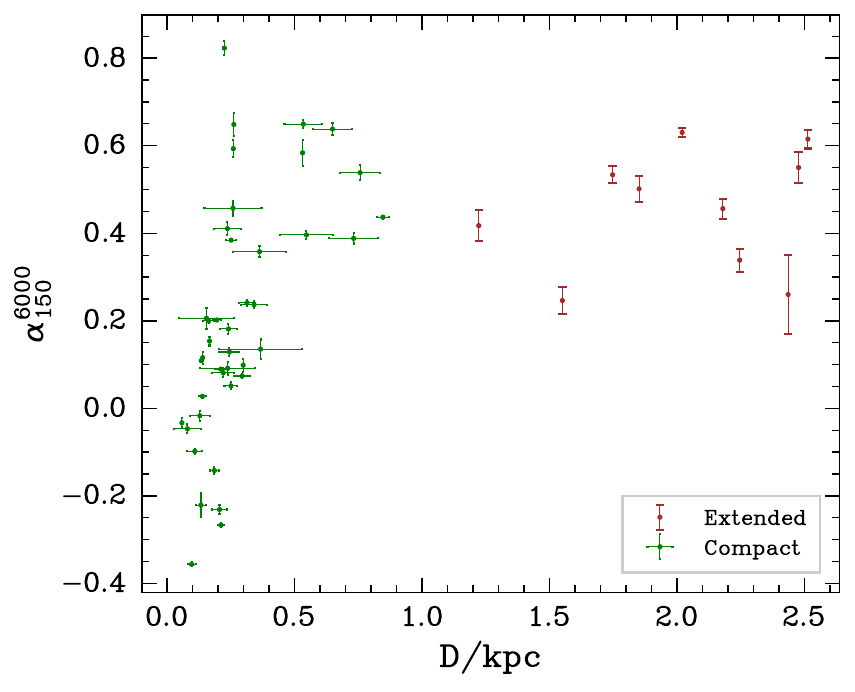}
    \caption{The spectral index/linear size distribution ($\alpha-D$ diagram) of both extended (brown points) and compact objects (green points) in our sample. Sizes of extended objects are maximum lengths while the sizes of compact sources are upper limits; the error bars shown in compact sources were obtained using {\scriptsize{IMFIT}}. VLA undetected objects are not included on this plot as their DR1 sizes (upper limits) are greater than 10 kpc. We see that the flat/inverted spectral indices are associated with compact sources.}
    \label{fig:spec_vs_size}
\end{figure}

In Fig.~\ref{fig:spec_vs_size}, we show the dependence of the spectral index on the total projected size (the $\alpha-D$ diagram) of the objects. As expected, a majority of the compact objects (34/42) show flat/inverted spectral indices, whereas 5/10 of the extended objects show steep spectral indices. We note two distinct categories of very compact flat spectrum populations and resolved steep spectrum populations. The latter are analogous to CSS objects, and may suggest that these objects are physically small in size, young and rapidly growing sources \citep[e.g.][]{2000MNRAS.319..445S,2021A&ARv..29....3O} that could potentially grow into more powerful and extended FRI/II objects.

\begin{figure}
    \centering
    \includegraphics[width=\columnwidth]{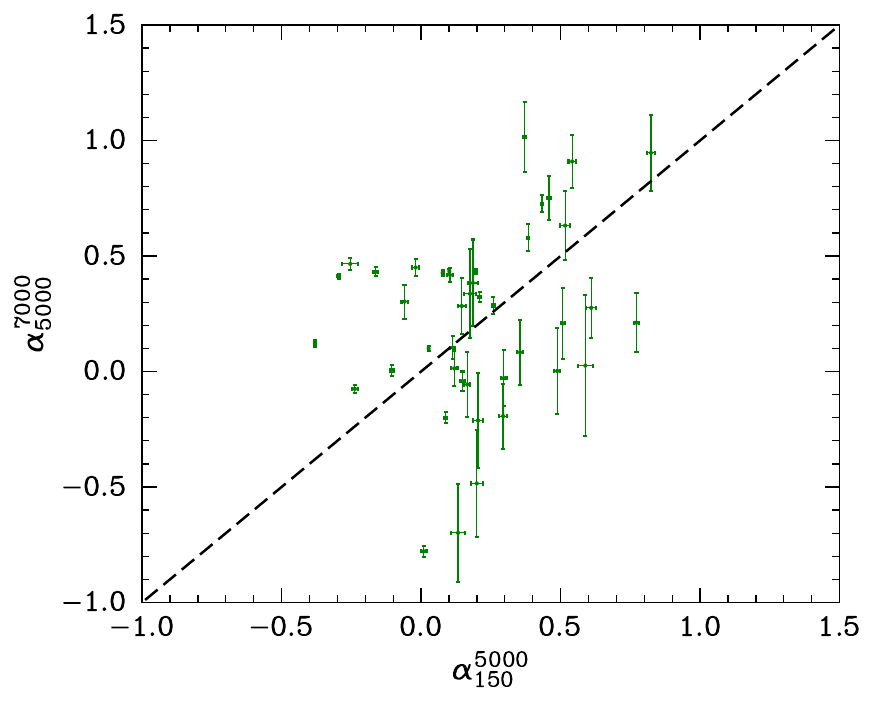}
    \caption{Radio colour-colour plots of 42 compact sources in our sample showing spectral indices from 150 MHz to 5 GHz against those from 5 to 7 GHz. The dashed black line indicates a power law; objects that lie above this line have spectral steepening at higher frequencies, while those below the line have spectral steepening at lower frequencies.}
    \label{fig:alpha_alpha}
\end{figure}

\begin{figure*}
\includegraphics[keepaspectratio,width=2\columnwidth]{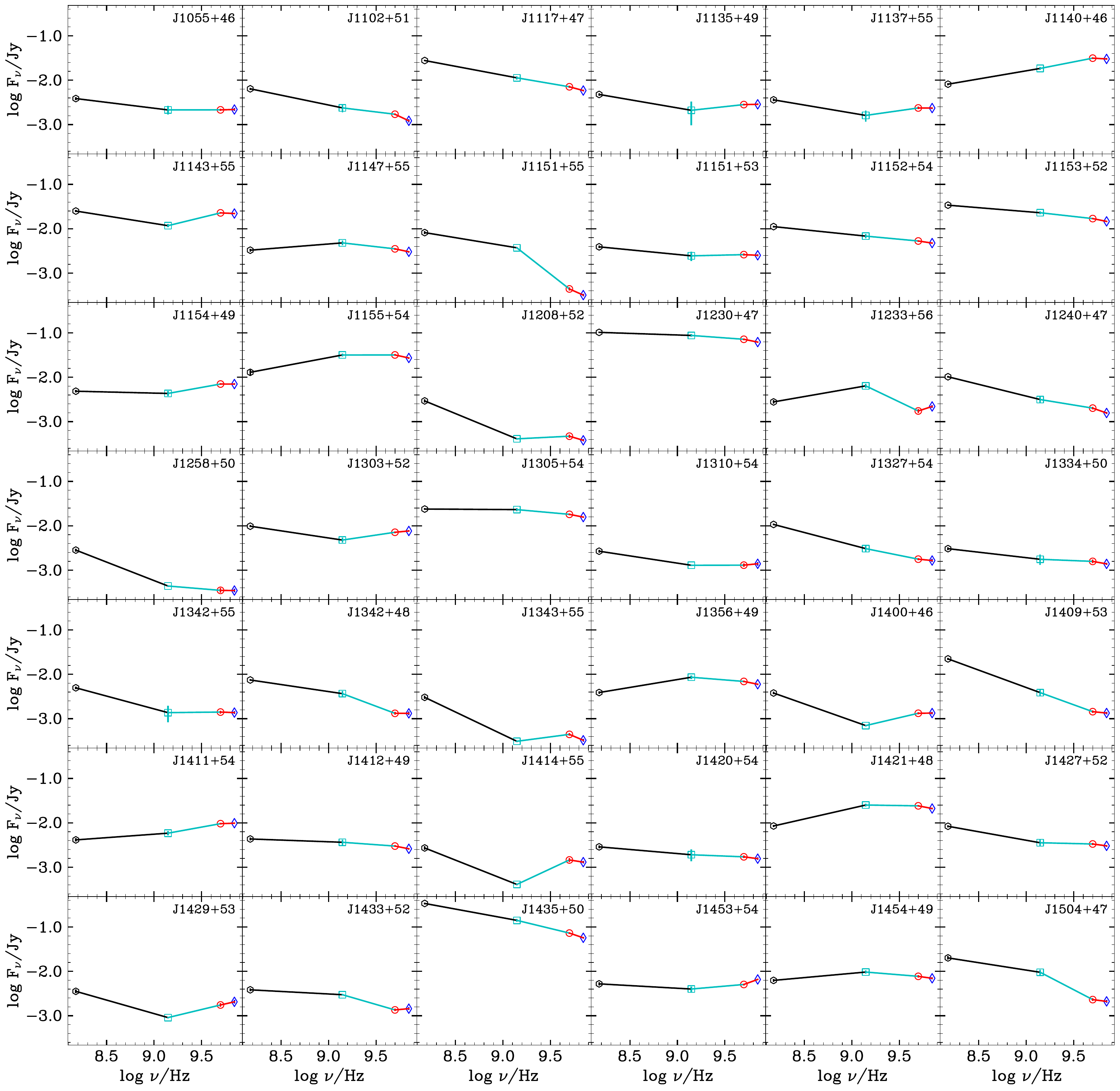}
\caption{Radio spectral energy distributions for the 42 compact objects in our sample, with data at 150 MHz (LOFAR; black points), 1.4 GHz (FIRST; cyan points), 5 GHz (VLA; red points), and 7 GHz (VLA; blue points). Black, cyan, and red lines indicate a power law with spectral indices estimated between 150 MHz and 1.4 GHz, 1.4 GHz and 5 GHz, and between 5 and 7 GHz respectively. Error bars are plotted but are too small to be visible on some of these plots. However, this information is tabulated in Tables~\ref{tb1:data} and \ref{tb:appendix1}.} 
\label{fig:SEDs}
\end{figure*}

We also obtained multi-frequency data (see additional data in Table~\ref{tb:appendix1}) to investigate the radio colour-colour plot of spectral indices ($\alpha_{5000}^{7000}$ versus $\alpha_{150}^{5000}$) for compact sources in our sample. We see in Fig.~\ref{fig:alpha_alpha} that a small fraction of our compact objects are steep-spectrum sources, with a large fraction of flat/inverted sources. Objects above the one-to-one correspondence exhibit spectral steepening at higher frequencies, while those below the line could indicate absorption at lower frequencies. This becomes apparent in Fig.~\ref{fig:SEDs} where we show radio spectral energy distributions (SEDs) for these compact objects. We notice a curvature in around 12\% (5/42) of these sources which is indicative of synchrotron self-absorption (SSA) and/or free-free absorption (FFA) at 150 MHz \citep[e.g.][]{2021A&ARv..29....3O}. We investigate self-absorption in the next section.

\subsection{Investigating self-absorption}
Here we investigate the effects of self-absorption for the 34 unresolved sources with flat/inverted radio spectra in the context of the two most favoured absorption models: synchrotron self-absorption (SSA) and free-free absorption (FFA). SSA and FFA models can be used respectively to estimate the physical properties of the radio-emitting and radio-absorbing region.

\subsubsection{Synchrotron self-absorption (SSA)}
\label{sec:SSA} %1984AJ.....89.1327H
When the radiation from a radio-emitting source interacts with the relativistic particles within the source, this can result in its absorption. This phenomenon is called synchrotron self-absorption \citep[SSA;][]{1966AuJPh..19..195K, 1974ApJ...188..353J}. We can use SSA models to estimate the magnetic field and the physical size \citep[e.g.][]{2022MNRAS.510..815K,2022ApJ...934...26P} of the compact radio-emitting source.

To estimate the physical properties of the radio-emitting region, we assume SSA at 150 MHz. We also assume that our sources have an ellipsoidal geometry, and are in a condition of equipartition of energy between synchrotron-emitting particles and magnetic fields. We start from the definition of the absorption coefficient of synchrotron radio emission presented by \citet{2011hea..book.....L}:
\begin{equation}
\label{eq:chi_SI}
\chi_{\nu}=3.354\times10^{-9}kB^{(p+2)/2}(3.54\times10^{18})^pb(p)\nu^{-(p+4)/2}\,\,\,\rm{m^{-1}}
\end{equation}

where $\nu$ is the observing frequency, $k$ is the normalization of the electron energy spectrum given by $N(E)=kE^{-p}$, $p$ is the electron energy index defined by $\alpha=(p-1)/2$ typically lying in the range $2-3$ for many radio-selected AGN. The quantity $b(p)$ is a constant that depends on the value of the exponent $p$. Assuming the low-frequency spectral index of $\alpha=0.5$ corresponds to $p=2$ (in the case of synchrotron emission from non-relativistic particles, a power-law index of 2 indicates that the number of particles decreases with increasing energy, which is characteristic of many particle acceleration processes), then $b(p)=0.269$ \citep{2011hea..book.....L}. 

We can estimate $B$ (the source magnetic field) using the derivations of \citet{2004ApJ...612..729H}:
\begin{equation}
\label{eq:b-field}
    B=\left(2\mu_0(1+\kappa) \frac{J(\nu)}{C}\nu^{(p-1)/2}I\right)^{2/(p+5)}.
\end{equation}

From Eq.~\ref{eq:b-field}, $J(v)$ is the synchrotron emissivity which can be determined from the radio luminosity and volume (i.e. $J(\nu)=L_{\rm{radio}}/V$). Eq.~\ref{eq:b-field} shows that $B$ fields depend on several observables for the source: flux density, redshift, observing frequency, and measured Gaussian angular size. We are using 6 GHz VLA image measurements for each source to evaluate Eq.~\ref{eq:b-field}. The $\kappa$ term is the ratio of the energy density in non-radiating to radiating particles given by $\zeta=1+\kappa$. In this case, $\zeta=1.0$ for no non-radiating particles and
\begin{equation}\label{eq:I}
    I=\begin{cases} 
    \rm{ln} (E_{max}/E_{min}), &\text{$p=2$}\\ 
    \rm{\frac{1}{(2-p)[E^{(2-p)}_{max}-E^{(2-p)}_{min}]}}, &\text{$p\ne2$} \end{cases}.
\end{equation}
From Eq.~\ref{eq:I} 
\begin{equation}
    C=c(p)\frac{e^3}{\epsilon_0 c m_e}\left(\frac{m^3_ec^4}{e}\right)^{-(p-1)/2}
\end{equation}
where the quantity $c(p)$ is dimensionless and is of order 0.050 for $p=2$. The value of $I$ exhibits only a weak dependency on the choices of $E\rm{_{min}}$ and $E\rm{_{max}}$ ($=\gamma$$m_e c^2$), which represent the lower and upper limits, respectively, of electron energies. These limits, set at $\gamma\rm{_{min}}=10$ and $\gamma\rm{_{max}}=10^6$, are determined based on considerations related to the low-energy cutoffs within the energy spectrum of relativistic particles in hotspots, and they also take into account adiabatic expansion assumptions related to the scale of hotspots and lobes \citep[see e.g.][for further details]{2004ApJ...612..729H}. Eq.~\ref{eq:chi_SI} tells us that SSA will be important if:
\begin{equation}
\label{eq:tau}
    \tau_{\rm{SSA}} = \chi_{\nu}.d\approx 1
\end{equation}

where $d$ is the depth of the source (in meters) and is taken as the angular minor axis scale of the source. Using the information presented in Tables~\ref{tb1:data} and \ref{tab:unResolved}, and utilising the {\sevensize{PYSYNCH}}\footnote{\url{https://github.com/mhardcastle/pysynch/}} code \citep{1998MNRAS.294..615H} to estimate $B$ fields, we find for the 34 flat/inverted spectrum (unresolved) sources that the measured sizes (given by upper limits) do not satisfy Eq.~\ref{eq:tau}, which implies one of two possibilities: (1) the objects are not self-absorbed at 150 MHz or (2) the sizes (upper limits) presented in Table~\ref{tab:unResolved} are overestimates. The first point could be true as we can see from Fig.~\ref{fig:alpha_alpha} that many of the compact objects have flat/inverted radio spectra between 150 MHz and 7 GHz, which would not be compatible with a simple homogeneous self-absorption model. We investigate the second point by shrinking the volume (and re-evaluate the $B$ fields) for each source until the condition set in Eq.~\ref{eq:tau} is met. The resulting predictions are presented in Table~\ref{tab:SSA}, where we see that the SSA predicted sizes for these sources are well below the measured sizes presented in Table~\ref{tab:unResolved}.

\begin{figure}
    \centering
    \includegraphics[width=\columnwidth]{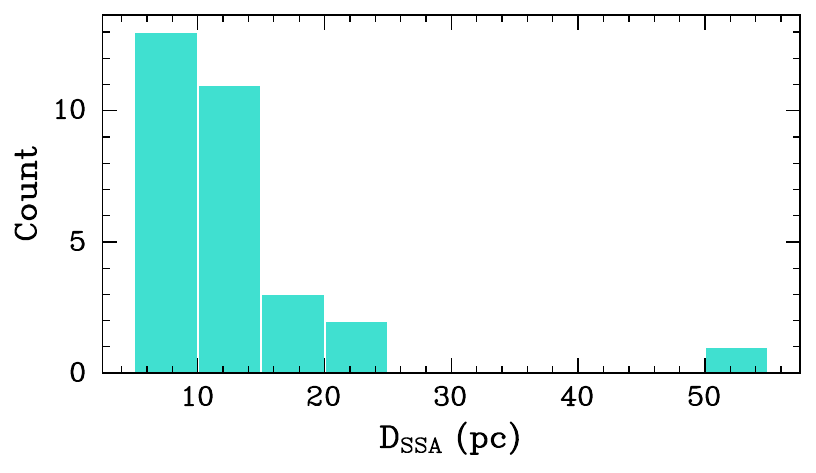}
    \includegraphics[width=\columnwidth]{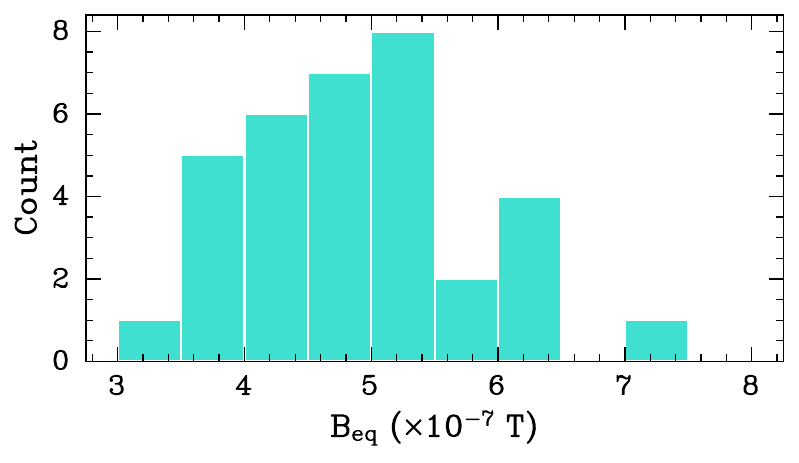}
    \caption{Distributions of source physical sizes (top panel) and source equipartition magnetic fields (bottom panel) assuming SSA at 150 MHz.}
    \label{fig:SSA}
\end{figure}

The distribution of the estimates of physical sizes and magnetic fields for the 34 flat/inverted spectrum objects are shown in Fig.~\ref{fig:SSA}. We find that the sizes of these objects have to be in the range of a few angular milliarcsecond scales ($\sim$2$-$30~mas, corresponding to $\sim$2$-$53 pc in linear scales) if they are to be self-absorbed at 150 MHz. Comparisons of these predicted sizes with actual sizes are not yet possible in our case, except for J1230+47, which is resolved at sub-parsec resolution in Very Long Baseline Interferometry (VLBI), Very Long Baseline Array (VLBA), European VLBI Network (EVN), and the Multi-Element Radio Linked Interferometer Network (eMERLIN) observations between 5$-$8.4 GHz \citep[][]{2018ApJ...863..155C,2021MNRAS.506.1609C,2021Galax...9..106B}; this object shows a two-sided jetted morphology extending up to a few pc in projected physical scales, comparable to our predictions. The angular sizes predicted for the compact sample in Table~\ref{tab:SSA} are well below the resolution of Dutch LOFAR (6 arcsec) and the VLA (0.35 arcsec), which means that they can only be resolved with long baseline interferometers such as eMERLIN, VLBA, and/or EVN at milli-arcsec resolution. Shown in Table~\ref{tab:SSA} and Fig.~\ref{fig:SSA} are estimates of the magnetic fields ($\sim$0.3$-0.75~\mu$T), an important property in constraining the evolution of young RLAGN \citep[e.g.][]{2002A&A...389..115D}. If accurate, these provide valuable information about the physical conditions in these sources.

Although the equipartion assumption provides a useful tool to estimate $B$ fields and particle densities in radio AGN, its limitations should be recognised as we know that it does not necessarily work for all sources \citep[e.g.][]{2008MNRAS.386.1709C}. Further multi-wavelength observations and modelling are necessary to gain a deeper understanding of the physical conditions in these sources.

\begin{table*}
    \centering
    \caption{SSA physical property predictions for the 34 flat/inverted spectrum compact objects.}
    \begin{tabular}{cccccccc}
    \hline\hline
        Source& Estimated\,physical\,size&$\rm{\theta_{SSA}}$& $B$ field & $k$ & $D_{\rm{SSA}}$ &$d_{\rm{SSA}}$& $\chi_v$ \\
        &&&&&&&\\
        &(pc)&(mas)&($\times10^{-7}$, T)&($\times10^{-9},\,\rm{m^{-3}}$)&(pc)&(pc)& ($\times10^{-16},\,\rm{m^{-1}}$)\\
        \hline
            J1055+46 &166 &5.9 $\times$ 2.3&5.1&9.1&9.8&3.8&0.08\\ %1
            J1102+51 &341 &7.9 $\times$ 2.2&5.4&10.0&10.4&2.9&0.10\\%2
            J1117+47 &251 &8.4 $\times$ 4.6&4.4&6.7&12.0&6.6&0.04\\%5
            J1135+49 &244 &4.8 $\times$ 3.7&5.0&8.6&4.9&3.8&0.07\\%7
            J1137+55 &220 &6.3 $\times$ 2.6&5.3&9.6&7.6&3.1&0.09\\%8
            J1140+46 &96 &13.2  $\times$ 12.7&3.6&4.5&13.8&13.2&0.02\\%9
            J1143+55 &139 &10.9 $\times$ 9.6&4.1&5.8&11.6&10.2&0.03\\%10
            J1147+55 &128 &6.4 $\times$ 3.2&5.3&9.6&6.4&3.2&0.09\\%11
            J1151+53 &250 &8.1 $\times$ 2.5&5.4&10.0&9.3&2.9&0.10\\%13
            J1152+54 &163 &7.8 $\times$ 4.0&4.8&7.9&9.1&4.6&0.06\\%14
            J1153+52 &195 &10.2 $\times$ 7.8&4.0&5.5&14.0&10.7&0.03\\%15
            J1154+49 &109 &8.7 $\times$ 4.3&4.9&8.4&9.1&4.5&0.07\\%16
            J1155+54 &133 &22.9 $\times$ 6.4&4.6&7.3&22.2&6.2&0.05\\%17
            J1230+47 &134 &21.6 $\times$ 15.1&3.8&5.0&16.8&11.7&0.02\\%22
            J1233+56 &367 &5.2 $\times$ 2.0&5.5&10.4&8.0&3.0&0.10\\%23
            J1240+47 &546 &3.8 $\times$ 3.7&4.4&6.7&6.9&6.8&0.04\\%24
            J1303+52 &294 &11.9 $\times$ 3.8&4.4&6.8&21.0&6.7&0.04\\%27
            J1305+54 &211 &11.4 $\times$ 8.5&3.6&4.5&19.2&14.4&0.02\\%30
            J1310+54 &298 &9.0 $\times$ 1.5&5.6&11.0&14.9&2.4&0.12\\%31
            J1327+54 &237 &3.8 $\times$ 3.1&6.1&12.9&2.5&2.1&0.16\\%34
            J1334+50 &362 &2.9 $\times$ 1.7&5.6&10.7&4.6&2.8&0.11\\%35
            J1342+48 &731 &6.2 $\times$ 1.8&5.5&10.4&10.5&3.0&0.11\\%37
            J1356+49 &185 &8.6 $\times$ 4.5&4.5&7.1&10.9&5.7&0.05\\%39
            J1400+46 &258 &2.4 $\times$ 1.7&6.3&13.8&2.5&1.7&0.19\\%40
            J1411+54 &205 &8.7 $\times$ 6.1&4.1&5.7&13.7&9.6&0.03\\%43
            J1412+49 &237 &4.8 $\times$ 3.9&4.6&7.3&6.4&5.3&0.05\\%44
            J1420+54 &140 &6.5 $\times$ 1.8&6.3&13.7&5.4&1.5&0.18\\%46
            J1421+48 &211 &14.1 $\times$ 8.5&3.9&5.1&19.2&11.7&0.03\\%47
            J1427+52 &313 &6.9 $\times$ 3.0&4.8&8.0&10.8&4.8&0.06\\%49
            J1429+53 &154 &9.1 $\times$ 1.1&7.4&18.8&7.7&0.9&0.34\\%50
            J1433+52 &241 &5.9 $\times$ 2.0&6.2&13.5&5.2&1.7&0.18\\%51
            J1435+50 &846 &28.8 $\times$ 12.6&3.2&3.6&52.9&23.1&0.01\\%52
            J1453+54 &58 &6.4  $\times$ 5.4&4.0&5.6&11.7&9.9&0.03\\%53
            J1454+49 &79 &7.2  $\times$ 5.5&5.0&8.7&5.3&4.0&0.07\\%54
        \hline
    \end{tabular}\\
    {\raggedright \textit{Notes:} Column description: (1) source name; (2) Estimated linear sizes from Table~\ref{tab:unResolved} for comparison with predicted SSA linear sizes; (3) Size estimates of objects assuming an ellipsoid shape; (4) SSA estimated magnetic fields; (5) Estimates of number density of electrons; (6) SSA maximum linear size prediction; (7) SSA predicted source depth; (8) Estimates of the absorption coefficient. Errors in predicted values are of the order 0.0001, and so are not shown here. \par}
    \label{tab:SSA}
\end{table*} 

\subsubsection{Free-free absorption (FFA)}
\label{sec:FFA}
Another important model that can explain self-absorption at low frequencies is free-free absorption \citep[FFA;][]{1966AuJPh..19..195K,1997ApJ...485..112B}. FFA, in principle, occurs when an electron in an ionized gas absorbs a photon and gains energy \cite[e.g.][]{2003AJ....126..723T,2014ApJ...780..178M,2018MNRAS.475.3493B}. Here we explore the probability of FFA at 150 MHz and we estimate the electron density (and path length of the absorber) of the radio-emitting region; a large ionized gas density (greater than 10$^{2}$~cm$^{-3}$) in compact RLAGN \citep[e.g. CSO/CSS/GPS sources;][]{2001ApJ...550..160M,2014ApJ...780..178M,2018MNRAS.475.3493B} is required for FFA to be important.

To estimate properties of the radio-absorbing region, we work with the equation presented by \citet{2016era..book.....C} that relates the FFA optical depth ($\tau_{\rm{FFA}}$) to the observing frequency ($v$; GHz), the electron temperature ($T_e; \rm{K}$), the free electron density ($n_e;~\rm{cm^{-3}}$), and the path length through the absorber ($S; \rm{pc}$): 

\begin{equation}\label{eq:tau_ffa}
    \resizebox{0.92\hsize}{!}{$\tau_{\rm{FFA}}\approx 3.28\times10^{-7}\left(\frac{T_e}{10^{4}~\rm{K}}\right)^{-1.35}\left(\frac{v}{\rm{GHz}}\right)^{-2.1}\int \left(\frac{n_e}{\rm{cm^{-3}}}\right)^2d\left(\frac{S}{\rm{pc}}\right)$}.
\end{equation}

The plasma ionisation condition for FFA requires that $T_e\gtrsim 10^{4}~\rm{K}$ and $n_e\gtrsim10^{3}~\rm{cm^{-3}}$ \citep[e.g.][]{2000PASJ...52..209K}. Assuming a homogeneous medium, we can therefore set $\tau_{\rm{FFA}}\approx1$ to obtain an emission measure of free electrons of:

\begin{equation}\label{eq:em}
    {\rm{EM}} \equiv \langle n_e\rangle^2 S\approx5.7\times10^4~\rm{cm^{-6}~pc}.
\end{equation}

Here, $\langle n_e\rangle$ is the average free electron density and we have assumed FFA is significant at 150 MHz. Eq.~\ref{eq:em} tells us that the emissivity from plasma scales as the square of the electron density; large values of $S$ imply a lower emissivity.

The inner regions of the AGN, for example the narrow-line regions (NLRs), are known to be ionised either by the AGN itself or by shocks from radio jets hence, if the radio source is seen through the NLR, it will be partly absorbed. In the local Universe, optical/UV observations of some AGN show spatially extended NLRs with electron densities in the range $\sim$10$^{2 - 4}~\rm{cm^{-3}}$ and sizes greater than 100 pc \citep[e.g.][]{peterson_1997,2011ApJ...732....9G}. However, most of the NLRs are compact with sizes spanning $1-3$ pc \citep{2013ApJ...779..109P}. If we adopt an electron density of $\langle n_e\rangle\approx10^{3}~\rm{cm^{-3}}$ \citep[e.g.][]{2011MNRAS.410.1527H}, we get a path length of the absorber of 0.06 pc (thus a free electron column depth of $N_e\approx2\times10^{20} \rm{cm^{-2}}$), which is well below the physical size estimates expected for NLRs. However, these estimates support the case for FFA at 150 MHz for a single line-emitting cloud \citep[with typical sizes given by the Str$\ddot{\rm{o}}$mgren depth $r\approx10^{18}n_e^{-1}$ cm;][]{peterson_1997} in the NLRs.

The above analysis suggests that FFA does not dominate in these sources but we note that models based on an inhomogeneous medium \citep[e.g. the ISM;][]{1997ApJ...485..112B,2018MNRAS.475.3493B} could provide a better constraint on estimating the properties of the absorber. However, modelling the jet-ISM interaction, for instance, strongly depends on being able to resolve our sources if we are to provide a strong case for FFA. We postpone this analysis to a future paper when VLBI observations of these compact objects are available.

\section{DISCUSSION}\label{sec:disc}
\subsection{Radio power/linear size diagram}
\label{sec:PD}
We discuss the relationship between the physical size and the radio luminosity of our sources with respect to the various types of compact and extended radio-selected AGN using a version of the so-called power/linear size diagram from \citet{2020NewAR..8801539H}.

The power/linear size ($P-D$) diagram is important in modelling the evolution of extended extragalactic radio sources \citep{1982IAUS...97...21B}. Its construction depends on observational measurements of the radio object's flux density, maximum angular size, and redshift \citep[e.g.][]{2016MNRAS.462.1910H}. These measurements directly result in estimates of the radio luminosity ($P$) and the maximum physical size ($D$), two important quantities that are often used to explain the evolution of RLAGN, that is, the radio luminosity increases with linear size \citep{2018MNRAS.475.2768H}.
%1995ApJ...451...76N

The \citet{1974MNRAS.167P..31F} type I and II (FRI/II) objects are traditionally thought to exhibit two distinct ranges of radio power with their radio emission extending to kpc scales and sometimes up to a few Mpc scales \citep[e.g.][]{2023A&A...672A.163O}. These objects were thought to cover two radio power regimes divided at around $\rm{10^{25}~W~Hz^{-1}}$ at 178 MHz \citep{1974MNRAS.167P..31F}. However, recent studies \citep[e.g.][]{2019MNRAS.488.2701M} have shown that FRI and FRII radio luminosities overlap at 150 MHz. CSO, GPS, and CSS sources display radio powers greater than $10^{25}~\rm{W~Hz^{-1}}$ at 1.4 GHz, and their sizes are typically smaller than 2 arcsec. Their projected physical sizes are typically less than 20 kpc for CSSs, less than 1 kpc for GPSs, and less than 500 pc for CSOs. However, low-radio-power (less than $10^{24}~\rm{W~Hz^{-1}}$) CSOs have also been observed with an overall
projected size of around 1 kpc \citep[e.g.][]{2012ApJ...760...77A}. Tracking these objects on the $P-D$ diagram, therefore, suggests that GPSs are the progenitors of CSSs which are in turn expected to grow into FRIIs; this is considered with the fact that GPS/CSS objects reveal morphologies similar to FRIIs but on sub-kpc scales \citep[e.g.][]{2021A&ARv..29....3O}. 
%1974Natur.250..625W,

\begin{figure*}
    \centering
    \includegraphics[width=2\columnwidth]{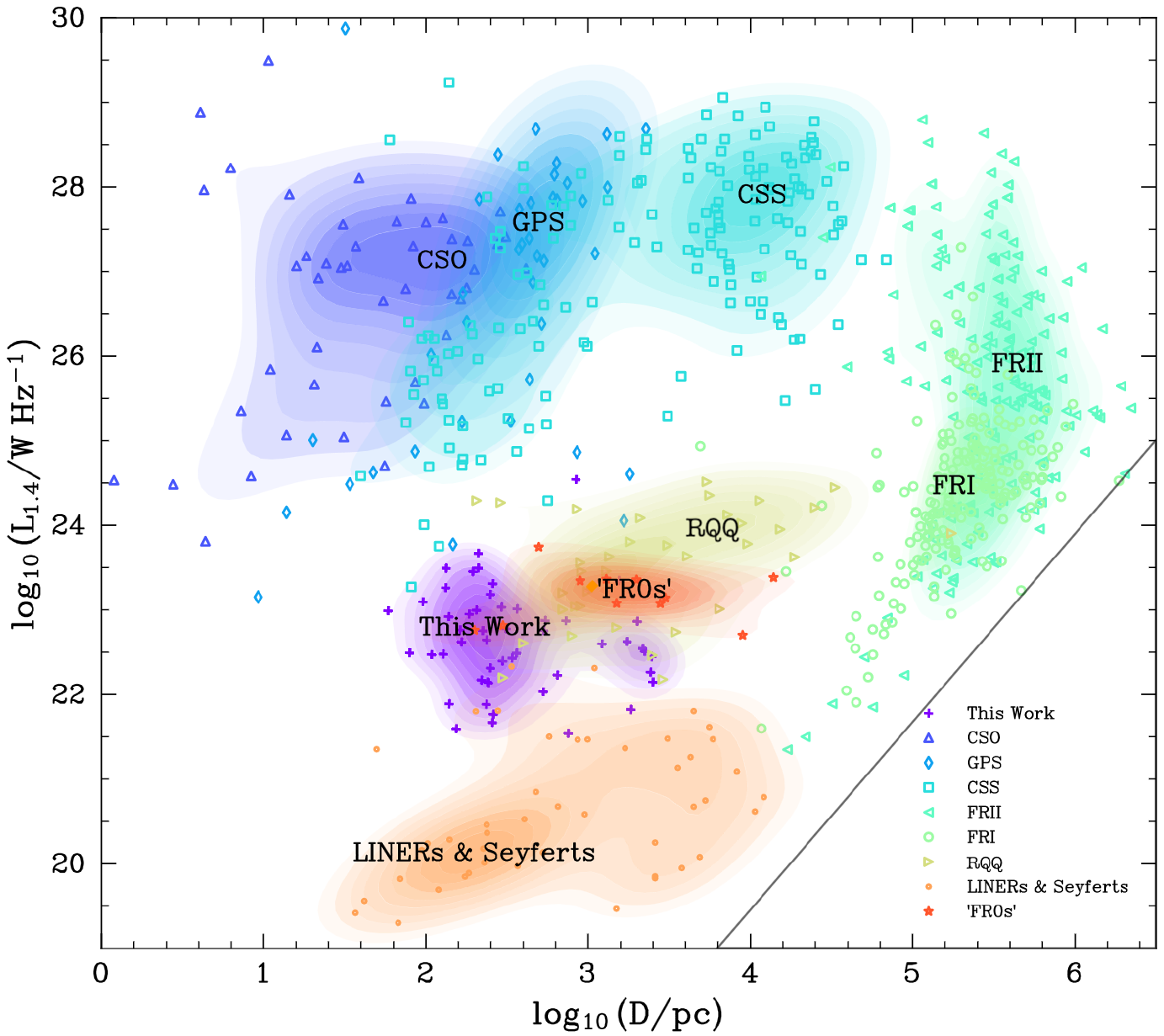}
    \caption{A power/linear-size ($P-D$) diagram adapted from \citet{2020NewAR..8801539H} indicating various types of radio-selected AGN \citep{2012ApJ...760...77A,2019MNRAS.485.2710J}. Plots of CSO, GPS, and CSS sources are adapted from \citet{2012ApJ...760...77A}. FRI and FRII plots shown are those classed by \citet{2019MNRAS.488.2701M}. RQQs are adapted from \citet{2019MNRAS.485.2710J} and \citet{1998MNRAS.297..366K} while LINERs and Seyferts are those classed by \citet{2018MNRAS.476.3478B} and \citet{2006AJ....132..546G}. The so-called FR0s are obtained from \citet{2016AN....337..114B,2019MNRAS.482.2294B}. The diagonal line (black) shows the area where radio sources are undetectable by any existing survey due to surface brightness limitations.}
    \label{fig:PD_diagram}
\end{figure*}

Fig.~\ref{fig:PD_diagram} shows a $P-D$ diagram consisting of different types of AGN overlaid with the position of compact and extended VLA objects presented in this study. Our objects (shown in purple contours and represented by `+' points) cover the range in radio powers between 10$^{21}$ and 10$^{25}$ W Hz$^{-1}$ at 1.4 GHz, with projected physical sizes spanning $1-3$ kpc for extended sources and less than $1$ kpc for compact objects; the latter sizes are upper limits from deconvolved source parameters provided my {\sevensize{IMFT}} (see Table~\ref{tab:unResolved}). We see that our objects do not occupy a special place on the $P-D$ diagram with regard to the different types of AGN shown. However, we see that some of them share the region populated by radio-quiet quasars (RQQs) and so-called FR0s. The fact that our sources share a region with RQQs is not surprising, given our selection choice. RQQs also display intermediate radio powers and show sub-kpc extended radio emission \citep{2019MNRAS.485.2710J} $-$ a detailed comparison with `FR0s' is given in Sec.~\ref{sec:FR0}. Unified models suggest that Seyferts are the low-luminosity counterparts of quasars \citep[e.g.][]{1993ARA&A..31..473A}, and seeing our objects lie between Seyferts and RQQs in terms of luminosity suggests that our sources represent a continuity between the populations of Seyferts and quasars. In terms of projected physical size alone, we can see that our compact sources share a region with CSOs \citep[e.g.][]{2012ApJ...760...77A} while those with extended radio emission span the range associated with CSS sources \citep[e.g.][]{1998PASP..110..493O}. 

The position of our sources on the radio-power/linear size plane could also indicate that these objects are the radio cores of low-luminosity large-scale structures such as those seen in FRI/IIs; this is also supported by the fact that a large number of our objects display flat/inverted radio spectra. To test this idea, and though not part of our sample, we included a low-redshift ($z\sim0.12$), low-radio-power (log$_{10}(L\rm{_{150})=23.5~W~Hz^{-1}}$) object (J1310+51; see Sec.~\ref{sec:appndx_4} and Fig.~\ref{fig:appndx})  classed as compact in DR1 that shows large-scale extended radio emission in DR2 with an FRI morphology. This object could potentially lie in the region of $P-D$ space that no existing survey is sensitive to due to surface brightness limitations (right-lower-end), and if this is the case, we may be underestimating the sizes for some of the population of these low-luminosity RLAGN.

\subsection{Comparison with similarly compact RLAGN}
\label{sec:FR0}
We now compare our RLAGN and those previously studied in other radio surveys at high resolutions in the local Universe \citep[e.g.][]{2010A&A...519A..48B,2014MNRAS.438..796S}. Those that remain unresolved are sometimes referred to as `FR0s' \citep{2011AIPC.1381..180G,2015A&A...576A..38B,2016AN....337..114B}. This terminology originates from the fact that they lack extended emission, making them a factor of $\sim$30 more core-dominated than typical FRIs \citep[see][]{2023A&ARv..31....3B}. However, as noted by \citet{2020NewAR..8801539H}, the `FR0' nomenclature can only be applied when observations capable of resolving these sources are available. Note that this is also the case for the \citeauthor{1974MNRAS.167P..31F} (FRI/II) morphological classification.

\begin{figure}
    \centering
    \includegraphics[width=\columnwidth]{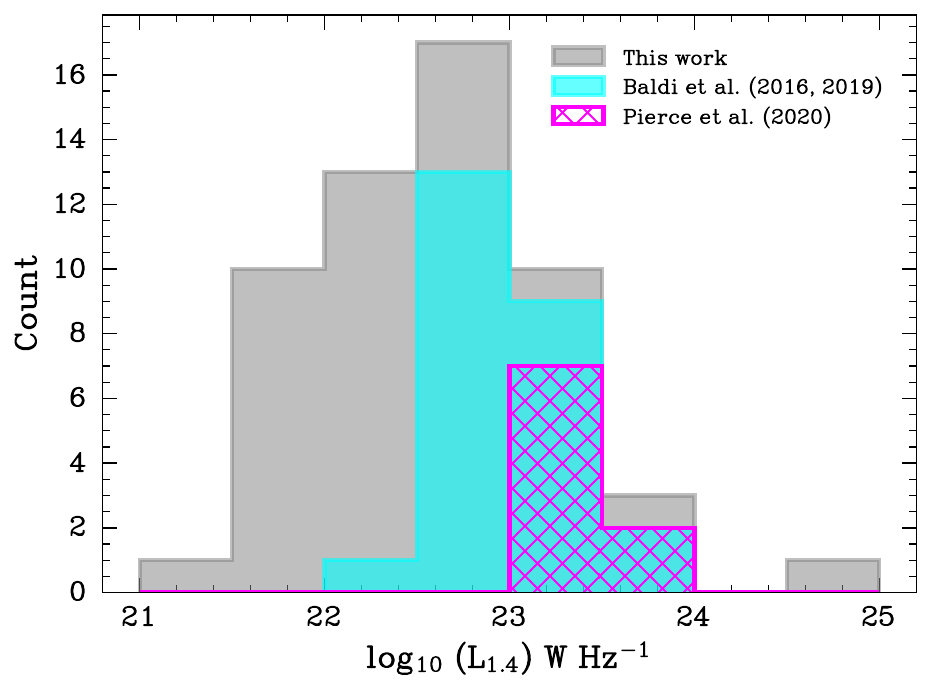}
    \caption{The distribution of radio luminosities at 1.4 GHz. Grey bins represent the sample studied here, the \citet{2016AN....337..114B,2019MNRAS.482.2294B} FR0 sample is shown in cyan, and the compact HERG sample of \citet{2020MNRAS.494.2053P} is shown in magenta (hatched bins).}
    \label{fig:Lum_dist}
\end{figure}

Most FR0s are spectroscopically classified as low-excitation radio galaxies \citep[LERGs; e.g.][]{2015A&A...576A..38B,2018A&A...609A...1B}, and are characterised by radio and host galaxy properties that are similar to those of FRI sources. The main observed difference thus far is the lack of radio extended emission. They are predominantly located in red massive ($\sim$10$^{11}$ M$_{\sun}$) early-type galaxies (ETGs) with SMBH masses greater than 10$^{8}$ M$_{\sun}$, and have radio-powers between $10^{22}$ and $10^{24}~\rm{W~Hz^{-1}}$ at 1.4 GHz \citep[][see Fig.~\ref{fig:Lum_dist}]{2015A&A...576A..38B}. High-resolution multi-frequency (1.4, 4.5, and 7.5 GHz) VLA observations of these objects show a range of radio properties on sub-kpc scales: steep and flat/inverted spectral indices and a range of radio morphologies \citep[core-jet, two-jet, and complex;][]{2015A&A...576A..38B}. 

Given that our RLAGN also lie in the local Universe and show similar radio powers and projected physical scales to the FR0 sample, our objects can be placed in the same class as FR0s. Therefore, a comparison between the two samples is useful for understanding their nature. The main difference between the two populations is that we selected only high excitation radio galaxies for this study \citep[HERGs; selected by adopting all four AGN-SF separation methods from][]{2019A&A...622A..17S}; however, we note that there also exits a large number of compact LERGs in DR1.

\citet{2016AN....337..114B,2019MNRAS.482.2294B} (hereafter \citetalias{2016AN....337..114B}\citetalias{2019MNRAS.482.2294B}) obtained A-array multi-frequency (1.5, 4.5, and 7.5 GHz) high-resolution VLA observations of 25 FR0s. These objects are compiled in the FR0 catalogue, FR0CAT \citep{2018A&A...609A...1B}. Radio spectral indices were derived for the 25 FR0s between 4.5 and 7.5 GHz spanning $-0.16<\alpha<1.33$. The spectral index distribution for FR0s between 4.5 and 7.5 GHz show 12/25 FR0s with a steep radio spectrum ($\alpha>0.5$) while 12/25 show a flat radio spectrum ($\alpha<0.5$). Only one FR0 shows a strongly inverted radio spectrum i.e. $\alpha=-0.16$. In comparison to the radio spectra of our sample (55 objects) derived between 5 and 7 GHz, 16/55 objects show steep radio spectra ($\alpha>0.5$), 26/55 show flat radio spectra ($\alpha<0.5$), and 13/55 show an inverted radio spectrum ($\alpha<0$). Median radio spectral indices between the two populations are $\Bar{\alpha}=0.30$ and $\Bar{\alpha}=0.41$ for our RLAGN and the FR0s respectively. Based on a two$-$sample Kolmogorov–Smirnov (KS) test, we find that the null hypothesis that our sources are drawn from the same spectral index distribution as the \citetalias{2016AN....337..114B}\citetalias{2019MNRAS.482.2294B} FR0s cannot be rejected.

We also compared our objects to the compact RLAGN sample (with `FR0-like' features) of \citet{2020MNRAS.494.2053P} (hereafter \citetalias{2020MNRAS.494.2053P}) selected from the local Universe ($z<0.1$). These objects share similar radio luminosities (Fig.~\ref{fig:Lum_dist}) and morphologies with our sources, and were also optically classified as HERGs \citep{2019MNRAS.487.5490P}. \citetalias{2020MNRAS.494.2053P} obtained high-resolution VLA observations of 16 HERGs at the same observing frequency bands as the \citetalias{2016AN....337..114B}\citetalias{2019MNRAS.482.2294B} FR0s, and it was found that 7/16 objects display clear extended radio emission ($\sim$2$-$19 kpc) in the VLA images, while the remaining 9/16 objects were compact down to the limiting angular resolution of the VLA ($\sim$0.3 arcsec), corresponding to projected physical sizes of $\sim$0.28$-$0.54 kpc.

\begin{figure}
    \centering
    \includegraphics[width=\columnwidth]{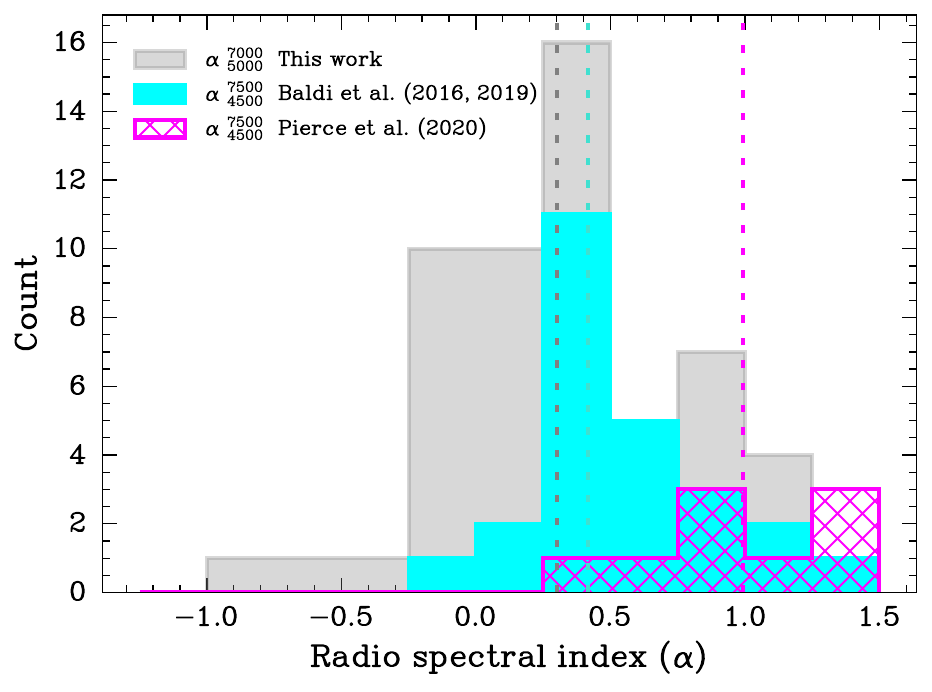}
    \caption{Comparison of the distribution of the radio spectral index between our sample (gray bins), the FR0 sample of \citet{2016AN....337..114B,2019MNRAS.482.2294B} (cyan bins), and the compact HERG sample of \citet{2020MNRAS.494.2053P} (magenta hatched bins). The dashed vertical lines are colour coded to represent the median spectral index values: in grey $\Bar{\alpha}=0.30$, in cyan $\Bar{\alpha}=0.41$, and in magenta   $\Bar{\alpha}=0.99$.}
    \label{fig:FR0_RLAGN_comp}
\end{figure}

The nine compact HERGs of \citetalias{2020MNRAS.494.2053P} display steeper radio spectral indices between 4.5 and 7.5 GHz ($0.39<\alpha<1.40$): eight objects have steep spectral indices ($\alpha>0.5$) and only one has a flat spectral index ($\alpha<0.5$). The median spectral index for the \citetalias{2020MNRAS.494.2053P} compact HERG sample is $\Bar{\alpha}=0.99$. Repeating the KS test as above, the null hypothesis that the \citetalias{2020MNRAS.494.2053P} compact HERG sample and our objects are drawn from the same spectral index distribution can be rejected at a high confidence level. This analysis, as well as that presented for the \citetalias{2016AN....337..114B}\citetalias{2019MNRAS.482.2294B} FR0s, is summarised in Fig.~\ref{fig:FR0_RLAGN_comp}, where we show the distribution of spectral indices for all three cases.

We attribute the differences in the spectral index statistics between our sample and the \citetalias{2020MNRAS.494.2053P} compact HERG sample to selection effects; the \citetalias{2020MNRAS.494.2053P} sample probes higher radio luminosities (median: log$_{10}(L_{1.4})=23.4$ W Hz$^{-1}$) than our sample (median: log$_{10}(L_{1.4})=22.6$ W Hz$^{-1}$), as shown in Fig.~\ref{fig:Lum_dist}.

We highlight similarities in projected physical sizes and radio morphologies between the \citetalias{2016AN....337..114B}\citetalias{2019MNRAS.482.2294B} extended FR0 sample and our extended RLAGN. In terms of physical scales, 7/25 FR0s show extended emission in the VLA images projected at $\sim$1$-$14 kpc in physical size. All extended objects in our sample lie within the range seen for extended FR0s (which includes the object J1213+50, which appears in both our sample and the \citetalias{2016AN....337..114B}\citetalias{2019MNRAS.482.2294B} FR0 sample, projected at $\sim$2 kpc in physical size). With regard to radio morphology, FR0s display a wide range of structures: two-sided jet, one-sided jet, and complex. These structures are similar to those seen in our sample (Fig.~\ref{fig:images}) and in the \citetalias{2020MNRAS.494.2053P} extended HERG sample. The object J1230+47 remains compact at all VLA frequencies in the FR0 sample as well as in our sample but shows parsec-scale extended emission in the VLBI, EVN, and eMERLIN studies of FR0s \citep{2018ApJ...863..155C,2021MNRAS.506.1609C,2021Galax...9..106B,2023A&A...672A.104G}. We note that the two overlapping objects J1213+50 and J1230+47 were classified as LERGs in the work of \citetalias{2016AN....337..114B}\citetalias{2019MNRAS.482.2294B}, following the \citet{2012MNRAS.421.1569B} AGN-SF selection criteria, while we classed them as HERGs, adopting the AGN-SF selection criteria from \citet{2019A&A...622A..17S}.

\begin{figure}
    \centering
    \includegraphics[width=\columnwidth]{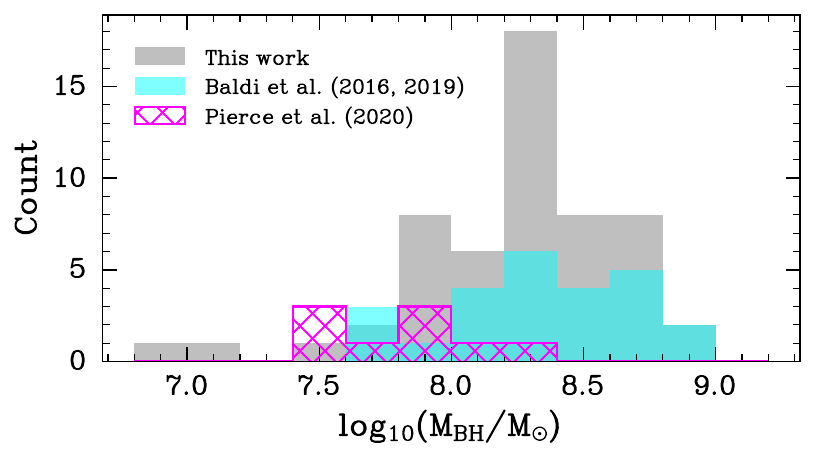}
    \caption{Distribution of black hole mass estimates for our sample (grey), \citet{2016AN....337..114B,2019MNRAS.482.2294B} FR0s (cyan), and the \citet{2020MNRAS.494.2053P} compact HERGs (magenta). We note that objects with BH masses less than $10^{7.5}~\rm{M_{\sun}}$ might overlap with radio-quiet AGN \citep[e.g.][]{2019NatAs...3..387P}, however, this does not affect the interpretation of our results.}
    \label{fig:BH_masses}
\end{figure}

Further similarities between our sample and \citetalias{2016AN....337..114B}\citetalias{2019MNRAS.482.2294B} FR0s can be seen in their host properties: stellar mass, black hole (BH) mass, and morphology. We inspected optical data from SDSS DR7 \citep{2009ApJS..182..543A}\footnote{The Sloan Digital Sky Survey seventh data release is available at \url{https://wwwmpa.mpa-garching.mpg.de/SDSS/DR7/}.} for our sample and found that all our objects are located in massive galaxies ($\rm{log_{10}(M_{*})>10.5~M_{\sun}}$) with the majority being hosted by elliptical galaxies (49/55) and a minority (6/55) hosted by spirals; this is consistent with their position on WISE colour-colour plots \citep[see e.g.][also Fig.~\ref{fig:wise_color}]{2010AJ....140.1868W,2016MNRAS.462.2631M,2019MNRAS.488.2701M}. Our sources have BH masses lying between $6.7<\rm{log_{10}(M_{BH}})<9.0~M_{\sun}$ (median: $\rm{log_{10}(M_{BH}})=8.3~M_{\sun}$)\footnote{BH mass estimates were derived using the $\rm{M_{BH}}-\sigma$ relation presented by \citet{2002ApJ...574..740T}, as was used by \citetalias{2016AN....337..114B}\citetalias{2019MNRAS.482.2294B} and \citetalias{2020MNRAS.494.2053P}. Velocity dispersion ($\sigma$) estimates were obtained from the MPA-JHU value-added catalogue.} while BH masses for the \citetalias{2016AN....337..114B}\citetalias{2019MNRAS.482.2294B} sample of FR0s lie between $7.6<\rm{log_{10}(M_{BH}})<8.9~M_{\sun}$ (median: $\rm{log_{10}(M_{BH}})=8.4~M_{\sun}$); see Fig.~\ref{fig:BH_masses} for comparisons of BH masses. Another parameter that can help us compare the host properties of our objects with those of the \citetalias{2016AN....337..114B}\citetalias{2019MNRAS.482.2294B} FR0s is the spectroscopic index, Dn(4000), which provides information about the stellar populations in galaxies \citep{1999ApJ...527...54B}. Older stellar populations have been found to be more biased towards higher values of the spectroscopic index (Dn(4000) $>1.7$) and are often associated with red massive ETGs, while lower values (Dn(4000) $<1.7$) are often associated with younger stars which contribute more to the blue region of the spectrum. We find for our sources that all but one source displays Dn(4000) $>1.7$, consistent with what has been found for most FR0s \citep[e.g.][]{2015A&A...576A..38B}.

In contrast, the \citetalias{2020MNRAS.494.2053P} compact HERG sample displays lower BH masses compared to our sample: $7.4<\rm{log_{10}(M_{BH}})<8.3~M_{\sun}$ (median: $\rm{log_{10}(M_{BH}})=7.8~M_{\sun}$) and the host morphologies are mixed, with many late-type galaxies existing in the sample \citep{2019MNRAS.487.5490P}. 

In summary, we find that the high core dominance observed in our RLAGN is similar to that observed for FR0s (\citetalias{2016AN....337..114B}\citetalias{2019MNRAS.482.2294B}) and the compact HERGs from \citetalias{2020MNRAS.494.2053P}. This property is also seen in a sample of FR0s studied in the DR2 area \citep{2020A&A...642A.107C}. Further high-resolution observations with long-baseline interferometers such as eMERLIN and/or VLBI will help in confirming if these galaxies can be placed in the same pool as the \citetalias{2016AN....337..114B}\citetalias{2019MNRAS.482.2294B} FR0s.

\section{SUMMARY \& CONCLUSION}
\label{sec:conclusion}
We have presented high-resolution 6 GHz VLA observations of a LOFAR sub-sample of 55 low-$L$ and low-$z$ high excitation radio galaxies (HERGs), taken in the A-configuration. We selected our AGN taking into account all four AGN-SF separation methods from \citet{2019A&A...622A..17S}, limiting our sources to peak flux densities greater than 2 mJy. These objects are compact at the resolution of LoTSS DR1, although some show extended radio emission in the improved images of LoTSS DR2. The summary of this paper is as follows:
\begin{itemize}
     \item We have found that 10/55 RLAGN show extended radio emission in the VLA images, while 42/55 objects are unresolved at the limiting angular resolution of 0.35 arcsec, corresponding to an upper limit of less than 1 kpc in projected physical size. The extended objects have projected physical sizes spanning $\sim$1$-$3 kpc, and show a range of radio morphologies: 
     two-jet (5), one-jet (4), and double-lobed (1). No radio emission was detected by the VLA for three objects.
    \item The radio spectra of these sources range between $-0.36<\alpha_{150}^{6000}<1.31$, showing similarities with various compact RLAGN types such as CSO, GPS, and CSS sources. 16/55 objects have steep radio spectra, consistent with optically thin synchrotron radio emission. 39/55 objects show flat/inverted radio spectra, likely due to self-absorption. We attribute the steepening in the radio spectra of these RLAGN to be analogous to that what is observed for young, rapidly growing CSS sources.
    \item We have estimated the physical properties of the radio-emitting region for 34 compact objects showing flat/inverted radio spectra, assuming SSA at 150 MHz. The SSA predicted sizes for these objects extend up to a few milliarcsec in angular size ($\sim$2$-$30~mas) or $\sim$2$-$53 pc in physical size. These angular sizes are well below the resolution of LOFAR and the VLA. We have estimated that source equipartition $B$ fields ($\sim$0.3$-$0.75~$\mu$T) are comparable to those estimated in young compact RLAGN such as CSSs. Alternatively using FFA, we have also estimated the free electron column depth, assuming a homogeneous absorber, of $\sim$2$\times10^{20}~\rm{cm}^{-2}$, supporting the case for FFA in a single line-emitting cloud in the NLRs; however, we note that the case for FFA is stronger for a non-uniform medium. We also note that our analysis favours SSA to be more dominant in our objects than FFA. 
    \item On the power/linear size ($P-D$) diagram, we find that some of our objects share a region with radio-quiet quasars (RQQs) and so-called FR0s in terms of radio luminosity and physical size. However, we find in terms of projected physical size that these objects share a region with CSS sources. Our objects in the $P-D$ space, therefore, do not occupy a special position with regard to other radio-selected AGN.
    \item We compared our sample with the FR0 sample from \citet{2016AN....337..114B,2019MNRAS.482.2294B}, and we highlighted similarities in the radio (spectra, morphology, projected physical size, and luminosity) and host (galaxy type, stellar mass, and black hole mass) properties of these galaxies. Additionally, we compared our results with those of \citet{2020MNRAS.494.2053P}; we note that although their FR0-like HERG sample display steeper spectral indices (attributed to selection effects), they share similar radio powers and projected physical sizes with our HERG sample.
\end{itemize} 

The majority of our sources ($\sim$76\%) remain unresolved or slightly resolved at high resolution. Most of these sources are highly core-dominated, however, a small number of extended objects ($\sim$18\%) show complex radio structures that cannot be well classified. This supports the idea that these sources are injecting energy back into their local environments on sub-kpc scales \citep[e.g.][]{2021MNRAS.503.4627U} and could play a crucial role in regulating star formation. There seems to be an emergence of a large population of low-luminosity and physically small RLAGN with flat/inverted radio spectra at the centres of galaxies. Based on the results of the current work and of previous studies, we know that many of them are genuinely very compact and must be having some effect on the central regions of their host galaxy. Future observations with milliarcsec high-resolution interferometers such as eMERLIN, VLBA, and/or EVN should reveal a range of radio morphologies in a majority of these objects on parsec scales.

\section*{Acknowledgements}
We acknowledge the anonymous referee's useful comments. JC would like to acknowledge financial support from the Newton-funded Development in Africa with Radio Astronomy (DARA) awarded by the UK’s Science and Technology Facilities Council (STFC) and a studentship from the University of Hertfordshire (UH). MJH and JCSP acknowledge support from the UK STFC [ST/V000624/1]. A portion of the work submitted for publication is derived from an MSc by research thesis conducted at the University of Hertfordshire by the first author. The full thesis is accessible at \url{https://uhra.herts.ac.uk/handle/2299/26489}. We made use of {\scriptsize{ASTROPY}}, a community-developed core {\scriptsize{PYTHON}} package for astronomy \citep{2013A&A...558A..33A} hosted at \url{http://www.astropy.org/}, {\scriptsize{APLPY}}, an open-source astronomical plotting package for {\scriptsize{PYTHON}} hosted at \url{https://aplpy.github.io/}, and of {\scriptsize{PYSYNCH}}, a {\scriptsize{PYTHON}} interface for synchrotron libraries hosted at \url{https://github.com/mhardcastle/pysynch/}.

\section*{Data Availability}
The data used in this article will be shared on reasonable request to the corresponding author(s).
%%%%%% REFERENCES
\bibliographystyle{mnras}
\bibliography{citations} 
%%%%%% APPENDICES
\appendix
\section{VLA objects with extended radio emission in LoTSS DR2}
\label{sec:appendix}
Here we present objects that show extended radio emission in the VLA and LoTSS DR2 images.

\subsection{LoTSS DR2 images and radio maps}\label{sec:dr2}

\subsubsection{J1111+55}
This source shows extended radio emission in both the VLA and DR2 images. A discussion is given in Sec.~\ref{sec:J1111}. 

\subsubsection{J1310+544}
\begin{figure*}
\includegraphics[keepaspectratio,width=0.9\columnwidth]{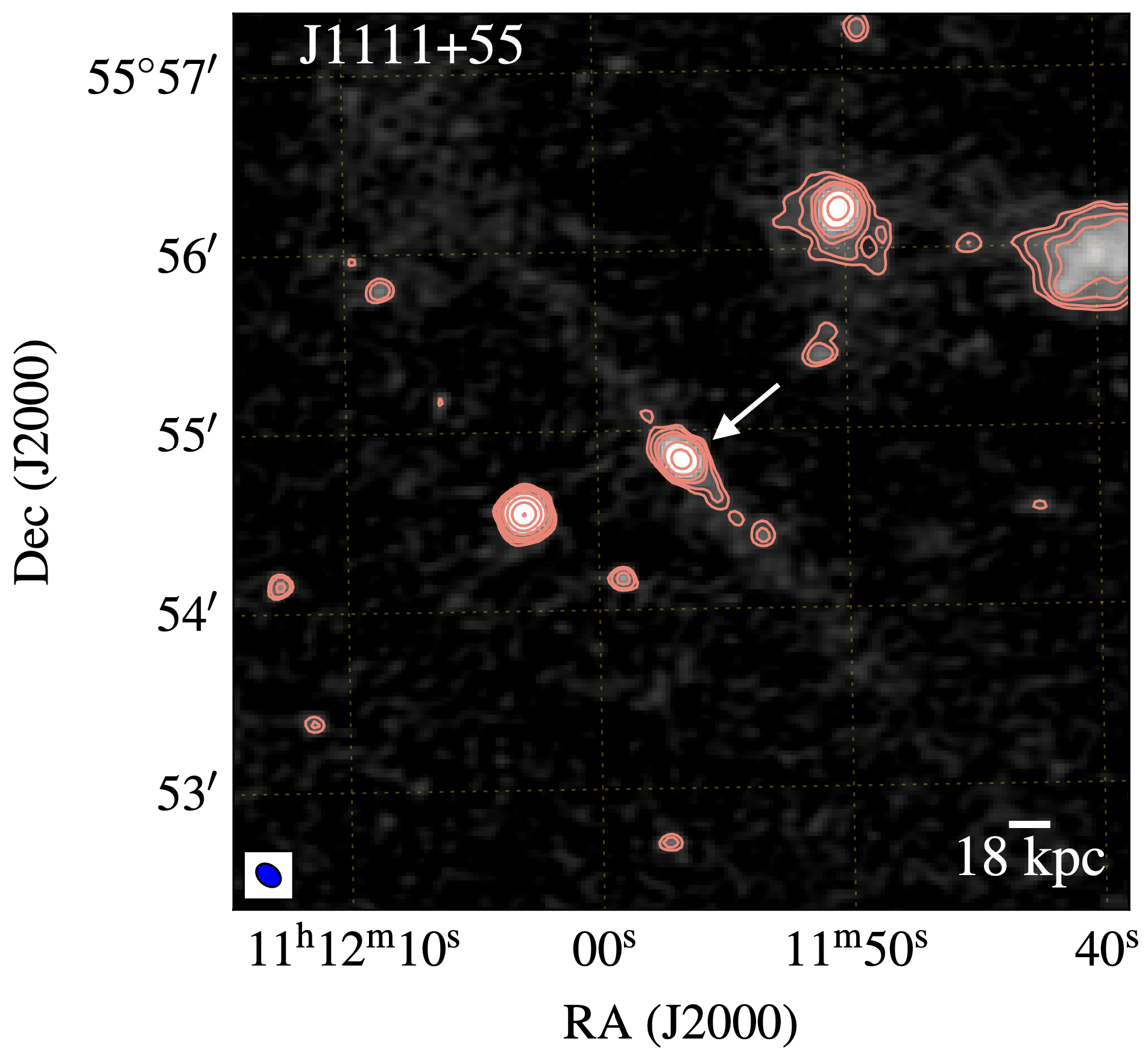}
\includegraphics[keepaspectratio,width=0.9\columnwidth]{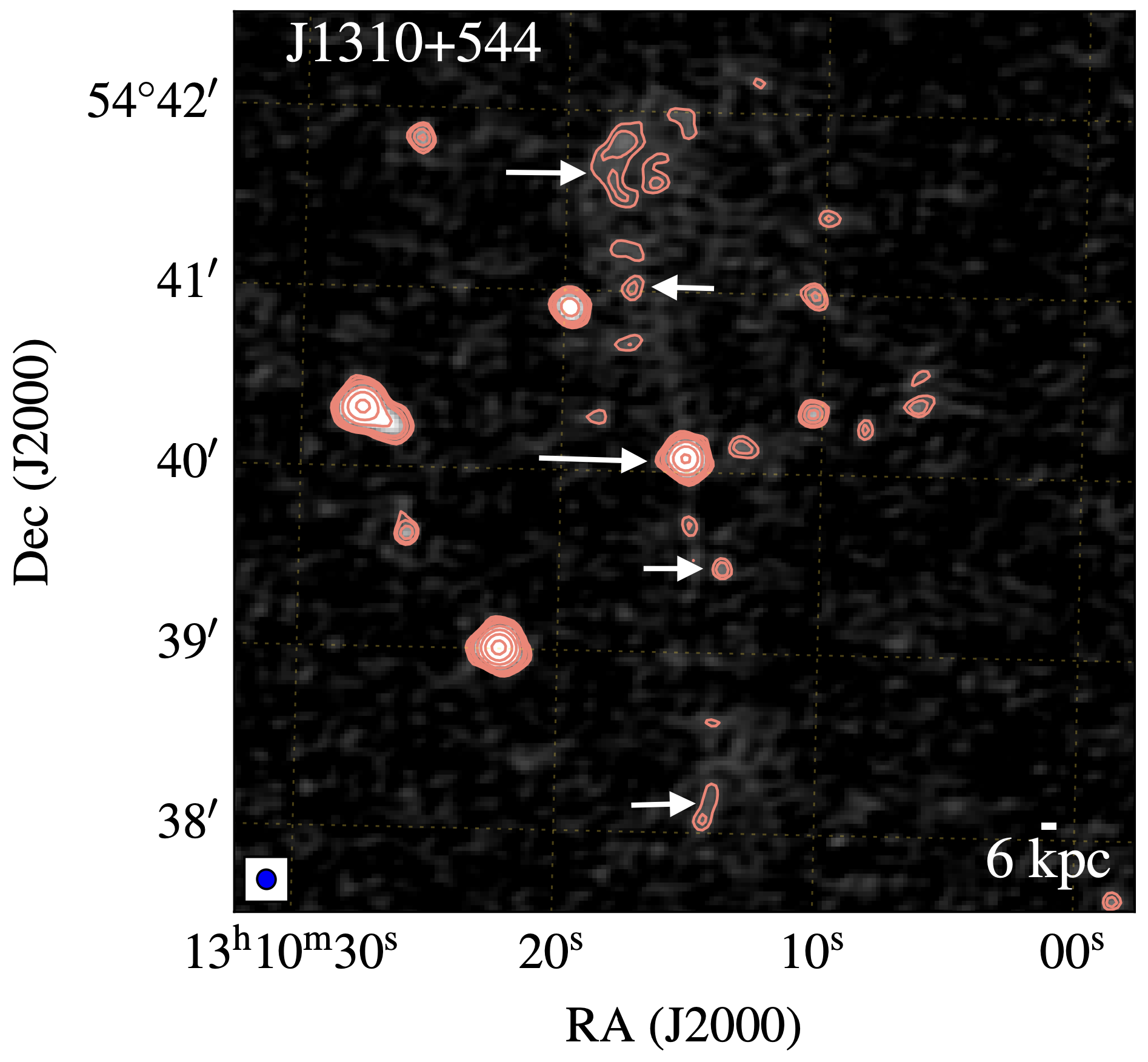}

\includegraphics[keepaspectratio,width=0.9\columnwidth]{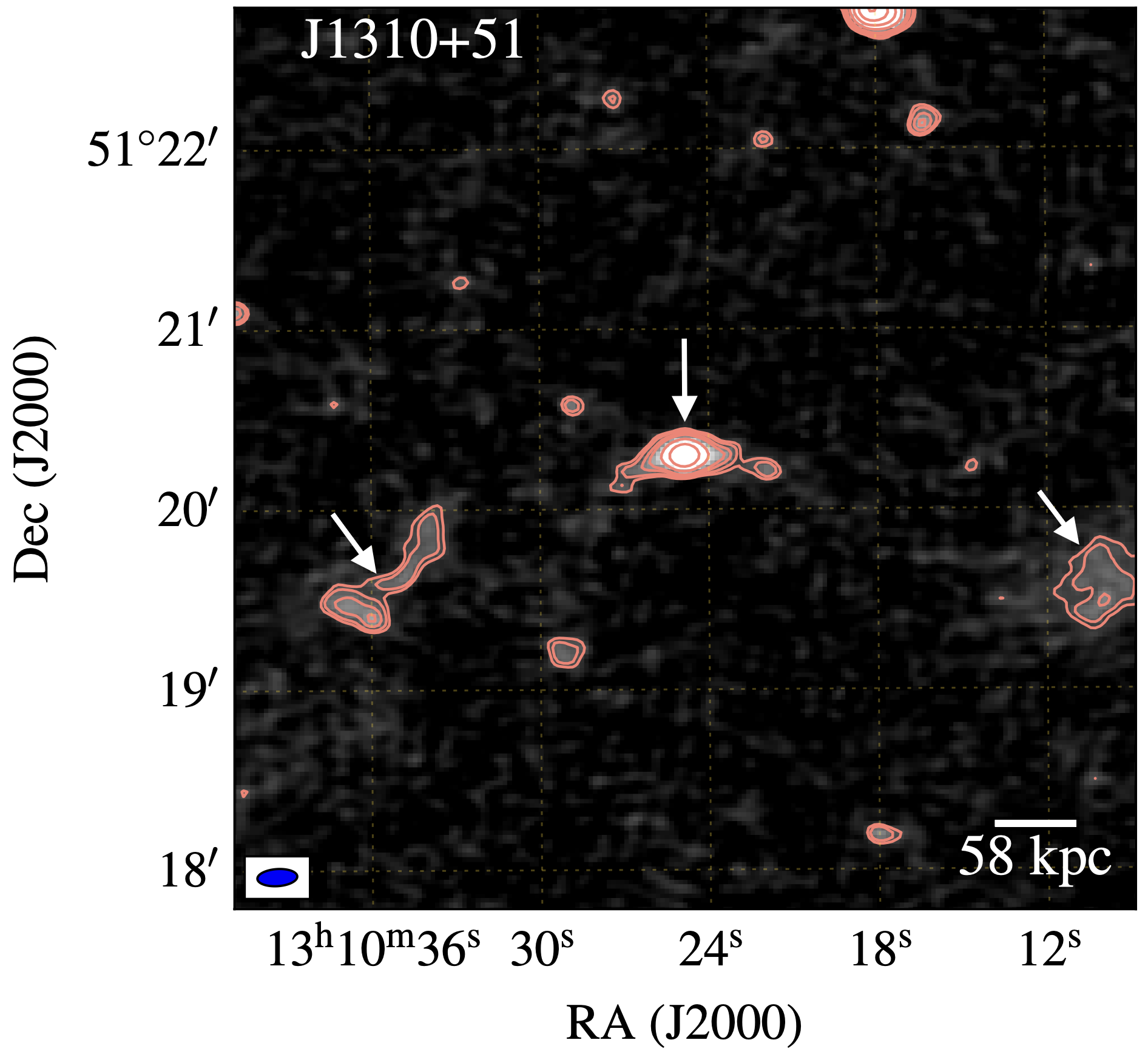}
\includegraphics[keepaspectratio,width=0.9\columnwidth]{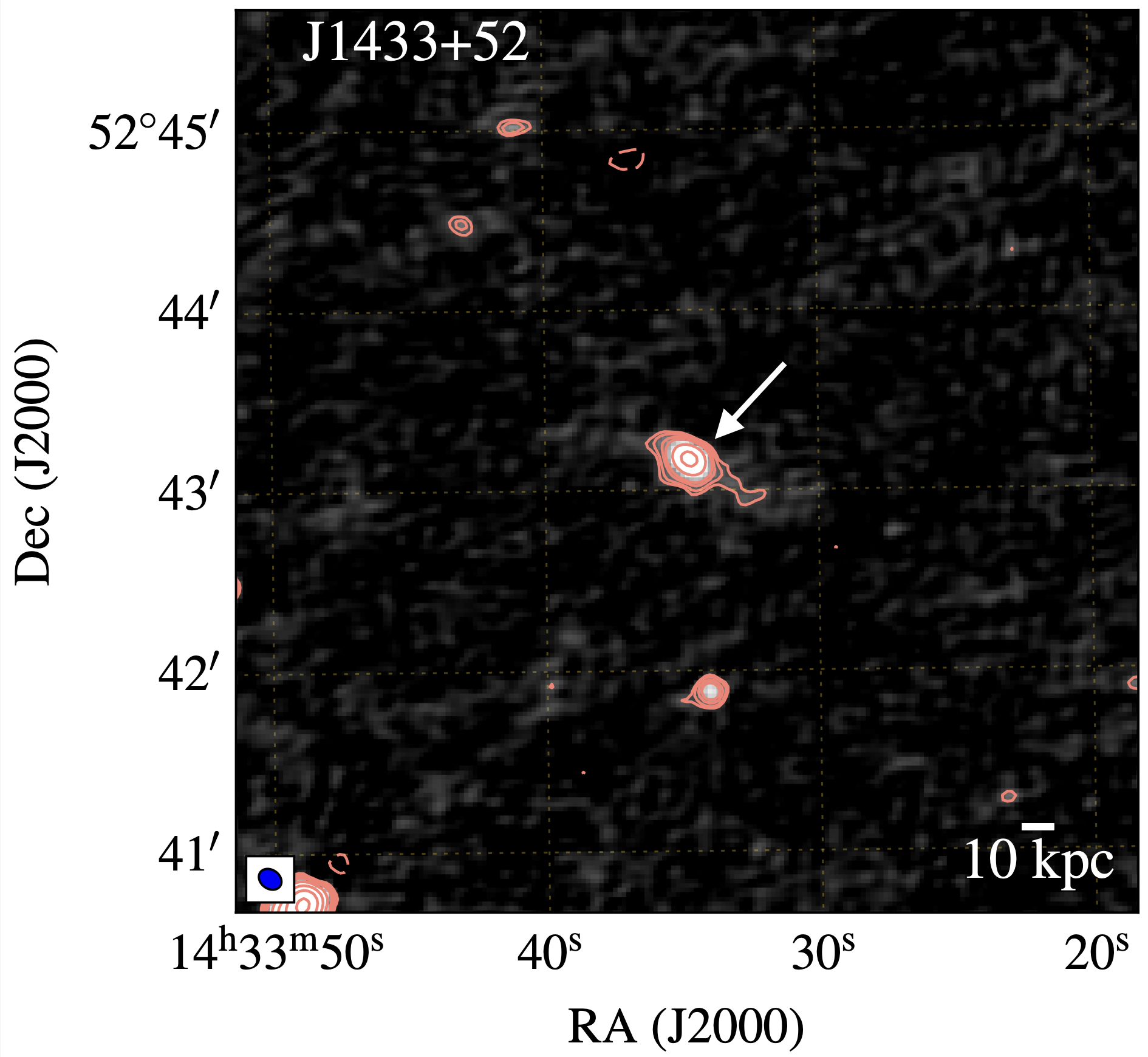}

\includegraphics[keepaspectratio,width=0.93\columnwidth]{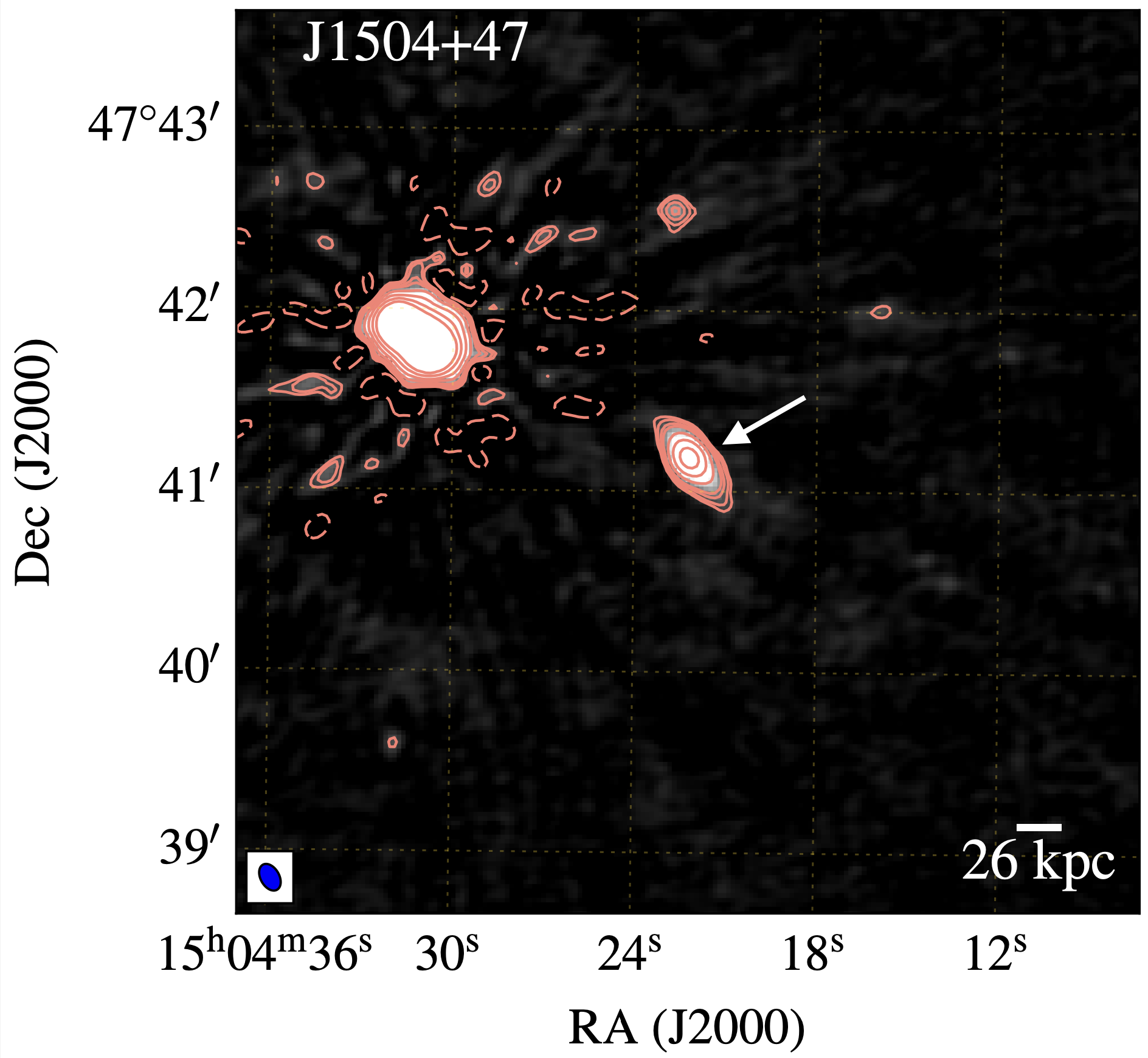}

\caption{DR2 images overlaid with 150 MHz contour maps in salmon. Contour levels are at [-3, 3, 5, 10, 20,...,640] $\times$ $\rm{\sigma_{rms}}$. Top panels from left to right: $\rm{\sigma_{rms}=42.8~and~45.5~\mu Jy~beam^{-1}}$ respectively. Middle panels from left to right: $\rm{\sigma_{rms}=37.7~and~46.5~\mu Jy~beam^{-1}}$ respectively. Bottom panel: $\rm{\sigma_{rms}=108.9~\mu Jy~beam^{-1}}$. Blue ellipses indicate the LOFAR beam. Note that source J1310+51 is not part of this paper's analysis as discussed in the text. Arrows on images highlight components associated with a jet, as suggested by the jet axis in the VLA images. Physical scales quoted here are LoTSS DR2 sizes given by upper limits.}
\label{fig:appndx}
\end{figure*}

This source shows extended radio emission in both the VLA and DR2 images. In the DR2 image (Fig.~\ref{fig:appndx}), the arrows shown for this object align with the jet axis suggested by the VLA image (Fig.~\ref{fig:images1}), confirming the emission to be real rather than image artefacts. A discussion is also given in Sec.~\ref{sec:J1310}.

\subsubsection{J1310+51}\label{sec:appndx_4}
This object is not part of our sample due to its redshift ($z\sim0.12$) falling outside our selection criteria ($z<0.1$) and is included for the sole purpose of displaying the brightness sensitivity of LoTSS DR2, as it shows extended radio emission in DR2 images, which is not visible in the DR1 image. We see two radio lobes on either side (shown by arrows) of the central emission. The central structure is projected at $\sim$58 kpc in physical size.

\subsubsection{J1433+52}
This source shows extended radio emission in DR2 and is compact in the VLA images. We see a bright and unresolved central radio emission with what appears like a one-sided jet and tails slightly away from the radio core to the south-west direction. The projected physical size of the source is estimated at $\sim$10 kpc.

\subsubsection{J1504+47}
The object is compact in the VLA images but shows extended radio emission in the DR2 images. The extended emission appears to be a two-sided jet projected at $\sim$26 kpc in physical scales.

\section{Additional VLA image information}
In order to compare our objects with the FR0 sample from \citetalias{2016AN....337..114B}\citetalias{2019MNRAS.482.2294B} and the compact HERG sample from \citetalias{2020MNRAS.494.2053P}, we split our VLA data to obtain 5 and 7 GHz image parameters, as was done in those works. This information is shown in Table~\ref{tb:appendix1}. 

\begin{table*}
     \caption{Additional VLA image information about the 55 objects at 5 and 7 GHz.}
    \begin{adjustbox}{width=1\textwidth}
    \begin{tabular}{lccccccccccc}
    \hline\hline
         Source&$\theta^{r}_{5.0}$&$\vartheta_{5.0}$&$S\rm{^{int}_{5.0}}$&$\sigma\rm{_{5.0}^{rms}}$&$\theta^{r}_{7.0}$&$\vartheta_{7.0}$&$S\rm{^{int}_{7.0}}$&$\sigma\rm{_{7.0}^{rms}}$&$\alpha_{0.15}^{5.0}$&$\alpha_{1.4}^{5.0}$&$\alpha_{5.0}^{7.0}$\\
         \hline
J1048+48	&	0.43	$\times$	0.39	&	40.8	&	7.03	$\pm$	0.24	&	15.5	&	0.29	$\times$	0.27	&	67.6	&	4.91	$\pm$	0.47	&	20.9	&0.64 $\pm$ 0.01&	0.77	$\pm$	0.05	&	1.07	$\pm$	0.30	\\
J1055+46	&	0.41	$\times$	0.35	&	70.8	&	2.15	$\pm$	0.05	&	22.7	&	0.43	$\times$	0.32	&	89.1	&	2.19	$\pm$	0.09	&	21.2 &0.17 $\pm$ 0.01	&	0.00	$\pm$	0.15	&	-0.06	$\pm$	0.14	\\
J1102+51	&	0.49	$\times$	0.40	&	71.8	&	1.71	$\pm$	0.02	&	8.8	&	0.30	$\times$	0.25	&	73.1	&	1.22	$\pm$	0.06	&	26.2&0.37 $\pm$ 0.01	&	0.26	$\pm$	0.13	&	1.01	$\pm$	0.15	\\
J1111+55	&	0.44	$\times$	0.36	&	65.7	&	0.58	$\pm$	0.03	&	26.4	&	0.30	$\times$	0.25	&	68.3	&	0.42	$\pm$	0.01	&	20.4&0.63 $\pm$ 0.02	&	0.56	$\pm$	0.34	&	0.96	$\pm$	0.34	\\
J1117+46	&	0.41	$\times$	0.39	&	77.5	&	2.71	$\pm$	0.32	&	41.9	&	0.36	$\times$	0.30	&	42.0	&	1.79	$\pm$	0.29	&	16.2&0.43 $\pm$ 0.01	&	0.40	$\pm$	0.08	&	1.23	$\pm$	0.60	\\
J1117+47	&	0.43	$\times$	0.39	&	66.7	&	7.11	$\pm$	0.10	&	18.8	&	0.30	$\times$	0.28	&	84.7	&	5.85	$\pm$	0.08	&	13.7&0.38 $\pm$ 0.00	&	0.36	$\pm$	0.04	&	0.58	$\pm$	0.06	\\
J1126+52	&	0.57	$\times$	0.51	&	-60.0	&	3.63	$\pm$	0.35	&	21.5	&	0.31	$\times$	0.25	&	71.9	&	2.46	$\pm$	0.09	&	22.4&0.55 $\pm$ 0.01	&	0.60	$\pm$	0.06	&	1.16	$\pm$	0.30	\\
J1135+49	&	0.41	$\times$	0.34	&	66.5	&	2.82	$\pm$	0.02	&	12.9	&	0.31	$\times$	0.26	&	74.9	&	2.86	$\pm$	0.03	&	17.1&0.15 $\pm$ 0.01	&	-0.23	$\pm$	0.41	&	-0.04	$\pm$	0.04	\\
J1137+55	&	0.41	$\times$	0.34	&	54.0	&	2.37	$\pm$	0.06	&	21.1	&	0.31	$\times$	0.27	&	62.8	&	2.36	$\pm$	0.02	&	17.3&0.12 $\pm$ 0.01	&	-0.30	$\pm$	0.20	&	0.01	$\pm$	0.08	\\
J1140+46	&	0.38	$\times$	0.34	&	37.7	&	31.40	$\pm$	0.12	&	23.5	&	0.26	$\times$	0.25	&	49.5	&	30.15	$\pm$	0.07	&	16.0&-0.38 $\pm$ 0.00	&	-0.42	$\pm$	0.03	&	0.12	$\pm$	0.01	\\
J1143+55	&	0.41	$\times$	0.34	&	53.9	&	22.77	$\pm$	0.05	&	19.6	&	0.28	$\times$	0.25	&	65.3	&	22.02	$\pm$	0.07	&	25.3&0.03 $\pm$ 0.00	&	-0.52	$\pm$	0.03	&	0.10	$\pm$	0.01	\\
J1147+55	&	0.42	$\times$	0.36	&	54.0	&	3.52	$\pm$	0.03	&	15.4	&	0.34	$\times$	0.29	&	74.8	&	3.03	$\pm$	0.03	&	25.0&-0.02 $\pm$ 0.01	&	0.24	$\pm$	0.07	&	0.45	$\pm$	0.04	\\
J1151+55	&	0.42	$\times$	0.36	&	63.6	&	0.44	$\pm$	0.02	&	13.5	&	0.31	$\times$	0.25	&	75.0	&	0.32	$\pm$	0.01	&	26.1&0.82 $\pm$ 0.02	&	1.68	$\pm$	0.04	&	0.95	$\pm$	0.16	\\
J1151+53	&	0.50	$\times$	0.39	&	76.9	&	2.61	$\pm$	0.03	&	18.9	&	0.32	$\times$	0.25	&	77.1	&	2.52	$\pm$	0.03	&	12.2&0.11 $\pm$ 0.01	&	-0.05	$\pm$	0.16	&	0.10	$\pm$	0.05	\\
J1152+54	&	0.42	$\times$	0.35	&	67.2	&	5.30	$\pm$	0.03	&	19.5	&	0.34	$\times$	0.30	&	89.4	&	4.75	$\pm$	0.02	&	17.9&0.21 $\pm$ 0.00	&	0.20	$\pm$	0.06	&	0.32	$\pm$	0.02	\\
J1153+52	&	0.52	$\times$	0.41	&	79.0	&	16.93	$\pm$	0.05	&	19.9	&	0.33	$\times$	0.25	&	84.1	&	14.64	$\pm$	0.04	&	27.8&0.20 $\pm$ 0.00	&	0.24	$\pm$	0.03	&	0.43	$\pm$	0.01	\\
J1154+49	&	0.39	$\times$	0.36	&	53.4	&	7.04	$\pm$	0.04	&	23.2	&	0.27	$\times$	0.25	&	65.5	&	7.03	$\pm$	0.04	&	20.6&-0.11 $\pm$ 0.01	&	-0.38	$\pm$	0.07	&	0.00	$\pm$	0.02	\\
J1155+54	&	0.43	$\times$	0.34	&	65.5	&	31.84	$\pm$	0.24	&	30.6	&	0.31	$\times$	0.24	&	70.8	&	27.22	$\pm$	0.14	&	33.2&-0.25 $\pm$ 0.03	&	0.00	$\pm$	0.03	&	0.47	$\pm$	0.03	\\
J1159+55	&	0.43	$\times$	0.35	&	69.7	&	$<$0.06			&	19.6	&	0.31	$\times$	0.25	&	77.1	&	$<$0.06			&	20.3&$-$	&	$-$	&	$-$		\\
J1203+51	&	0.47	$\times$	0.35	&	80.7	&	$<$0.07			&	22.4	&	0.34	$\times$	0.25	&	86.2	&	$<$0.07			&	22.5&$-$	&	$-$	& $-$		\\
J1208+52	&	0.47	$\times$	0.36	&	80.0	&	0.47	$\pm$	0.02	&	15.1	&	0.34	$\times$	0.26	&	84.6	&	0.38	$\pm$	0.01	&	28.2&0.52 $\pm$ 0.02	&	-0.11	$\pm$	0.03	&	0.63	$\pm$	0.15	\\
J1213+50	&	0.40	$\times$	0.35	&	58.2	&	78.40	$\pm$	1.10	&	51.8	&	0.28	$\times$	0.25	&	70.4	&	71.39	$\pm$	0.71	&	57.2&0.25 $\pm$ 0.00	&	0.21	$\pm$	0.02	&	0.28	$\pm$	0.05	\\
J1230+47	&	0.42	$\times$	0.36	&	66.2	&	71.73	$\pm$	0.24	&	25.5	&	0.30	$\times$	0.26	&	75.4	&	61.99	$\pm$	0.07	&	24.7&0.10 $\pm$ 0.00	&	0.16	$\pm$	0.03	&	0.43	$\pm$	0.01	\\
J1233+56	&	0.47	$\times$	0.34	&	76.3	&	1.74	$\pm$	0.11	&	32.6	&	0.33	$\times$	0.25	&	81.4	&	2.20	$\pm$	0.07	&	10.7&0.13 $\pm$ 0.02	&	1.02	$\pm$	0.08	&	-0.70	$\pm$	0.21	\\
J1240+47	&	0.55	$\times$	0.46	&	70.5	&	2.02	$\pm$	0.05	&	25.4	&	0.38	$\times$	0.32	&	77.2	&	1.57	$\pm$	0.03	&	14.2&0.46 $\pm$ 0.01	&	0.35	$\pm$	0.10	&	0.75	$\pm$	0.10	\\
J1243+54	&	0.45	$\times$	0.35	&	75.6	&	1.36	$\pm$	0.11	&	22.2	&	0.32	$\times$	0.25	&	81.7	&	1.01	$\pm$	0.13	&	23.3&0.45 $\pm$ 0.01	&	0.24	$\pm$	0.17	&	0.88	$\pm$	0.45	\\
J1258+50	&	0.46	$\times$	0.36	&	79.3	&	0.35	$\pm$	0.03	&	15.9	&	0.34	$\times$	0.26	&	86.4	&	0.35	$\pm$	0.02	&	25.3&0.59 $\pm$ 0.03	&	0.18	$\pm$	0.07	&	0.03	$\pm$	0.31	\\
J1303+52	&	0.51	$\times$	0.42	&	77.8	&	7.15	$\pm$	0.05	&	25.8	&	0.31	$\times$	0.25	&	80.1	&	7.65	$\pm$	0.03	&	21.9&0.09 $\pm$ 0.00	&	-0.32	$\pm$	0.07	&	-0.20	$\pm$	0.02	\\
J1304+49	&	0.47	$\times$	0.35	&	80.7	&	$<$0.06			&	21.6	&	0.34	$\times$	0.26	&	85.6	&	$<$0.07			&	22.3&$-$	&	$-$&	$-$		\\
J1304+55	&	0.41	$\times$	0.35	&	55.5	&	0.66	$\pm$	0.02	&	21.6	&	0.29	$\times$	0.26	&	70.1	&	0.52	$\pm$	0.01	&	11.6&0.50 $\pm$ 0.01	&	-0.37	$\pm$	0.02	&	0.71	$\pm$	0.11	\\
J1305+54	&	0.44	$\times$	0.34	&	70.9	&	18.12	$\pm$	0.05	&	30.2	&	0.32	$\times$	0.26	&	79.2	&	15.70	$\pm$	0.05	&	20.2&0.08 $\pm$ 0.00	&	0.19	$\pm$	0.03	&	0.43	$\pm$	0.01	\\
J1310+54	&	0.46	$\times$	0.36	&	73.4	&	1.30	$\pm$	0.05	&	29.3	&	0.32	$\times$	0.25	&	77.9	&	1.39	$\pm$	0.08	&	23.5&0.20 $\pm$ 0.02	&	0.00	$\pm$	0.03	&	-0.21	$\pm$	0.21	\\
J1310+54	&	0.43	$\times$	0.37	&	56.0	&	1.75	$\pm$	0.21	&	29.2	&	0.30	$\times$	0.26	&	78.4	&	1.44	$\pm$	0.27	&	25.3&0.33 $\pm$ 0.01	&	0.06	$\pm$	0.02	&	0.58	$\pm$	0.66	\\
J1327+54	&	0.47	$\times$	0.40	&	76.5	&	1.77	$\pm$	0.04	&	17.8	&	0.39	$\times$	0.27	&	71.4	&	1.65	$\pm$	0.08	&	19.9&0.51 $\pm$ 0.01	&	0.43	$\pm$	0.10	&	0.21	$\pm$	0.15	\\
J1334+50	&	0.47	$\times$	0.37	&	83.1	&	1.58	$\pm$	0.09	&	44.8	&	0.33	$\times$	0.26	&	-90.0	&	1.39	$\pm$	0.03	&	16.4&0.19 $\pm$ 0.02	&	0.09	$\pm$	0.18	&	0.38	$\pm$	0.19	\\
J1342+55	&	0.44	$\times$	0.36	&	68.7	&	1.41	$\pm$	0.05	&	24.8	&	0.32	$\times$	0.25	&	82.6	&	1.37	$\pm$	0.04	&	22.6&0.35 $\pm$ 0.01	&	-0.02	$\pm$	0.29	&	0.08	$\pm$	0.14	\\
J1342+48	&	0.46	$\times$	0.36	&	83.4	&	1.32	$\pm$	0.05	&	23.1	&	0.33	$\times$	0.26	&	-85.6	&	1.32	$\pm$	0.07	&	22.5&0.49 $\pm$ 0.01	&	0.80	$\pm$	0.09	&	0.00	$\pm$	0.19	\\
J1343+55	&	0.43	$\times$	0.35	&	67.4	&	0.44	$\pm$	0.04	&	16.8	&	0.31	$\times$	0.26	&	75.8	&	0.33	$\pm$	0.03	&	18.9&0.54 $\pm$ 0.01	&	-0.28	$\pm$	0.02	&	0.91	$\pm$	0.37	\\
J1356+49	&	0.46	$\times$	0.36	&	86.3	&	6.92	$\pm$	0.03	&	17.5	&	0.35	$\times$	0.27	&	-83.6	&	5.98	$\pm$	0.03	&	14.6&-0.16 $\pm$ 0.01	&	0.17	$\pm$	0.04	&	0.43	$\pm$	0.02	\\
J1400+46	&	0.48	$\times$	0.37	&	-87.9	&	1.33	$\pm$	0.03	&	18.6	&	0.36	$\times$	0.26	&	-80.9	&	1.34	$\pm$	0.04	&	27.8&0.30 $\pm$ 0.01	&	-0.50	$\pm$	0.02	&	-0.03	$\pm$	0.12	\\
J1409+53	&	0.45	$\times$	0.35	&	72.7	&	1.44	$\pm$	0.05	&	25.6	&	0.31	$\times$	0.25	&	81.0	&	1.34	$\pm$	0.04	&	21.7&0.77 $\pm$ 0.01	&	0.78	$\pm$	0.10	&	0.21	$\pm$	0.13	\\
J1411+55	&	0.49	$\times$	0.43	&	54.8	&	5.77	$\pm$	0.43	&	16.8	&	0.33	$\times$	0.29	&	70.2	&	4.41	$\pm$	0.33	&	20.3&0.43 $\pm$ 0.01	&	0.31	$\pm$	0.05	&	0.80	$\pm$	0.31	\\
J1411+54	&	0.41	$\times$	0.34	&	59.1	&	9.58	$\pm$	0.05	&	27.6	&	0.29	$\times$	0.24	&	64.0	&	9.82	$\pm$	0.03	&	22.1&-0.24 $\pm$ 0.01	&	-0.39	$\pm$	0.05	&	-0.08	$\pm$	0.02	\\
J1412+49	&	0.54	$\times$	0.43	&	85.9	&	3.00	$\pm$	0.02	&	10.1	&	0.34	$\times$	0.28	&	-89.2	&	2.60	$\pm$	0.02	&	14.9&0.10 $\pm$ 0.01	&	0.15	$\pm$	0.09	&	0.42	$\pm$	0.03	\\
J1414+55	&	0.43	$\times$	0.35	&	53.5	&	1.46	$\pm$	0.09	&	39.4	&	0.30	$\times$	0.25	&	63.8	&	1.30	$\pm$	0.04	&	34.0&0.18 $\pm$ 0.02	&	-1.00	$\pm$	0.05	&	0.34	$\pm$	0.19	\\
J1420+54	&	0.43	$\times$	0.35	&	63.2	&	1.72	$\pm$	0.05	&	26.6	&	0.31	$\times$	0.25	&	69.4	&	1.56	$\pm$	0.05	&	25.6&0.15 $\pm$ 0.01	&	0.08	$\pm$	0.21	&	0.28	$\pm$	0.12	\\
J1421+48	&	0.45	$\times$	0.35	&	85.6	&	24.16	$\pm$	0.05	&	30.1	&	0.33	$\times$	0.25	&	-89.3	&	21.03	$\pm$	0.05	&	21.7&-0.29 $\pm$ 0.01	&	0.03	$\pm$	0.03	&	0.41	$\pm$	0.01	\\
J1422+47	&	0.46	$\times$	0.35	&	86.2	&	0.51	$\pm$	0.06	&	14.6	&	0.33	$\times$	0.25	&	-88.3	&	0.37	$\pm$	0.05	&	21.1&0.54 $\pm$ 0.01	&	1.17	$\pm$	0.02	&	0.95	$\pm$	0.51	\\
J1427+52	&	0.53	$\times$	0.43	&	83.3	&	3.34	$\pm$	0.03	&	16.8	&	0.31	$\times$	0.25	&	80.5	&	3.03	$\pm$	0.02	&	15.5&0.26 $\pm$ 0.01	&	0.05	$\pm$	0.11	&	0.29	$\pm$	0.04	\\
J1429+53	&	0.43	$\times$	0.34	&	68.9	&	1.75	$\pm$	0.10	&	15.0	&	0.31	$\times$	0.25	&	79.1	&	2.06	$\pm$	0.11	&	28.4&0.20 $\pm$ 0.02	&	-0.51	$\pm$	0.04	&	-0.48	$\pm$	0.23	\\
J1433+52	&	0.44	$\times$	0.35	&	75.3	&	1.35	$\pm$	0.04	&	20.7	&	0.32	$\times$	0.25	&	80.8	&	1.45	$\pm$	0.05	&	30.1&0.29 $\pm$ 0.01	&	0.62	$\pm$	0.02	&	-0.19	$\pm$	0.14	\\
J1435+50	&	0.45	$\times$	0.36	&	82.3	&	72.89	$\pm$	0.36	&	24.6	&	0.38	$\times$	0.28	&	-74.7	&	57.08	$\pm$	0.62	&	55.9&0.43 $\pm$ 0.00	&	0.52	$\pm$	0.02	&	0.73	$\pm$	0.04	\\
J1453+54	&	0.48	$\times$	0.34	&	83.4	&	5.07	$\pm$	0.04	&	20.8	&	0.34	$\times$	0.25	&	86.7	&	6.58	$\pm$	0.02	&	19.9&0.01 $\pm$ 0.01	&	-0.19	$\pm$	0.08	&	-0.78	$\pm$	0.02	\\
J1454+49	&	0.48	$\times$	0.38	&	-82.0	&	7.73	$\pm$	0.17	&	27.5	&	0.34	$\times$	0.27	&	-81.0	&	6.98	$\pm$	0.08	&	28.1&-0.06 $\pm$ 0.01 &	0.17	$\pm$	0.04	&	0.30	$\pm$	0.07	\\
J1504+47	&	0.50	$\times$	0.36	&	-88.5	&	2.31	$\pm$	0.07	&	40.9	&	0.35	$\times$	0.25	&	-85.2	&	2.11	$\pm$	0.06	&	24.3&0.61 $\pm$ 0.02	&	1.11	$\pm$	0.08	&	0.28	$\pm$	0.13	\\
         \hline
    \end{tabular}
    \label{tb:appendix1}
    \end{adjustbox}
     {\raggedright \textit{Notes:} Column description: (1) Source name; (2) Restoring beam (arcsec) at 5 GHz; (3) Position angle (PA; degrees) at 5 GHz; (4) Integrated flux density at 5 GHz (mJy); (5) RMS values at 5 GHz ($\mu$Jy beam$^{-1}$); (6) Restoring beam (arcsec) at 7 GHz; (7) Position angle (PA; degrees) at 7 GHz; (8) Integrated flux density at 7 GHz (mJy); (9) RMS values at 7 GHz ($\mu$Jy beam$^{-1}$); (10) Spectral index estimated between 150 MHz and 5 GHz; (11) Spectral index estimated between 1.4 and 5 GHz; (12) Spectral index estimated between 5 and 7 GHz. Upper limits are represented by the `$<$' sign. The `$-$' sign represents null values for the three undetected sources at 1.4, 5, and 7 GHz. \par}
\end{table*}

%%%% Don't change these lines
\bsp	% typesetting comment
\label{lastpage}
\end{document}